\documentclass[twocolumn]{aastex631}

\makeatletter
\let\frontmatter@title@above=\relax
\makeatother

\usepackage{apjfonts}
\usepackage{rotating}
\usepackage{amsmath,color,graphicx,tablefootnote,enumitem}
\usepackage{graphics,subfigure,float,multirow}
\usepackage[rightcaption]{sidecap}

\newcommand{\CH}[1]{\colhead{#1}}
\newcommand{\D}{$^{\dagger}$}

\newcommand{\CM}{\checkmark}
\newcommand{\CC}{{\checkmark\kern-0.6em\checkmark}}
\newcommand{\HC}{{\checkmark}\textsuperscript{\textcolor{black}{\kern-0.60em{\bf$-$}}}}

\newcommand{\W}{$\lambda$}
\usepackage{rotating}
\newcommand{\rot}[1]{\begin{turn}{90}#1\enspace\end{turn}}

\submitjournal{ApJ}  

\shorttitle{AURORA He}
\shortauthors{Berg et al.}

\begin{document}

\shortauthors{Berg et al.}
\title{The AURORA Survey: Robust Helium Abundances at High Redshift Reveal A Subpopulation of Helium-Enhanced Galaxies in the Early Universe}

\author[0000-0002-4153-053X]{Danielle A. Berg}
\affiliation{Department of Astronomy, The University of Texas at Austin, 2515 Speedway, Stop C1400, Austin, TX 78712, USA}
\author[0000-0003-4792-9119]{Ryan L. Sanders}
\affiliation{Department of Physics and Astronomy, University of Kentucky, 505 Rose Street, Lexington, KY 40506, USA}
\author[0000-0003-3509-4855]{Alice E. Shapley}
\affiliation{Department of Physics \& Astronomy, University of California, Los Angeles, 430 Portola Plaza, Los Angeles, CA 90095, USA}
\author[0000-0001-8426-1141]{Michael W. Topping}
\affiliation{Steward Observatory, University of Arizona, 933 N Cherry Avenue, Tucson, AZ 85721, USA}
\author[0000-0001-9687-4973]{Naveen A. Reddy}
\affiliation{Department of Physics \& Astronomy, University of California, Riverside, 900 University Avenue, Riverside, CA 92521, USA}
\author[0000-0003-0605-8732]{Evan D.\ Skillman}
\affiliation{Minnesota Institute for Astrophysics, University of Minnesota, 116 Church St. SE, Minneapolis, MN 55455, USA}
\author[0009-0006-2077-2552]{Erik Aver}
\affiliation{Department of Physics, Gonzaga University, 502 E Boone Ave., Spokane, WA 99258, USA}
\author[0000-0002-3736-476X]{Fergus Cullen}
\affiliation{Institute for Astronomy, University of Edinburgh, Royal Observatory, Edinburgh EH9 3HJ, UK}
\author[0000-0002-7622-0208]{Callum T. Donnan}
\affiliation{NSF's National Optical-Infrared Astronomy Research Laboratory, 950 N. Cherry Ave., Tucson, AZ 85719, USA}
\author[0000-0002-1404-5950]{James S. Dunlop}
\affiliation{Institute for Astronomy, University of Edinburgh, Royal Observatory, Edinburgh EH9 3HJ, UK}
\author[0000-0001-5860-3419]{Tucker Jones}
\affiliation{Department of Physics and Astronomy, University of California, Davis, 1 Shields Avenue, Davis, CA 95616, USA}
\author[0000-0002-0101-336X]{Ali Ahmad Khostovan}
\affiliation{Department of Physics and Astronomy, University of Kentucky, 505 Rose Street, Lexington, KY 40506, USA}
\author[0000-0003-4368-3326]{Derek J. McLeod}
\affiliation{Institute for Astronomy, University of Edinburgh, Royal Observatory, Edinburgh EH9 3HJ, UK}
\author[0000-0002-7064-4309]{Desika Narayanan}
\affil{Department of Astronomy, University of Florida, 211 Bryant Space Sciences Center, Gainesville, FL 32611 USA}
\affil{Cosmic Dawn Center at the Niels Bohr Institute, University of Copenhagen and DTU-Space, Technical University of Denmark}
\author[0000-0001-5851-6649]{Pascal A. Oesch}
\affiliation{Department of Astronomy, University of Geneva, Chemin Pegasi 51, CH-1290 Versoix, Switzerland}
\affiliation{Cosmic Dawn Center (DAWN), Niels Bohr Institute, University of Copenhagen, Jagtvej 128, DK-2200 K\o{}benhavn N, Denmark}
\author[0000-0003-4464-4505]{Anthony J. Pahl}
\altaffiliation{Carnegie Fellow}
\affiliation{The Observatories of the Carnegie Institution for Science, 813 Santa Barbara Street, Pasadena, CA 91101, USA}
\author[0000-0002-5139-4359]{Max Pettini}
\affiliation{Institute of Astronomy, University of Cambridge, Madingley Road, Cambridge, CB3 0HA, UK}
\author[0000-0003-4264-3381]{N. M. F\"{o}rster Schreiber}
\affiliation{Max-Planck-Institut für extraterrestrische Physik (MPE), Giessenbachstr.1, D-85748 Garching, Germany}
\author[0000-0001-6106-5172]{Daniel P. Stark}
\affiliation{Department of Astronomy, University of California, Berkeley, Berkeley, CA 94720, USA}


\begin{abstract}
We present the first robust helium (He) abundance measurements in star-forming galaxies 
at redshifts $1.6\lesssim z\lesssim 3.3$ using deep, moderate-resolution JWST/NIRSpec spectroscopy
 from the AURORA survey. 
We establish a High$-z$ He Sample consisting of 20 galaxies with multiple high-S/N ($>5\sigma$) 
\ion{He}{1} emission-line detections, including the critical near-infrared \W10833 line.
This is the first study at high redshift leveraging \W10833 to break degeneracies 
between temperature, electron density, optical depth, and He$^+$/H$^+$, enabling reliable 
He abundance determinations in the early universe. 
We use a custom MCMC framework incorporating direct-method electron temperature 
priors, extended optical depth ($\tau_{\lambda3890}$) model grids up to
densities of $10^6$~cm$^{-3}$, and simultaneous fits of the physical conditions
and \ion{He}{1}/\ion{H}{1} line ratios to derive ionic He$^+$/H$^+$ abundances. 
Most of the AURORA galaxies follow the extrapolated $z\sim0$ He/H--O/H trend,
indicating modest He enrichment by $z\sim2-3$.
However, we identify a subpopulation of four galaxies that exhibit elevated He mass fractions 
($\Delta Y>0.03$) without corresponding enhancements in N/O or $\alpha$-elements ($\sim20$\%\ of the sample). 
This abundance pattern is inconsistent with enrichment from asymptotic giant branch stars,
but favors early He enrichment from very massive stars (VMSs; $M\gtrsim100\ M_\odot$), 
which can eject He-rich, N-poor material via stellar winds and binary stripping in young stellar populations.
We speculate that these elevated-He systems may represent an early phase of globular cluster (GC) formation 
where N enrichment is still lagging behind He production. 
This work demonstrates the power of JWST multi-line \ion{He}{1} spectroscopy for 
tracing early stellar feedback, enrichment pathways, and GC progenitor signatures 
in the high-redshift universe.
\end{abstract}



\section{Introduction}\label{sec:intro}

Historically, helium (He) abundance studies have concentrated primarily
on the determination of the primordial He abundance ($Y_p$) and
the post-Big Bang chemical enrichment of helium ($\Delta$Y/$\Delta$Z).
The primordial He abundance is a cornerstone of modern cosmology, 
providing one of the most powerful observational tests of standard Big Bang 
nucleosynthesis (SBBN) and setting key constraints on both the cosmic baryon 
density and the number of relativistic species in the early universe 
\citep[e.g.,][]{walker91,steigman07,cyburt16, fields20}. 
In the local universe, the primordial He mass fraction ($Y_p$) has been well
constrained to within a few percent using precise measurements of He 
abundances from emission lines in metal-poor \ion{H}{2} regions
\citep[e.g., $Y_{\rm p}=0.2448\pm0.0033$,][]{aver22}.
This value agrees with predictions from the combination of \citet{planck20} 
cosmic microwave background (CMB) observations and SBBN theory. 

Constraints on $Y_{\rm p}$ can come directly from CMB observations alone 
\citep[e.g.,][]{komatsu11, planck14, planck20}, but, to date,
these constraints have not been competitive.
\citet{cooke18} pioneered the study of \ion{He}{1} absorption 
lines from nearly-pristine intergalactic clouds at $z\approx1.7$ to 
determine a consistent primordial He abundance (with large uncertainties).
However, the required alignment of bright background sources with pristine gas
clouds makes absorption line studies challenging.
Thus, nearly all He  
abundance measurements to date have been confined to emission-line studies from nearby galaxies.
As a result, He abundances in the high-redshift universe and, subsequently,
the observed redshift evolution of He abundances, are almost completely unexplored 
in this context.

As a result of efforts to measure $Y_p$, there exist a wealth of He abundance 
studies relying on observations of low-metallicity dwarf galaxies that benefit 
from bright \ion{He}{1} recombination lines, minimal chemical enrichment effects, 
and well-constrained physical conditions. 
The pioneering work by \citet{peimbert74} was followed by many studies that 
successively improved upon the sample size, the quality of the spectra, 
and the sophistication of the analysis
\citep[e.g.,][]{pagel92, izotov98, olive04, peimbert07}. 
Because of the large enrichment of He by the Big Bang and the
relatively small degree of enrichment since 
\citep[e.g., the initial solar He abundance is only $\approx$ 10\% greater than the primordial abundance;][]{serenelli10, magg22}, 
for measurements of He to have meaning, the uncertainty must be reduced to 
the level of a few percent.
Thus, in pursuit of a precise measurement of $Y_p$, various researchers have 
established the observational framework for deriving He abundances from optical 
\ion{He}{1} emission lines, using careful corrections for reddening, collisional 
excitation, underlying stellar absorption, and optical depth effects. 
While these corrections are small, typically a few percent or less, they are all 
critical for making a meaningful measurement.

In the last decade, there have been several improvements in the treatment of 
uncertainties and parameter degeneracies in He/H abundance determinations.
For one, modern studies now include multiple \ion{He}{1} emission lines rather 
than relying on just one or two, allowing for better constraints on the physical 
conditions in the ionized gas (e.g., $T_e$, $n_e$, and optical depth ($\tau_{\lambda3890}$)).
Second, improved \ion{He}{1} emissivities incorporate collisional excitation, 
recombination, and radiative transfer effects and include the rest-frame 
near-infrared (near-IR) \W10833 line \citep[e.g.,][]{porter05,porter09,porter12,aver13}.
The addition of the \W10833 line to He abundance determinations helps 
break the primary temperature/density degeneracy and/or the secondary optical 
depth/density degeneracy \citep[e.g.,][]{izotov14,aver15,hsyu20}.
Improvements have also been made to stellar absorption accounting, 
such as modeling underlying absorption explicitly using stellar population synthesis 
\citep[e.g.,][]{gonzalez-delgado05, aver21}.
Finally, sophisticated statistical approaches have been incorporated, 
including Markov Chain Monte Carlo (MCMC) and Bayesian methods \citep[e.g.,][]{aver11,fernandez19,hsyu20}.

It is important to extend He/H measurements in galaxies to high redshifts
because it allows for a direct test of the universality of SBBN predictions 
across cosmic time and environment, probing whether early chemical enrichment 
and stellar populations evolved as expected, and provides an empirical benchmark 
on the buildup of He from stellar nucleosynthesis at an epoch near the peak 
of cosmic star formation \citep[$z\sim2$;][]{madau14}. 
Current models predict that He/H will only enrich mildly beyond the primordial 
value, with the He mass fraction ($Y$) increasing linearly with metallicity ($Z$)
as $\Delta Y/\Delta Z\sim1-3$, depending on the assumptions for stellar yields, 
gas inflows, and outflows \citep[e.g.,][]{peimbert07,matteucci85,romano10}. 
However, observations of local globular clusters, which contain some of the oldest 
stellar systems in the Universe, show anomalously high He abundances in their 
second-generation populations \citep[$Y\gtrsim0.3$; e.g.,][]{piotto07,milone18} 
that must be `baked in' during early enrichment episodes.
As a result, the unique abundance templates of globular clusters raise fundamental 
questions about the origin and timing of such early enrichment. 
Therefore, measuring He abundances at $z\sim 2-3$ offers a rare opportunity to test 
chemical evolution models in the early universe and to investigate whether the physical 
processes responsible for He enhancement in metal-rich globular clusters were already in place 
during the peak epoch of star formation ($z\sim2-3$), when globular cluster formation 
may have been most active \citep[e.g.,][]{elbadry19,kruijssen19}.

Despite its importance, measuring the He/H abundance ratio in high-redshift galaxies 
has remained out of reach until recently due to a combination of observational and 
methodological challenges, including the faintness of \ion{He}{1} emission lines, 
the need for precise reddening corrections, and the complexities of accounting for 
electron temperature, density, and optical depth effects for distant, unresolved 
sources. 
Fortunately, JWST's unprecedented combination of sensitivity, spectral resolution, 
and wavelength coverage in the near-IR has enabled a new era of nebular abundance 
studies in galaxies across cosmic time.
Recently, \citet{yanagisawa24} presented the first high-redshift emission-line 
measurements of He/H using NIRSpec spectroscopy of three lensed $z\simeq 6$ galaxies. 
This study marked a pioneering step, but the analysis was based on a 
limited set of \ion{He}{1} emission lines and lacked coverage of the critical 
rest-frame near-IR \W10833 line due to the high redshifts of the sources in the sample. 
As a result, the He/H determinations were subject to strong degeneracies between 
electron density, optical depth, and He$^+$ abundance. 
Other JWST spectra have revealed detections of the \W10833 line for 
moderate-redshift galaxies, such as the 23 $z=1.4-4.0$ AURORA galaxies 
in \citet{shapley25a} and seven $z=1.16-2.74$ galaxies from
SMACS0723 ERO (PID 2736) identified in \citet{carnall23} and \citet{brinchmann23},
opening a new path for robust He abundance determinations at $z>1$.

Building on these JWST \ion{He}{1} studies, the AURORA survey enables the next major 
advance in high-redshift He/H abundances by incorporating \ion{He}{1} \W10833.
Here we present the first robust determination of He/H abundances in a sample of 
galaxies at $1.6\lesssim z\lesssim 3.3$, using deep rest-frame optical and near-IR 
spectroscopy from the AURORA survey.
We apply a state-of-the-art MCMC analysis that simultaneously fits multiple 
\ion{He}{1} and \ion{H}{1} recombination lines and uses expanded optical depth grids 
to account for possible density evolution.
This allows us to self-consistently account for variations in underlying physical 
conditions and observational uncertainties, breaking the longstanding 
degeneracies between $n_e$--$T_e$ and $n_e$--$\tau_{\lambda3890}$. 
This analysis not only opens a new observational window onto the He enrichment 
of galaxies in the early universe but also establishes a critical framework for 
testing the predictions of cosmological and chemical evolution models across cosmic time.

The remainder of this paper is organized as follows.
The AURORA High$-z$ He Sample is described in Section~\ref{sec:sample},
along with a brief description of the observations and emission line fits.
The MCMC He abundance code is presented in Section~\ref{sec:HeH},
along with discussion of the \ion{He}{1} line sensitivities in \S~\ref{sec:sens},
details of the fitting framework and derived He$^+$/H$^+$ abundances in \S~\ref{sec:Framework},
and the total He/H abundance calculation in \S~\ref{sec:tot}.
For completeness, we also present several tests of the MCMC modeling in 
Section~\ref{sec:tests} of the Appendix.
We present the He abundance results and compare to the local He/H--O/H trend in 
Section~\ref{sec:HeEvol} and discuss the evolution of He/H abundances in 
Section~\ref{sec:enrichment}.
Finally, we provide recommendations for measuring He/H abundances 
at high redshifts based on our lessons learned in Section~\ref{sec:best} and 
present our conclusions in Section~\ref{sec:conclusions}.
The AURORA survey adopts cosmological parameters of $H_0=70$ km s$^{-1}$ Mpc$^{-1}$, 
$\Omega_{\rm m} = 0.30$, and $\Omega_\Lambda = 0.7$, a \citet{chabrier03} 
initial mass function (IMF), and a solar oxygen abundance of $12+\log({\rm O/H})=8.69$
from \citet{asplund21}.
\vspace{5ex}

\section{AURORA Observations\label{sec:sample}}

\subsection{JWST/NIRSpec Observations and Data Reduction}\label{sec:obs}
The Assembly of Ultradeep Rest-optical Observations Revealing Astrophysics (AURORA) 
survey \citep[PID: 1914][]{shapley25a,shapley25b,sanders24b} is a JWST/NIRSpec Cycle-1 
program designed with the primary goal of detecting faint auroral emission lines and 
characterizing direct chemical abundances for $z\simeq 1.5-4.5$ star-forming galaxies. 
Galaxies satisfying multiple additional science goals were targeted as well, 
spanning $1.5 \leq z \leq 10$ in redshift.
AURORA observations were obtained using the NIRSpec Micro Shutter Assembly (MSA) in two 
fields, COSMOS and GOODS-N, targeting a total of 97 galaxies. 
Observations employed the medium-resolution gratings G140M/F100LP, G235M/F170LP, and 
G395M/F290LP, providing continuous spectral coverage from $1-5\ \mu$m with a 
resolution of $R\approx1000$. 
Integration times of 12.3, 8.0, and 4.2 hours were chosen in the G140M, G235M, and G395M 
gratings, respectively, to achieve a uniform emission-line sensitivity across all three 
gratings, corresponding to a measured 5$\sigma$ limiting line flux of 
$5\times10^{-19}$~erg~s$^{-1}$~cm$^{-2}$. 
A 3-point nod pattern was adopted for each observation, and 3 microshutters were assigned 
for each MSA ``slit".

The two-dimensional (2D) spectra were reduced using a custom pipeline built around 
the standard JWST Science Calibration Pipeline (\texttt{calwebb}), with steps added 
for improved cosmic ray rejection, $1/f$ noise correction using \texttt{nsclean} 
\citep[][]{rauscher24}, optimal 
extraction, and transient artifact masking. 
Flat-fielding, wavelength calibration, and flux calibration were performed using updated 
Calibration Reference Data System (CRDS) reference files. 
Slit-loss corrections were computed by modeling the spatial light distribution of each 
galaxy based on NIRCam (or, in the few cases lacking NIRCam coverage, HST/WFC3-IR) imaging, 
convolved with wavelength-dependent PSFs, and accounting for the microshutter geometry 
following the methodology of \citet{reddy23}.

Flux calibration proceeded in two stages. 
First, grating-to-grating flux density offsets were corrected by matching continuum and 
line fluxes in overlapping spectral regions, adopting G235M as the reference grating. 
Second, absolute calibration was performed for each target by comparing synthetic 
photometry derived from the 1D spectra with wide- and medium-band imaging from JWST NIRCam 
and HST, and scaling all three gratings by the median of the imaging-to-spectrum ratios 
across all available filters that overlapped the NIRSpec coverage \citep[][]{sanders24b}. 
Full details of the observations, sample selection, data reduction, flux 
calibration and emission-line measurements are presented in \citet{shapley25a}, 
\citet{shapley25b}, and \citet{sanders24b}.


\begin{figure*}
\begin{center}
	\includegraphics[width=1.0\linewidth,trim=20mm 5mm 25mm 10mm,clip]{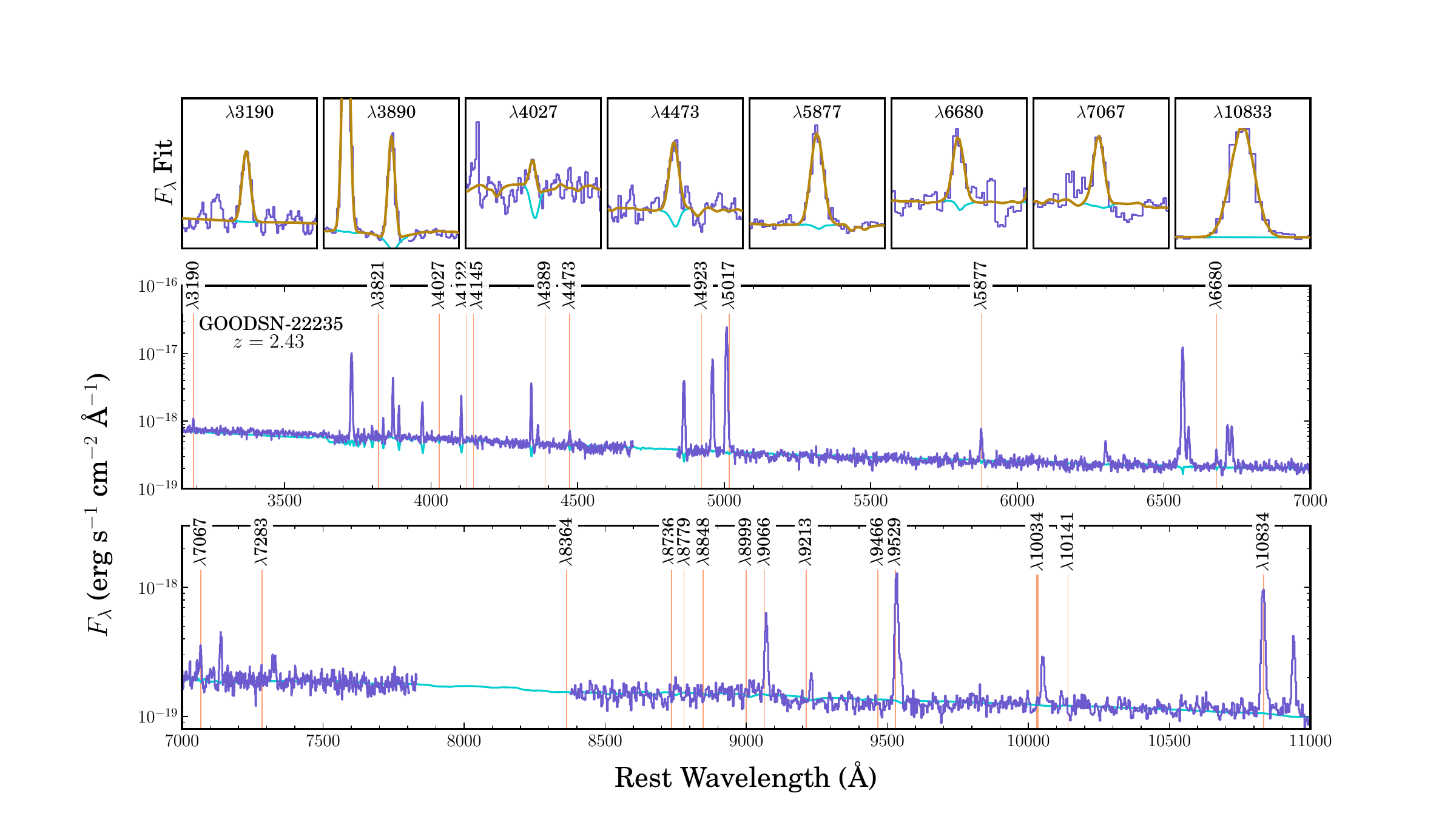}
\caption{
{\it JWST}/NIRSpec rest-frame optical and near-IR spectrum of GOODSN-22235,
one of the 20 galaxies in the High$-z$ He Sample used in this work, 
with at seven significant \ion{He}{1} line detections.
The stellar continuum derived from the SED fitting is plotted in blue.
The orange vertical lines indicate the vacuum-wavelengths of \ion{He}{1} emission lines.
The top row of panels shows zoom-ins on the significant \ion{He}{1} emission-line 
detections used to determine the He/H abundance. \label{fig:spec}}
\end{center}
\end{figure*}

\subsection{Emission Line Measurements}\label{sec:emline}
A detailed description of the emission-line measurements is given in \citet{sanders24b}.
We outline the main points here.
Emission-line fluxes were measured by fitting Gaussian profiles to the fully reduced 
JWST/NIRSpec 1D spectra.
Redshifts and velocity widths were determined from the brightest detected emission line 
for each galaxy, with kinematic constraints applied to all subsequent fits. 

The underlying continuum was modeled as the sum of the best-fit stellar population 
model (from SED fitting) and a nebular continuum component tied to the H$\beta$ intensity, 
both convolved to match the instrumental resolution. 
As described in detail in \citet{shapley25a}, multi-wavelength photometry was
cataloged for all AURORA targets. 
For most galaxies analyzed in this work, HST and JWST photometry was
assembled in the DAWN JWST Archive \citep{heintz2025}. 
For three sources in COSMOS (4156, 5283, 5571) that were not covered by JWST/NIRCam, 
we modeled the photometric SED assembled by the 3DHST survey \citep{skelton2014}. 
We used the \texttt{FAST} program \citep{kriek09} to fit the multi-wavelength optical 
through IR SED for each galaxy, and adopted the flexible stellar population synthesis 
models of \citet{conroy09}. 
A \citet{chabrier03} initial mass function (IMF) was assumed for all galaxies, 
coupled with a ``delayed-$\tau$" star-formation history, and a combination of either 
$1.4\ Z_\odot$ models and the \citet{calzetti00} dust attenuation law or 
$0.27\ Z_\odot$ models with the SMC extinction curve of \citet{gordon03}. 
The dust+metallicity combination was chosen to yield the lowest $\chi^2$.

In the High$-z$ He Sample described below in Section~\ref{sec:hesample}, 
19 galaxies were  modeled assuming 1.4 $Z_{\odot}+$Calzetti, and one 
galaxy with 0.27 $Z_{\odot}$+SMC. 
Before SED fitting, all photometric bands were corrected for the contributions 
from nebular line and continuum emission based on measurements from the spectra, 
as described in \citet{sanders24b}.
Since the stellar continuum modeling and nebular emission-line measurements are 
interdependent, the process of nebular line measurement, photometric correction, 
and SED fitting was performed iteratively until convergence.
This approach to estimating the galaxy continuum emission naturally accounts for 
stellar absorption features, which are critical for accurate Balmer line and 
\ion{He}{1} line measurements. 
Line-flux uncertainties were estimated through Monte Carlo resampling of the 1D 
spectra, according to their error spectra.

The final line catalog contains a multitude of emission lines, 
including \ion{H}{1} Balmer and Paschen series lines and 14 \ion{He}{1} lines
of particular relevance to this study: \W3190, \W3890, \W4027, \W4145, \W4389, 
\W4473, \W4923, \W5017, \W5877, \W6680, \W7067, \W7283, \W10833, and 
\W12788\footnote{Note that all emission line wavelengths are given in vacuum 
following the common practice for spectra observed in space.}.
Upon quality examination, five \ion{He}{1} lines were removed from analysis for 
the purpose of this work.
\ion{He}{1} \W4145 was removed due to a lack of reliable emissivity calculations.
\ion{He}{1} \W5017 was determined to be unreliable due to blending issues
with the wing of the bright [\ion{O}{3}] \W5008 line at the observed resolution. 
Additionally, low-angular-momentum singlet lines like \ion{He}{1} \W5017 and \W7283 
have been called into question due to their possible departures from Case B 
\citep[][]{izotov07,mendez-delgado24}; thus, we do not include \ion{He}{1} \W7283 
in our analysis either.
Additionally, low-angular-momentum singlet lines like \ion{He}{1} \W5017 and 
\W7283 have been called into question due to their possible departures from 
Case B \citep[][]{izotov07,mendez-delgado24}; thus, we do not include 
\ion{He}{1} \W7283 in our analysis either.
Note, however, that for the one galaxy in our sample with a \W7283 detection
the He abundance analysis returned negligible differences when the line was included.
Finally, \ion{He}{1} \W4923 was removed due to challenges in fitting this
weak feature near the strong [\ion{O}{3}] line, as well as \ion{He}{1} 
\W12788 due to its lack of use in other He abundance studies.

An example of the high-quality AURORA spectra used in this work is shown in 
Figure~\ref{fig:spec}.
The spectrum shows the rest-frame wavelength coverage of $\sim3150-11000$~\AA\ for 
GOODSN-22235 at $z=2.43$ that enabled significant detections ($>5\sigma$) of 
seven \ion{He}{1} emission lines, including the infrared \W10833 line.


\begin{figure*}
\begin{center}
	\includegraphics[width=1.0\linewidth]{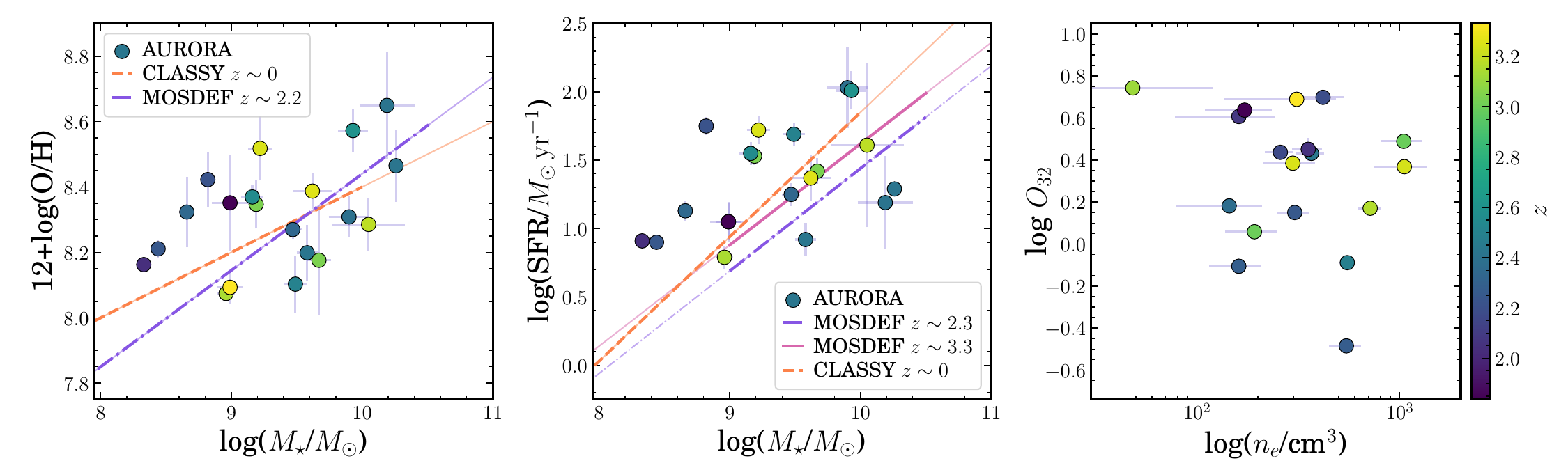}
\caption{Sample properties for the AURORA High$-z$ He Sample.
The left panel shows the mass-metallicity relationship (MZR) for the 20 galaxies 
used in this work, color-coded by redshift.
For this trend, we use direct-method metallicities and stellar masses determined 
from the SED fits.
To compare to a similar sample of direct metallicities at $z\sim2-3$, we plot
the $z\sim2.2$ MZR from the MOSDEF Survey \citep[SFR-corrected trend from][dashed-dotted purple line]{sanders20} 
and the $z\sim0$ MZR of the CLASSY Survey \citep[][dashed orange line]{berg22}, 
which has enhanced SFRs that likens the sample to $z\sim2-3$ galaxies. 
The middle panel shows H$\alpha$-derived SFR versus stellar mass relationship compared 
to both the CLASSY Survey and the $z\sim2.3$ (dashed-dotted purple line) and $z\sim3.3$ (pink 
line) trends from the MOSDEF Survey \citep{sanders21}.
The right panel shows the ionization, represented by $O_{32}=I_{\lambda5008}/I_{\lambda3728}$,
versus the nebular electron density determined from the low-ionization [\ion{S}{2}] ratio.
\label{fig:samp_props}}
\end{center}
\end{figure*}

\subsection{Dust reddening, temperature, density, and abundances}
Nebular dust reddening ($E(B-V)$), electron temperatures ($T_e$), electron densities ($n_e$),
and direct-method oxygen abundances were obtained using the methods described fully in 
\citet{sanders25}, which we briefly outline here.
$E(B-V)$, $T_e$, and $n_e$ were iteratively determined until convergence among all 
three parameters was reached.
All Balmer and Paschen series \ion{H}{1} recombination lines detected at $S/N\ge5$ were 
simultaneously fitted to determine $E(B-V)$, assuming the \citet{cardelli89} extinction law 
and intrinsic \ion{H}{1} intensity ratios relative to H$\beta$ calculated with \texttt{pyneb} 
\citep{luridiana15} at the derived $T_e$ and $n_e$ for each object.

The high-ionization $T_e$([\ion{O}{3}]) was calculated from the dust-corrected 
[\ion{O}{3}] \W4363/\W5007 ratio, assuming collision strengths from \citep{aggarwal99}.
The low-ionization $T_e$([\ion{O}{2}]) was derived from the dust-corrected 
[\ion{O}{2}] \W\W7320,7330/[\ion{O}{2}] \W\W3726,3729 ratio, adopting the collision 
strengths of \citet{kisielius2009}.
The dust-corrected ratio [\ion{S}{3}] \W6314/\W9533 was used to infer $T_e$([\ion{S}{3}]) 
with the collision strengths of \citet{hudson2012}.
Of the 20 AURORA galaxies that are the focus of this study (see Sec.~\ref{sec:hesample}), 
14 have measurements of both $T_e$([\ion{O}{3}]) and $T_e$([\ion{O}{2}]); 
4 have $T_e$([\ion{O}{2}]) but lack a [\ion{O}{3}] \W4363 detection; and 
2 have $T_e$([\ion{O}{3}]) but lack a [\ion{O}{2}] \W\W7320,7330 detection.
For galaxies lacking a $T_e$ constraint in one of the ionic zones, the missing $T_e$ 
was inferred using the $T_e$([\ion{O}{3}])--$T_e$([\ion{O}{2}]) relation of \citet{campbell86}.
Electron density was derived from the [\ion{S}{2}]\W6716/\W6731 ratio using the 
collision strengths of \citet{tayal2010}.

Ionic abundance ratios were computed using \texttt{pyneb}.
The O$^{+2}$/H abundance ratio was determined from the measured [\ion{O}{3}] \W5007 ratio 
using $T_e$([\ion{O}{3}]), while O$^+$/H was derived from the dust-corrected 
[\ion{O}{2}] \W\W3726,3729/H$\beta$ ratio using $T_e$([\ion{O}{2}]).
The total oxygen abundance is taken to be the sum of O$^+$/H and O$^{+2}$/H.
The N$^+$/O$^+$ abundance ratio was inferred from the dust-corrected 
[\ion{N}{2}] \W6584/[\ion{O}{2}] \W\W3726,3729 ratio using $T_e$([\ion{O}{2}]) with the 
N$^+$ collision strengths of \citet{tayal11}.
Total N/O was inferred using an ionization correction factor (ICF) determined by fitting a 
grid of \texttt{cloudy} photoionization models \citep{ferland17} to the strong 
rest-optical line ratios for each source, and using the ionic population fractions of 
N and O in the best-fit model (Sanders et al., in prep.).
Because of the similar ionization potentials of N$^+$ and O$^+$, the ICFs 
were small ($<0.1$~dex) for all objects included in this analysis.
The resulting O/H and N/O abundances are listed in Table~\ref{tbl1}.

\subsection{High--{\it z} He Sample}\label{sec:hesample}
We selected AURORA galaxies that have at least 6 significant ($>5 \sigma$) \ion{He}{1} 
line detections, including \ion{He}{1} \W10833 to ensure sufficient density sensitivity.
Of these galaxies, we removed any that are flagged as an AGN due to the detection of 
broad H$\alpha$ emission, resulting in the {\it High$-z$ He Sample} of 20 galaxies.
The entire High$-z$ He Sample has detections ($>3 \sigma$) of at least one auroral 
line (i.e., [\ion{O}{3} \W4364, [\ion{O}{2}] \W\W7320,7330, or [\ion{S}{3}] \W6314), 
enabling direct-method metallicity determinations.
The entire sample also has $>5 \sigma$ [\ion{N}{2}] \W6585 detections, enabling analysis 
of the relative N/O abundance in the low-ionization gas and any trends with He/H abundances.
The properties of the High$-z$ He Sample are listed in Table~\ref{tbl1} and shown in 
Figure~\ref{fig:samp_props}.

The High$-z$ He Sample spans a redshift range of $1.6\lesssim z\lesssim3.3$ and consists 
of moderately massive star-forming galaxies ($10^{8.63} < \log(M_\star/M_\odot) < 10^{10.62}$).
The left panel of Figure~\ref{fig:samp_props} shows the mass-metallicity relationship 
(MZR) for the High$-z$ He Sample using direct-method gas-phase abundances and stellar 
masses determined from SED fitting from \citet{sanders25}.
We note that there is a general dearth of direct-method MZRs at $z>1$ with the exception 
of the MOSFIRE Deep Evolution Field (MOSDEF) survey.
For comparison, we plot the MOSDEF $z\sim2.2$ direct-method MZR from \citet{sanders20}.
We used the SFR-corrected version \citep[Equation 10 in ][]{sanders20} to account for 
the fact that the subsample of MOSDEF galaxies with direct metallicities has elevated 
SFRs compared to the full $z\sim2.2$ population; note that the SFR-corrected direct-method 
MZR is consistent with the $z\sim2.2$ MOSDEF strong-line MZR from \citet{sanders21}.
We also plot the $z\sim0$ direct-method MZR from the COS Legacy Archive Spectroscopic 
SurveY \citep[CLASSY; ][]{berg22}, which has been likened to $z\sim2-3$ galaxies due 
to its similarly high SFRs.
This shows that the galaxies in the High$-z$ He Sample have typical enrichment levels for their stellar 
masses at $z\sim2-3$.

The middle panel of Figure~\ref{fig:samp_props} shows the star-forming main sequence for 
High$-z$ He Sample. 
Instantaneous SFRs were estimated using the dust-corrected H$\alpha$ luminosities with 
the conversion from \citet{hao11}, adjusted to the \citeauthor{chabrier03} IMF 
(${\rm SFR} (M_{\odot} {\rm yr}^{-1})= 10^{-41.33} L({\rm H}\alpha) ({\rm erg }{\rm s}^{-1})$).
For comparison, we also plot the CLASSY sample and the $z\sim2.3$ and $z\sim3.3$ MOSDEF 
trends reported in \citet{sanders21}.
The CLASSY sample has enhanced star-formation rates relative to typical $z\sim0$ galaxies, 
agreeing well with the $z\sim3.3$ trend from MOSDEF.
The High$-z$ He Sample, while spanning a range of $1.6\lesssim z\lesssim3.3$, is in closer 
agreement with the $z\sim3.3$ trend than the $z\sim2.3$ trend. 
Thus, the sample generally has slightly elevated star-formation rates for $z\sim2-3$ 
galaxies, indicative of bursty star formation.

The right panel of Figure~\ref{fig:samp_props} shows that gas ionization, as characterized 
by the O$_{32}=$ [\ion{O}{3}] \W5008/[\ion{O}{2}] \W3728 emission line ratio, versus the 
ionized gas density, determined from the low-ionization [\ion{S}{2}] \W6718/\W6733 emission-line ratio.
The High$-z$ He Sample probes a broad range of ionization parameter, but a fairly narrow 
range of densities of $10^2\lesssim n_e\ ({\rm cm}^{-3})\lesssim 10^3$.
Local star-forming galaxies typically have average (low-ionization) gas densities near 
the low density limit ($\sim100$ cm$^{-3}$), while galaxies near cosmic noon have elevated 
average densities in comparison \citep[e.g.,][median $n_e\sim250\ {\rm cm}^{-3}$]{sanders16}.
Consistent with this result, the High$-z$ He Sample has a median low-ionization gas density 
of $n_e\sim 300\pm60\ {\rm cm}^{-3}$.
 

\begin{deluxetable*}{rlcccccc}
\tabletypesize{\small}
\tablecaption{AURORA He Emission Line Sample}
\tablehead{
\CH{}    & \CH{}      & \CH{}           & \CH{}    & \CH{log $M_\star$} & \CH{log SFR} & \CH{12+} & \CH{}\\ [-2ex]
\CH{\#}  & \CH{Name}  & \CH{R.A., Decl.} & \CH{$z$} & \CH{($M_\odot$)}  & \CH{($M_\odot$ yr$^{-1}$)} & \CH{log(O/H)} & \CH{log(N/O)}}
\startdata 
\ \ 1. & COSMOS-8442   & 10:00:32.500, +02:15:55.361 & 1.605 &  8.99$\pm$0.14 & 1.05$\pm$0.14 & 8.35$\pm$0.15 & $\ldots$    \\
\ \ 2. & COSMOS-4205   & 10:00:33.581, +02:13:12.724 & 1.837 &  8.22$\pm$0.01 & 0.91$\pm$0.05 & 8.16$\pm$0.02 & $-1.43\pm0.04$ \\
\ \ 3. & GOODSN-25004  & 12:36:46.168, +62:15:51.650 & 2.049 &  8.82$\pm$0.01 & 1.75$\pm$0.06 & 8.42$\pm$0.08 & $-1.21\pm0.03$ \\
\ \ 4. & COSMOS-4029   & 10:00:36.665, +02:13:07.413 & 2.076 &  8.44$\pm$0.04 & 0.90$\pm$0.05 & 8.21$\pm$0.03 & $-1.55\pm0.05$ \\
\ \ 5. & COSMOS-5283   & 10:00:44.795, +02:13:55.049 & 2.174 &  9.47$\pm$0.07 & 1.25$\pm$0.08 & 8.27$\pm$0.03 & $-1.42\pm0.03$ \\
\ \ 6. & COSMOS-4156   & 10:00:43.035, +02:13:11.155 & 2.190 &  9.01$\pm$0.19 & 0.71$\pm$0.13 & 8.32$\pm$0.11 & $-1.72\pm0.06$ \\  
\ \ 7. & GOODSN-30053  & 12:36:47.409, +62:17:28.750 & 2.245 &  9.90$\pm$0.15 & 2.03$\pm$0.30 & 8.31$\pm$0.06 & $-1.07\pm0.04$ \\
\ \ 8. & GOODSN-27876  & 12:36:29.706, +62:16:45.304 & 2.271 & 10.19$\pm$0.21 & 1.19$\pm$0.34 & 8.65$\pm$0.16 & $-0.90\pm0.06$ \\
\ \ 9. & COSMOS-5571   & 10:00:45.911, +02:14:09.294 & 2.278 & 10.13$\pm$0.28 & 1.69$\pm$0.70 & 8.46$\pm$0.11 & $-1.08\pm0.04$ \\
 10.   & GOODSN-19067  & 12:36:43.550, +62:14:08.988 & 2.281 &  9.58$\pm$0.08 & 0.92$\pm$0.12 & 8.20$\pm$0.09 & $-1.30\pm0.05$ \\
 11.   & GOODSN-21522  & 12:36:48.900, +62:14:51.189 & 2.363 &  9.49$\pm$0.08 & 1.69$\pm$0.08 & 8.10$\pm$0.09 & $-1.30\pm0.06$ \\
 12.   & GOODSN-22235  & 12:36:33.414, +62:15:04.790 & 2.430 &  9.16$\pm$0.08 & 1.55$\pm$0.09 & 8.37$\pm$0.04 & $-1.57\pm0.03$ \\
 13.   & GOODSN-30564  & 12:36:48.978, +62:17:39.629 & 2.483 &  9.93$\pm$0.12 & 2.01$\pm$0.14 & 8.57$\pm$0.07 & $-0.94\pm0.03$ \\
 14.   & GOODSN-19848  & 12:36:38.511, +62:14:21.935 & 2.992 &  9.19$\pm$0.04 & 1.53$\pm$0.01 & 8.35$\pm$0.07 & $-1.10\pm0.06$ \\
 15.   & GOODSN-22384  & 12:36:38.757, +62:15:07.255 & 2.993 &  9.67$\pm$0.09 & 1.42$\pm$0.09 & 8.18$\pm$0.17 & $-0.99\pm0.06$ \\
 16.   & GOODSN-21033  & 12:36:50.784, +62:14:44.566 & 3.112 &  8.96$\pm$0.06 & 0.79$\pm$0.09 & 8.07$\pm$0.02 & $-1.10\pm0.05$ \\
 17.   & COSMOS-4740   & 10:00:38.107, +02:13:33.824 & 3.155 & 10.05$\pm$0.28 & 1.61$\pm$0.60 & 8.29$\pm$0.08 & $-1.09\pm0.05$ \\
 18.   & GOODSN-28209  & 12:36:45.222, +62:16:52.249 & 3.233 &  9.22$\pm$0.09 & 1.72$\pm$0.10 & 8.52$\pm$0.10 & $-1.24\pm0.08$ \\
 19.   & COSMOS-8363   & 10:00:32.740, +02:15:52.863 & 3.248 &  9.62$\pm$0.15 & 1.37$\pm$0.17 & 8.39$\pm$0.05 & $-1.10\pm0.05$ \\
 20.   & GOODSN-22932  & 12:36:51.842, +62:15:15.704 & 3.331 &  8.99$\pm$0.09 & 1.05$\pm$0.13 & 8.09$\pm$0.05 & $-1.38\pm0.04$ 
\enddata	
\tablecomments{
Properties of the AURORA He emission-line sample, in order of increasing redshift: galaxies with several significant \ion{He}{1}
lines, including the near-IR \W10833 line.
Columns 1 and 2 list the sample numbers and names of the galaxies in the High$-z$ He Sample, ordered by
redshift.
The corresponding coordinates and redshifts are listed in Columns 3 and 4.
The stellar masses listed in Column 5 were derived from SED fitting using EASY,
while the star-formation rates in Column 6 were determined from the H$\alpha$ fluxes.
Columns 7--8 lists the metallicity of the galaxy, as characterized by the direct-method gas-phase 
oxygen abundance, and the relative N/O abundance derived from the emission-line fits reported in \citet{sanders25}.}
\label{tbl1}
\end{deluxetable*}

\section{He Abundance Determinations}\label{sec:HeH}

He abundance determinations from emission-line spectra require careful 
treatment of temperature, density, and radiative transfer effects. 
We tested several fitting strategies (see Sections~\ref{sec:priors}-\ref{sec:10833} 
in the Appendix) before adopting a configuration that maximizes the 
diagnostic power of our high-S/N AURORA spectra while minimizing degeneracies and biases. 
The physical motivation and our adopted methodology are described below. 


\begin{deluxetable*}{cccccccc}
\tabletypesize{\small}
\setlength{\tabcolsep}{3pt}
\tablecaption{\ion{He}{1} Line Properties}
\tablehead{
\CH{}    & \CH{Relative}   & \CH{Term}       & \CH{Spin}         & \CH{Dominant} & \CH{$n_e-$}      & \CH{$T_e-$}      & \CH{RT} \\ [-2ex]
\CH{Line}& \CH{Emissivity} & \CH{Transition} & \CH{Multiplicity} & \CH{Process}  & \CH{sensitivity} & \CH{sensitivity} & \CH{Effects} }
\startdata 
\W3188.669  & 0.97  & $4\ ^3P\rightarrow2\ ^3S$ & Triplet & Recom.        & mod.   & low    & high        \\      
\W3889.749  & 2.44  & $3\ ^3P\rightarrow2\ ^3S$ & Triplet & Recom. + Coll.& mod.   & low    & high        \\      
\W4027.329  & 0.48  & $3\ ^3P\rightarrow2\ ^3S$ & Triplet & Recom. + Coll.& low    & low    & mod.        \\
\W4472.735  & 1.00  & $4\ ^3D\rightarrow2\ ^3P$ & Triplet & Recom.        & low    & low    & mod.        \\      
\W5877.250  & 2.79  & $3\ ^3D\rightarrow2\ ^3P$ & Triplet & Recom.        & mod.   & mod.   & mod.        \\
\W6679.995  & 0.79  & $3\ ^1D\rightarrow2\ ^1P$ & Singlet & Recom.        & mod.   & mod.   & low         \\
\W7067.138  & 0.61  & $3\ ^3S\rightarrow2\ ^3P$ & Triplet & Recom. + Coll.& high   & mod.   & significant \\
\W10833.305 & 10.04 & $2\ ^3P\rightarrow2\ ^3S$ & Triplet & Recom. + Coll.& highest& high   & most sign.
\enddata	

\tablecomments{Properties of the main \ion{He}{1} lines used in this work.
Column 1 lists the vacuum wavelength in \AA. 
Column 2 lists the emissivity, normalized to \W4473, for the median temperature ($T_e = 1.15\times10^4$ K) 
and density ($n_e=3.0\times10^2$ cm$^{-3}$) of the High$-z$ He Sample.
Columns 3 and 4 list the term symbol transitions, $n\ ^{2S+1} L$ where $n$ is the principal 
quantum number, $S$ is the total spin, and $L$ = total orbital angular momentum, and 
spin multiplicity, where $2S+1 = 1$ signifies singlet states and $2S+1 = 3$ for triplet
states.
Dominant emission processes are listed in Column 5, where ``Recom." stands for recombination, 
and ``Coll." stands for collisional excitation.
Each line's sensitivity to density, temperature, and radiative transfer effects 
(i.e., optical depth), are given in Columns 6, 7, and 8, respectively.}
\label{tbl2}
\end{deluxetable*}

\subsection{\texorpdfstring{\ion{He}{1}}{} Line Sensitivities}\label{sec:sens}

\begin{figure*}
\begin{center}
	\includegraphics[width=0.975\linewidth, trim=0 0mm 0 0, clip]{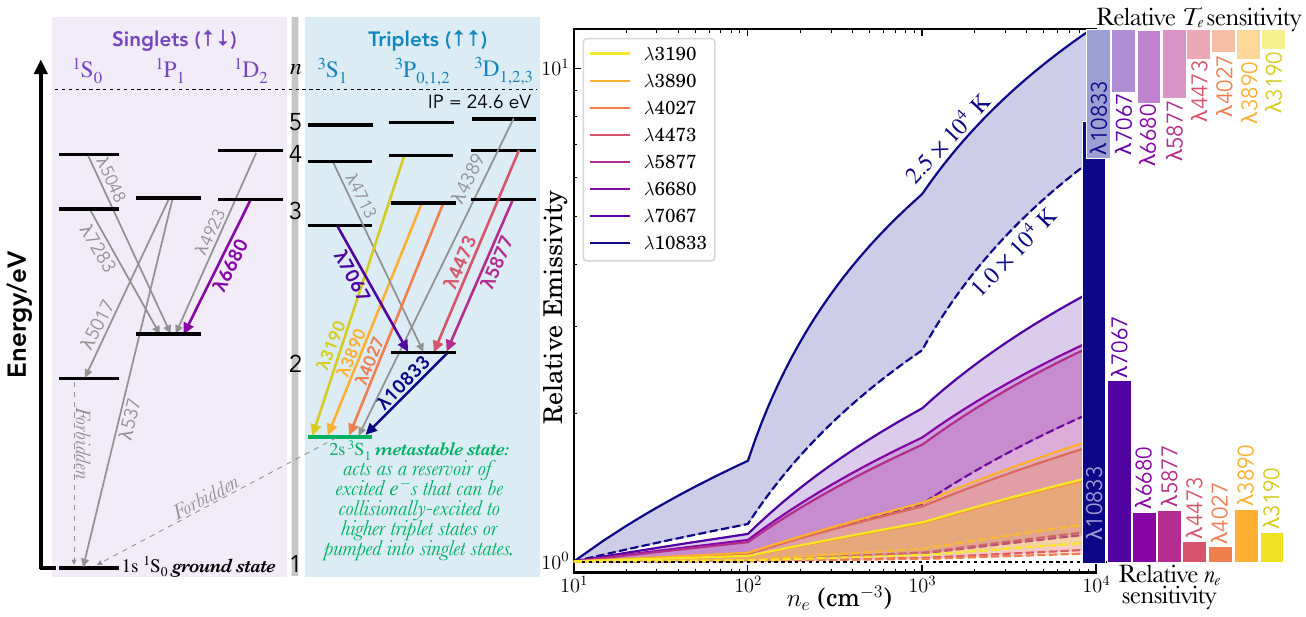}
\caption{
{\it Left:} Partial energy diagram for \ion{He}{1}.
The eight main emission lines used in this work are indicated by colored, bold
transitions and corresponding wavelengths.
The metastable $2\ ^3S$ level (marked green) is an important reservoir of 
electrons for collisional-excitation at higher densities.
{\it Right:} Relative emissivities of the eight \ion{He}{1} lines versus 
electron density.
Each trend is normalized to the value at $n_e=10^1 {\rm\ cm}^{-3}$.
The primary dependence of most lines is density, but the secondary temperature
dependence can also be significant, as shown by the shaded regions spanned by 
assuming a constant temperature of $T_e = 1.0\times10^4$ K (dashed lines) and
$T_e = 2.5\times10^4$ K (solid lines). 
The \ion{He}{1} \W10833 line is by far the most sensitive to both 
density and temperature, so provides important constraints on He abundance fits.
\ion{He}{1} \W7067 is the second most sensitive line to $n_e$, but the other 
lines are difficult to distinguish owing to their similar $T_e$ and $n_e$ 
sensitivities. 
The relative density sensitivities are shown by the bar chart at the bottom 
right, where the bar height corresponds to the maximum change in emissivity 
over $n_e = 10^1 \rightarrow10^4 {\rm\ cm}^{-3}$ for the sample 
median $T_e = 1.15\times10^4$ K.
Similarly, relative temperature sensitivity bars (top right) show the 
maximum change in emissivity over $T_e = 1.10\times10^4 \rightarrow 
2.5\times10^4$ K for the sample median $n_e = 3\times10^2 {\rm\ cm}^{-3}$.
Note that radiative transfer effects are not included here. 
\label{fig:linesens}}
\end{center}
\end{figure*}

The transitions producing the optical and infrared \ion{He}{1} lines used in this 
work are shown in the left-hand panel of Figure~\ref{fig:linesens}, with their 
properties given in Table~\ref{tbl2}.
These transitions arise primarily from He$^{+}$ recombination (Column 5 of Table~\ref{tbl2}), 
followed by a cascade of radiative decays.
For typical nebular conditions, the dominant process is radiative recombination,
which sets up level populations according to the statistical weights. 
Such recombinations produce lines that are relatively insensitive to temperature 
(recombination rate $\propto T_e$) and density, and are optically thin, leading to
minimal radiative transfer effects (i.e., sensitivity to optical depth, $\tau$).
The minimal temperature sensitivity of most \ion{He}{1} lines over the nebular 
temperature range of $1-2.5\times10^4$~K is shown in the right-hand panel of 
Figure~\ref{fig:linesens}.
We compare line strengths relative to the moderately strong \W4473 line for the 
median temperature ($T_e = 1.15\times10^4$ K) and density ($n_e\sim300{\rm\ cm}^{-3}$) 
of the High$-z$ He Sample. 
The \W5877 and \W10833 lines are the strongest, as expected, with line strengths of 
approximately 3$\times$ and $10\times$ the \ion{He}{1} \W4473 flux, respectively. 
The \ion{He}{1} \W3189, \W4027, \W6680, and \W7067 lines have strengths similar to 
\W4473 (48-97\%).

At higher densities and temperatures, collisional effects become important.
The relevance of collisional excitation is determined by populating the $2\ ^3S$ 
metastable level of \ion{He}{1} (radiative decay lifetime $\approx$2~hours), 
which serves as a reservoir of excited electrons
that can be collisionally excited to higher triplet states. 
Because the collisional rates from lower level $i$ to upper level $j$ are 
\begin{align}
    C_{ij} &\propto n_e n_i T_e^{-1/2} e^{-\Delta E_{ij}/kT_e} 
\end{align}
collisional excitation begins to compete with recombination at temperatures of 
$T_e > 1.5\times10^4$ K, significantly, boosting line strengths.
This temperature sensitivity is especially important for primordial He abundances,
where the extremely metal-poor galaxies have high electron temperatures.
The galaxies in the High$-z$ He Sample display [\ion{O}{3}] temperatures of 
$0.73-1.49 \times10^4$ K, suggesting their emission is not dominated by 
collisional excitation, but the temperature-related effects are also not negligible 
for those objects with $T_e > 1.0\times10^4$ K.
Concerning density, recent studies have suggested that star-forming galaxies may evolve to 
higher typical electron densities with increasing redshift \citep[e.g.,][]{topping25b}, 
where contributions from collisional excitation and radiative transfer effects can be strengthened.

Figure~\ref{fig:linesens} shows the density and temperature sensitivity of the 
\ion{He}{1} lines connected to the $2\ ^3S$ metastable state. 
The \W10833 line has the strongest sensitivities of these lines, as it can be excited directly
from the metastable state or from collisional-excitation to higher triplet states
followed by a radiative cascade.
Therefore, it is important to disentangle the temperature-density degeneracy in order to
understand the collisional excitation contributions to \ion{He}{1} emission.

At low densities ($n_e < 10^2$ cm$^{-3}$), the $2\ ^3S$ level is rarely populated, 
and so \ion{He}{1} emission is mainly due to the recombination radiative cascade.
Here we can assume Case B recombination, where the \ion{He}{1} lines in Table~\ref{tbl2} 
are optically thin such that they all behave like singlet transitions.
Any electrons excited to the $2\ ^3S$ level have long radiative lifetimes because decay 
to the $1\ ^1S$ ground state (singlet) is a forbidden electric dipole transition.
Instead, electrons in the $2\ ^3S$ level decay very slowly via forbidden two-photon emission 
or are depopulated by collisions (the probability of which increases with density).

At high densities ($n_e > n_{\rm crit.}\sim A/q \sim10^3$ cm$^{-3}$, where $A$ 
is the Einstein-$A$ coefficient describing the spontaneous radiative decay rate 
and $q$ is the collisional de-excitation coefficient.),
the $2\ ^3S$ level will be substantially populated.
At such densities, the timescale for subsequent collisions becomes comparable 
to or shorter than the spontaneous emission timescale.
The $2\ ^3S$ electrons can then be redistributed into higher triplet levels or 
pump the population into the singlet cascade via collisional coupling, 
enhancing lines like \W7283.
At $n_e>10^4$ cm$^{-3}$, collisional excitation strongly enhances triplet 
and even singlet lines.

For intermediate electron densities between the low-density and high-density 
regimes of He ($n_e\sim10^2-10^4$ cm$^{-3}$), the 
metastable $2\ ^3S$ level becomes significantly 
populated, making \ion{He}{1} triplet lines increasingly 
sensitive to both collisional excitation and radiative 
transfer effects. 
In particular, higher level triplet lines (e.g., \W3890, 
\W7067, and \W10833) will start to be strongly affected by 
collisional excitation from the metastable $2\ ^3S$ level, 
leading to resonant scattering as these transitions become 
optically thick ($\tau \gg 1$).
This alters both the observed line strengths and profile 
shapes (e.g., line broadening), especially in regions with 
modest bulk motions or complex velocity fields.

The radiative transfer of metastable \ion{He}{1} triplet 
lines is complex
\citep[for a detailed description, see Ch. 4.6 in][]{osterbrock06}.
Optically-thick line photons are trapped and scattered 
many times, enhancing excitation of the upper levels via 
resonance fluorescence and pumping.
In turn, multiple scatterings can alter the escape 
probability such that the the net emission of these 
\ion{He}{1} lines is enhanced or decreased from their 
Case B values. 
\citet{benjamin02} performed \ion{He}{1} calculations 
that incorporated the effects of scattering, pumping, 
and level populations and used them to fit radiative 
transfer correction factors that are parameterized as 
\begin{equation}
    f_\tau(\lambda)= 1 + \frac{\tau}{2}[{\rm line-specific\ term}].
\end{equation}
These corrections are functions of the optical depth at 3890 \AA\ 
($\tau_{\lambda3890}$), $n_e$, and $T_e$, and can increase with 
optical depth, particularly in the regime where the $2\ ^3S$ level 
is heavily populated.
Thus, while $\tau_{\lambda 3890}$ and $n_e$ are clearly 
degenerate, the dependence of $\tau_{\lambda 3890}$ on 
$n_e$ is not straightforward, as the relevant density for 
optical depth effects of \ion{He}{1} is the population 
density of the $2\ ^3S$ level, which depends on temperature 
and collisional coupling.
These radiative transfer corrections are critical for 
accurately deriving He$^+$/H$^+$ abundances in moderate- 
to high-density regions, and help break degeneracies 
between $\tau_{\lambda 3890}$ and other physical 
parameters when multiple \ion{He}{1} lines are modeled 
simultaneously.


\subsection{Fitting Framework and Physical Assumptions}\label{sec:Framework}
Our framework for deriving He/H abundance ratios is motivated by 
work performed on local low-metallicity dwarf galaxies constraining 
the primordial He abundance 
\citep[e.g.,][]{olive04,aver12,aver15,aver21,izotov14,hsyu20}.
Modern He abundance analyses have converged on an approach that 
models the \ion{H}{1} and \ion{He}{1} spectrum using 8 free parameters:
electron temperature ($T_e$); electron density ($n_e$); optical depth 
at the \ion{He}{1} \W3890 line center ($\tau_{\lambda 3890}$) 
parameterizing radiative transfer effects; dust reddening ($E(B-V)$); 
underlying stellar absorption for \ion{H}{1} and \ion{He}{1} lines 
($a_{\mathrm{H}}$ and $a_{\mathrm{He}}$); neutral hydrogen fraction 
($\xi$); and the singly-ionized helium abundance by number density 
($y^+=\mbox{He}^+/\mbox{H}^+$).
Best-fit parameters are obtained by fitting this model to the measured 
\ion{He}{1}/H$\beta$ line flux ratios and sampling the posterior 
probability density functions using a Markov Chain Monte Carlo (MCMC) 
approach \citep[e.g.,][]{aver12}.

We employ a simplified version of this model with a reduced number 
of free parameters motivated by the arguments described below.
First, we do not include parameters for stellar \ion{H}{1} and 
\ion{He}{1} absorption because it is already included in the 
continuum model used during emission-line flux measurements 
(see Sec.~\ref{sec:emline}).
Second, we do not include the neutral hydrogen fraction parameter. 
The neutral hydrogen fraction determines the amount of collisional 
excitation contributing to \ion{H}{1} line emission. 
The collisional-to-recombination ratio is a strong function of $T_e$, 
such that below $1.5\times10^4$~K the contribution from collisional 
excitation to Balmer and Paschen \ion{H}{1} lines is negligible 
\citep[$<$1\%;][]{aver21}.
Finally, we do not include $E(B-V)$ as a free parameter in the 
model but instead obtain $E(B-V)$ by simultaneously fitting all 
\ion{H}{1} lines detected at S/N$\ge$5, assuming the 
\citet{cardelli89} extinction curve, and dust correct all 
\ion{He}{1}/H$\beta$ line ratios before fitting.
Our model thus has 4 free parameters 
($T_e$, $n_e$, $\tau_{\lambda 3890}$, and $y^+$).

The intrinsic \ion{He}{1}/H$\beta$ ratios were modeled using
\begin{equation}\label{eq:he}
    \frac{{I_{\rm HeI,\lambda}}}{{I_{\rm H\beta}}}=
    y^+ \frac{{\epsilon_{\rm HeI,\lambda}}(T_e,n_e)}{{\epsilon_{\mathrm{H}\beta}}(T_e,n_e)} 
    f_{\tau,\lambda}(T_e,n_e,\tau_{\lambda 3890})
\end{equation}
where $\epsilon_\lambda$ is the emissivity of the \ion{He}{1} line, 
$\epsilon_{\mathrm{H}\beta}$ is the emissivity of H$\beta$, and 
$f_{\tau,\lambda}$ applies both radiative transfer and collisional 
effects to the \ion{He}{1} lines.
Emissivities of \ion{H}{1} lines were computed by linearly 
interpolating the grid calculated by \citet{storey95}.
\ion{He}{1} emissivities were linearly interpolated from the 
precomputed tables of \citet{porter12} and \citet{porter13}.
Given its importance but very red relative wavelength,
extra care was taken with the \ion{He}{1} \W10833 line.
To ensure the \ion{He}{1} \W10833/H$\beta$ line ratio was accurate, 
we determined the \ion{He}{1} \W10833/Pa$\gamma$ \W10941 ratio and 
scaled it by the theoretical Pa$\gamma$/H$\beta$ emissivity ratio, 
which is well-constrained by the measured electron temperature 
($T_e$) and density ($n_e$).

The $f_{\tau,\lambda}$ term is of particular importance as it 
encapsulates the radiative transfer and collisional effects on 
the \ion{He}{1} triplet transitions, and is a function of 
$T_e$, $n_e$, and $\tau_{\lambda 3890}$.
\citet{benjamin02} computed $f_{\tau,\lambda}$ grids 
(called $f_{\mathrm{line}}$ in that work) and provided coefficients 
for analytic fitting functions for near-UV, optical, and near-IR 
\ion{He}{1} lines (see \citealt{aver15} for an equivalent treatment 
for \W10833).
However, these $f_{\tau,\lambda}$ fitting functions were only 
calibrated over limited temperature, density, and 
$\tau_{\lambda 3890}$ ranges of $1.2\times10^4{\rm\ K}<T_e<2.0\times10^4{\rm\ K}$, 
$1{\rm\ cm}^{-3}<n_e<300{\rm\ cm}^{-3}$, and 
$0\le \tau_{\lambda 3890}\le 2$, appropriate for primordial He 
abundance studies of nearby metal-poor galaxies.
Applications to higher-density high-redshift galaxies with larger 
$\tau_{\lambda 3890}$, or high-metallicity galaxies with 
$T_e<1.2\times10^4{\rm\ K}$, would thus require potentially unreliable 
extrapolations.
Consequently, we employed the same Fortran code created by 
\citet{benjamin02} to recompute $f_{\tau,\lambda}$ for all 
\ion{He}{1} lines used in this study over a 3D grid spanning 
$\log(n_e/{\rm cm}^{-3}) = 0$ to 6 in 0.25~dex steps; 
$T_e = 5,000$ to 20,000~K in 500~K steps; and 
$\tau_{\lambda 3890}=0$ to 15 in steps of 1.
In our model, $f_{\tau,\lambda}$ at a given $T_e$, $n_e$, and 
$\tau_{\lambda 3890}$ is determined by linearly interpolating 
these grids.
These expanded grids enable accurate He abundance determinations 
for the wide range of galaxy properties spanned by high-redshift 
galaxies at $1.5\lesssim z \lesssim3.5$ and beyond.

Using this model, we defined the likelihood function based 
on the $\chi^2$ statistic:
\begin{equation}
    \chi^2 = \sum_\lambda \frac{\left(\frac{{I_{\rm HeI,\lambda}}}{{I_{\rm H\beta}}}_{\rm obs} - 
    \frac{{I_{\rm HeI,\lambda}}}{{I_{\rm H\beta}}}_{\rm mod}\right)^2}{\sigma_\lambda^2}
\end{equation}
where the numerator is the square of the difference between the 
dust-corrected \ion{He}{1}/H$\beta$ intensity ratio from observations 
and the ratio predicted by our model (equation~\ref{eq:he}), 
$\sigma_\lambda$ is the uncertainty on the dust-corrected 
\ion{He}{1}/H$\beta$ intensity ratio, and the sum is over all 
significantly-detected \ion{He}{1} lines for each object, 
listed in Table~\ref{tbl3}.

We performed an initial $\chi^2$ minimization using the \texttt{curve\_fit} 
optimizer from the \texttt{SciPy} package to obtain an initial fit.
We examined the $\chi^2$ values for the set of lines used for each galaxy.
Following the methodology of \citet{hsyu20}, we cut lines that deviate from
the minimization by more than $2\sigma$.
Half of the High$-z$ He Sample required no line cuts, while we cut one 
or two line for the the other half.
The singlet \W6680 line was the most commonly discrepant line, 
followed by \W4473. 
Upon visual examination, the most discrepant cases were due to weak lines
that could be significantly affected by systematic uncertainties such as
uncertainties on the He absorption from the stellar continuum fit, 
scatter in the relative grating to grating flux calibration, 
uncertainties in the wavelength-dependent slit losses, etc.
In particular, no error budget was allotted to the continuum fit, 
likely resulting in underestimated \ion{He}{1} flux uncertainties,
which in turn would drive the $\chi^2$ values higher.
GOODSN-27876 has the largest $\chi^2$ value in the High$-z$ He sample, but 
it also has some of the strongest Balmer absorption features among the the 
AURORA galaxies, suggesting it also has significant \ion{He}{1} 
absorption and associated uncertainties that would significantly reduce
its $\chi^2$ value.

Using the initial fit as a starting point, we employed the \texttt{emcee} 
\texttt{python} package to explore the parameter space using MCMC sampling.
A Gaussian prior was applied to $T_e$ with center and standard deviation 
defined by the measured values and uncertainties of $T_e$([\ion{O}{3}]), 
while flat priors were applied to the other parameters within the 
ranges $1\le \log(n_e/{\rm cm}^{-3})\le 6$, 
$0\le \tau_{\lambda 3890} \le 15$, and $0.05 \le y^+ \le 0.20$.
The effects of different prior assumptions are investigated in 
Appendix~\ref{sec:priors}.
We used 32 walkers, a burn-in period of 500 steps that are discarded, 
and a main run of 500 steps.
The resulting posterior distributions allowed us to investigate 
covariance between parameters.
Best-fit values for each parameter were taken to be the 
50$^{\rm th}$ percentile of the marginalized 1D posterior distributions, 
while the 16$^{\rm th}$ and 84$^{\rm th}$ percentiles were taken to be 
the lower and upper 1$\sigma$ uncertainty bounds, respectively.

We report the initial fit assessment of the $\chi^2$ line minimization and 
the best-fit parameters ($T_e$(\ion{He}{1}), $\log n_e$(\ion{He}{1}),
$\tau_{\lambda3890}$, $y^+$) in Table~\ref{tbl3}.
We used the standard 95\% confidence level (CL) statistical threshold to examine
the initial model match to the suite of \ion{He}{1} emission lines used in each fit.
Comparing the degrees of freedom (number of He/H$\beta$ fluxes minus the four model
parameters) and corresponding 95\% CL values to the $\chi^2$ values, we find that 
the models are well matched to our data for 17 of the galaxies.
The three galaxies with large $\chi^2$ values are marked by a dagger in Table~\ref{tbl3}.
Note that the derived $y^+$ values with and without applying the 2$\sigma$
line cut are consistent within the uncertainty.

Following \citet{aver15}, we flagged any unphysical fit parameters.
One galaxy in the High$-z$ He Sample, COSMOS-4205, has a large optical depths of 
$\tau_{\lambda3890} > 4$, which can lead to resonant broadening of the \ion{He}{1} 
triplet lines relative to the singlet lines.
We inspected the observed spectra of COSMOS-4205 and found all line widths 
to be consistent and well fit by the velocity-constrained emission line model. 
This suggests that the \ion{He}{1} lines for COSMOS-4205 are minimally affected by 
optical depth effects, indicating that the current analysis is valid.


\begin{deluxetable*}{r cc cc cc cc cc cccc r@{}lccc}
\tabletypesize{\small}
\setlength{\tabcolsep}{3pt}
\tablecaption{AURORA He Fitting Results}
\tablehead{
\CH{} & \multicolumn{9}{c}{5$\sigma$ \ion{He}{1} Line Detections} &&  \multicolumn{3}{c}{Initial Fit Assessment} && \multicolumn{5}{c}{Best He Fit Parameters}   \\
\cline{2-10} \cline{12-14} \cline{16-20}
\CH{Name} & \CH{\#} & \rot{\W3190} & \rot{\W3890} & \rot{\W4027} & \rot{\W4473} & \rot{\W5877} & \rot{\W6680} & \rot{\W7067} & \rot{\W10833}
&& \CH{DOF} & \CH{95\% CL} & \CH{$\chi^2$}
&& \multicolumn{2}{c}{$T_e$(\ion{He}{1}) (K)} & \CH{$\log n_e$(\ion{He}{1})} & \CH{$\tau_{\lambda3890}$} & \CH{$y^+$}}
\startdata        
COSMOS-8442   & 6 &     & \CC &     & \CC & \CC & \CC & \CC & \CC && 2 & 5.99 &  2.3 &&  9900&$\pm$900  & 2.29$\pm$0.10 & 1.41$\pm$1.11 & 0.081$\pm$0.004 \\ 
COSMOS-4205   & 6 &     & \CC & \CC & \CC & \CC & \CM & \CC & \CC && 2 & 5.99 &  3.8 && 12800&$\pm$100  & 2.14$\pm$0.03 & 4.36$\pm$0.71 & 0.098$\pm$0.002 \\ 
GOODSN-25004  & 7 & \CC & \CM & \CC & \CC & \CC & \CC & \CC & \CC && 3 & 7.82 &  7.6 && 10700&$\pm$400  & 2.22$\pm$0.04 & 0.26$\pm$0.25 & 0.088$\pm$0.002 \\ 
COSMOS-4029   & 6 &     & \CC & \CC & \CM & \CC & \CC & \CC & \CC && 2 & 5.99 &  5.2 && 12300&$\pm$200  & 2.33$\pm$0.03 & 1.14$\pm$0.62 & 0.088$\pm$0.002 \\ 
COSMOS-5283   & 7 & \CC & \CC & \CC & \CM & \CC & \CC & \CC & \CC && 3 & 7.82 & 11.6$^\dagger$ 
                                                                                     && 11900&$\pm$200  & 2.14$\pm$0.03 & 0.34$\pm$0.29 & 0.088$\pm$0.002 \\ 
COSMOS-4156   & 5 & \CC & \CC & \CM & \CM & \CC & \CM & \CC & \CC && 1 & 3.84 &  4.6 && 14900&$\pm$200  & 2.38$\pm$0.02 & 0.21$\pm$0.21 & 0.078$\pm$0.002 \\ 
\bf{GOODSN-30053}&5&    & \CC & \CC & \CM & \CC & \CM & \CC & \CC && 1 & 3.84 & 0.22 && 11500&$\pm$700  & 1.95$\pm$0.12 & 0.84$\pm$0.47 & 0.112$\pm$0.004 \\ 
GOODSN-27876  & 6 &     & \CC &     & \CC & \CC & \CC & \CC & \CC && 2 & 5.99 & 28.5$^\dagger$ 
                                                                                     &&  8700&$\pm$700  & 2.21$\pm$0.13 & 0.33$\pm$0.39 & 0.100$\pm$0.007 \\ 
\bf{COSMOS-5571}&5&     & \CC &     & \CC & \CC & \CM & \CC & \CC && 1 & 3.84 &  4.8 &&  8700&$\pm$600  & 1.44$\pm$0.58 & 0.48$\pm$0.44 & 0.113$\pm$0.005 \\ 
GOODSN-19067  & 6 & \CC & \CC &     & \CM & \CC &     & \CC & \CC && 1 & 3.84 &  0.7 && 13600&$\pm$1100 & 1.88$\pm$0.21 & 1.19$\pm$0.91 & 0.096$\pm$0.006 \\ 
GOODSN-21522  & 6 &     & \CC &     & \CC & \CC & \CC & \CC & \CC && 2 & 5.99 &  5.8 && 12800&$\pm$600  & 2.18$\pm$0.06 & 2.35$\pm$1.00 & 0.084$\pm$0.004 \\ 
GOODSN-22235  & 6 & \CM & \CC & \CC & \CM & \CC & \CC & \CC & \CC && 2 & 5.99 &  1.2 && 11200&$\pm$200  & 2.33$\pm$0.04 & 3.36$\pm$0.60 & 0.082$\pm$0.002 \\ 
GOODSN-30564  & 7 &     & \CC & \CC & \CC & \CC & \CC & \CC & \CC && 3 & 7.82 &  4.8 &&  8700&$\pm$300  & 2.43$\pm$0.05 & 2.26$\pm$0.62 & 0.108$\pm$0.002 \\ 
GOODSN-19848  & 6 & \CM & \CC & \CC & \CM & \CC & \CC & \CC & \CC && 2 & 5.99 &  10.5$^\dagger$
                                                                                     && 11600&$\pm$300  & 2.65$\pm$0.05 & 0.96$\pm$0.76 & 0.088$\pm$0.003 \\ 
\bf{GOODSN-22384}&6&\CC & \CC &     & \CC & \CC & \CC &     & \CC && 2 & 5.99 &  3.9 && 12800&$\pm$1300 & 2.19$\pm$0.08 & 0.45$\pm$0.52 & 0.102$\pm$0.004 \\ 
GOODSN-21033  & 8 & \CC & \CC & \CC & \CC & \CC & \CC & \CC & \CC && 4 & 9.50 &  9.4 && 12800&$\pm$200  & 2.39$\pm$0.04 & 2.76$\pm$0.85 & 0.082$\pm$0.002 \\ 
\bf{COSMOS-4740}&6&     & \CC &     & \CC & \CC & \CC & \CC & \CC && 2 & 5.99 &  4.3 && 14400&$\pm$900  & 2.15$\pm$0.05 & 0.72$\pm$0.54 & 0.107$\pm$0.003 \\ 
GOODSN-28209  & 7 & \CC & \CC & \CC &     & \CC & \CC & \CC & \CC && 3 & 7.82 &  5.4 &&  9800&$\pm$500  & 2.74$\pm$0.09 & 3.83$\pm$1.42 & 0.096$\pm$0.004 \\ 
COSMOS-8363   & 6 & \CC & \CC &     & \CC & \CC & \CM & \CC & \CC && 2 & 5.99 &  4.1 &&  9600&$\pm$400  & 2.51$\pm$0.07 & 2.94$\pm$1.09 & 0.082$\pm$0.003 \\ 
GOODSN-22932  & 8 & \CC & \CC & \CC & \CC & \CC & \CC & \CC & \CC && 4 & 9.49 &  5.6 && 13300&$\pm$500  & 2.19$\pm$0.06 & 0.82$\pm$0.75 & 0.085$\pm$0.004  
\enddata	
\tablecomments{
Column 2 lists the number of \ion{He}{1} emission lines used in this work that are detected 
at a significance of $5\sigma$.
Columns 3-10 identify the eight \ion{He}{1} lines that are used in the He abundance analysis
and which lines are detected for each galaxy.
A single check-mark denotes a 5$\sigma$ detection, while a second check-mark marks the 
lines that were within 2$\sigma$ of the initial chi-squared minimization and used in the He fit.
Columns 11-13 are used to evaluate the goodness of the initial line fit;
Columns 11 and 12 list the degrees of freedom (DOF) of each fit and the corresponding 
95\% confidence level (CL) and Column 13 lists the chi-squared ($\chi^2$) of the initial 
\ion{He}{1} emission line minimization.
Seventeen galaxies have $\chi^2$ values that are less than or close to the 95\% CL,
indicative of a good fit. 
The final best-fit parameter values from the He fitting code using the measured 
$T_e$([\ion{O}{3}]) values as priors are listed in Columns 14-17:
Columns 14 and 15 list the electron temperature, $T_e$(\ion{He}{1}), in K
and the electron density, $\log n_e$(\ion{He}{1}), in cm$^{-3}$,
Column 16 gives the optical depth characterized for the \ion{He}{1} \W3890
line, $\tau_{\lambda3890}$, and Column 17 lists the final He$^+$/H$^+$ abundance,
$y^+$.\\
$^\dagger$Three galaxies with high $\chi^2$ values correspond to slightly lower
confidence levels. }\label{tbl3}
\end{deluxetable*}


\begin{figure*}
\begin{center}
	\includegraphics[width=0.99\linewidth, trim=0 0mm 0 0, clip]{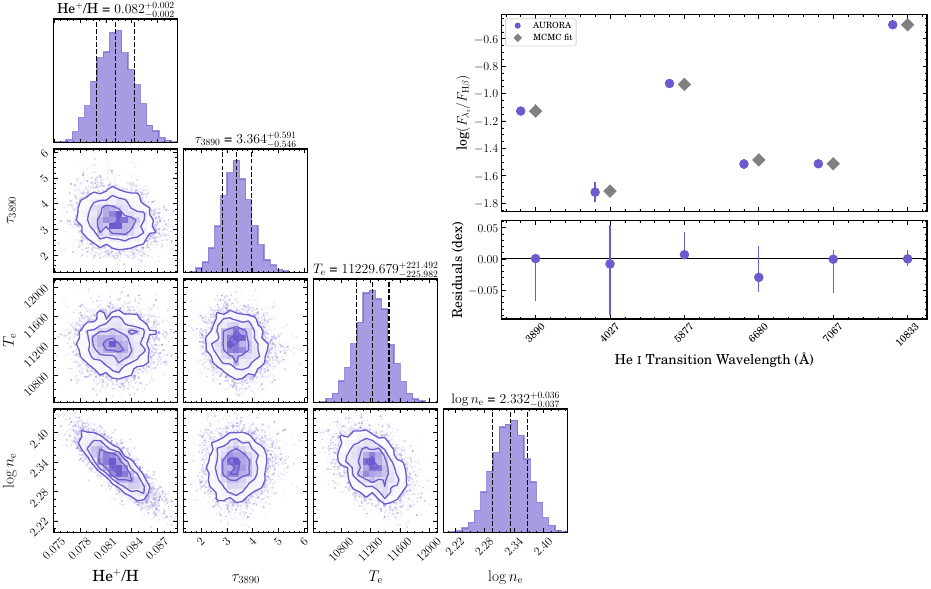}
\caption{{\it Left:} Example of a corner plot for GOODSN-22235 generated by the MCMC 
analysis of the He$^+$/H$^+$ abundance using six \ion{He}{1} lines, including \W10833, 
and a $T_e$([\ion{O}{3}]) prior.
The marginalized posterior distributions for the key parameters -- 
He$^+$/H$^+$, electron temperature ($T_e$), electron density ($n_e$), and optical depth 
at 3890 \AA\ ($\tau_{\lambda3890}$) -- are displayed along the diagonal of the corner plot, 
with the two-dimensional covariances between parameters shown in the off-diagonal panels. 
Owing to the temperature prior used, there is only a small residual covariance 
between temperature and density, while the negative covariance in the bottom left 
panel reflects the possible degeneracy between density and He$^+$/H abundance in the 
presence of collisionally-enhanced \ion{He}{1} emission.
{\it Right:} The top panel shows the observed \ion{He}{1}/H$\beta$ flux ratios for
GOODSN-22235 that passed the 2$\sigma$ criteria (purple circles) compared to the model 
values (gray diamonds) and the bottom panel plots the residuals. \label{fig:cp}}
\end{center}
\end{figure*}


An example of the remarkably good fit of the MCMC code to the \ion{He}{1} line 
fluxes observed for the AURORA High$-z$ He Sample is shown in the right-hand panel
of Figure~\ref{fig:cp}.
The six emission lines measured for GOODSN-22235 that passed the 2$\sigma$ cut
all agree with the model within their uncertainties.
The corresponding posterior distributions and covariances among the 
key physical parameters derived from the \ion{He}{1} lines of GOODSN-22235 
is shown in the left-hand panel of Figure~\ref{fig:cp}.
The derived He$^+$/H$^+$ abundance is tightly constrained at 0.088$\pm$0.002, 
reflecting the power of deep {\it JWST} spectroscopy and our multi-line approach 
in recovering robust, precise He abundances even at high redshift. 
The electron temperature is tightly constrained at 
$T_e=(1.12\pm0.02)\times10^4$ K, a consequence of the $T_e$([\ion{O}{3}]) prior, 
and the electron density is determined to be $\log n_e=2.33\pm0.04$ cm$^{-3}$, 
with the \W10833 line providing critical leverage for high-precision density.
The optical depth at \W3890 line center 
($\tau_{\lambda3890}=3.36_{-0.59}^{+0.55}$) is moderately constrained.

The corner plot in Figure~\ref{fig:cp} also reveals covariances between 
some parameters.
In the test run configuration using no priors, there was a strong covariance 
between $T_e$ and $n_e$; this covariance is significantly reduced here using a 
temperature prior, although some covariance persists. 
Instead, the strongest (negative) covariance is between He$^+$/H$^+$ and $n_e$, 
reflecting the sensitivity of the \ion{He}{1} triplet lines -- especially \W10833.
At higher densities, these lines are collisionally-enhanced (excited from the 
metastable $2\,^3S$ level), allowing the model to match the observed fluxes with 
a lower intrinsic helium abundance. 
Conversely, at lower densities, less collisional boosting requires a higher 
He$^+$/H$^+$ abundance to explain the same observed line strengths. 
This degeneracy underscores the importance of simultaneously fitting physical 
conditions and abundances, and both singlet and triplet \ion{He}{1} lines.
While singlet lines provide a crucial anchor on the abundance, triplet lines 
are needed to constrain the density.
Further, since the He$^+$/H ratio is computed after correcting for density, 
optical depth, and temperature, residual degeneracies between He$^+$/H and 
$\log n_e$ reflect how uncertainties in density propagate through to the 
abundance solution.
This density-sensitivity of the derived He$^+$/H abundance cautions 
against imposing strong density priors, which can unintentionally 
bias the inferred abundance.

\subsection{He{\sc i} Densities}\label{sec:Density}
\begin{figure}[ht]
\begin{center}
	\includegraphics[width=0.9\linewidth, trim=0 4mm 0 0, clip]{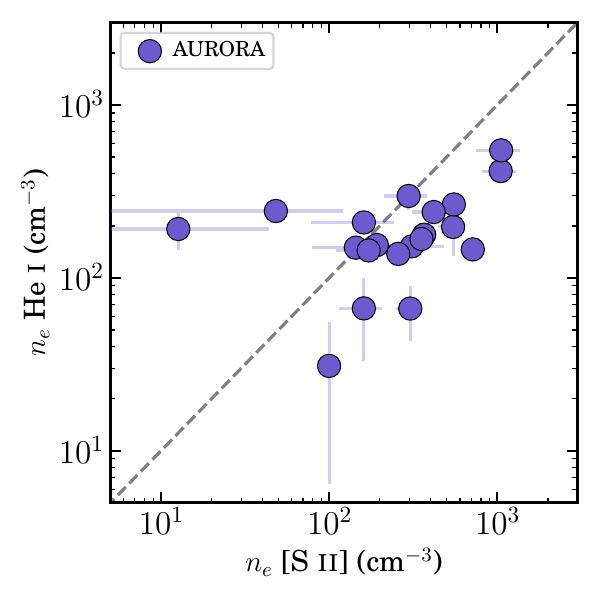} 
\caption{Comparison of the electron density ($n_e$) determined from the \ion{He}{1} fitting 
code versus the properties measured directly from the [\ion{S}{2}] emission lines
for the AURORA High$-z$ He Sample.
The [\ion{S}{2}] and \ion{He}{1} densities, which trace low- and high-ionization
gas, respectively, are in good agreement.
This trend suggests that the gas density is fairly uniform across ionization zones
for this sample and demonstrates the power of multi-line \ion{He}{1} studies to 
provide an independent measure of the nebular electron density.
\label{fig:density}}
\end{center}
\end{figure}

Rather than use the [\ion{S}{2}] density measurements as a prior constraint for the 
He abundance determinations, the multiple \ion{He}{1} line detections for the 
High$-z$ He Sample galaxies enable an independent density estimate. 
In Figure~\ref{fig:density} we compare the MCMC-derived $n_e$(\ion{He}{1}) with the 
measured low-ionization density, $n_e$([\ion{S}{2}]), for the AURORA High$-z$ He Sample.
The \ion{He}{1} and [\ion{S}{2}] densities lie close to the 1-to-1 line showing 
good general agreement.
On average, the \ion{He}{1} densities of the High$-z$ He Sample are 43\% lower 
($\sim$160 cm$^{-3}$) than the [\ion{S}{2}] densities.
There are two galaxies that visually diverge from this trend, with 
$n_e$([\ion{S}{2}])$<n_e$(\ion{He}{1}), but the physical difference is negligible 
and insignificant as the [\ion{S}{2}] densities lie in the insensitive 
low-density limit of the diagnostic line ratio.

Comparing densities from different ions is interesting in the context of 
nebular structure.
The simple three-zone ionization model of \ion{H}{2} regions defines 
the low-ionization zone by the N$^+$ ionization range (I.P. $=14.53-29.160$ eV), 
the intermediate-ionization zone by the S$^{+2}$ ionization range (I.P. $=23.33-34.83$ 
eV), and the high-ionization zone by the O$^{+2}$ ionization range ($35.12-54.93$ eV).
Given the low-ionization potential of S$^+$ ($10.36-23.33$ eV), 
the [\ion{S}{2}] density characterizes the electron density of the 
low-ionization gas, while the higher He$^+$ ionization potential 
($24.59-54.42$ eV) more closely maps to the intermediate- and high-ionization gas.
This means that the $n_e$(\ion{He}{1}) and $n_e$([\ion{S}{2}]) densities 
trace unique ionization regions of gas in the nebula and so have the 
potential to be very different.

Alternatively, \ion{C}{3}] densities, which characterize 24.38--47.89 eV 
intermediate- to high-ionization gas, are expected to more closely match 
\ion{He}{1} densities than the low-ionization [\ion{S}{2}] diagnostic.
Unfortunately, the lower redshifts of the galaxies in the High$-z$ He Sample 
precludes \ion{C}{3}] observations with NIRSpec, so we cannot compare 
\ion{He}{1} and \ion{C}{3}] densities directly.
Alternatively, we consider the \ion{C}{3}] densities reported for higher 
redshift AURORA galaxies by \citet{topping25b}, who find that the median
\ion{C}{3}] density is roughly 30 times higher than the median [\ion{S}{2}] density.
As a result, this work argues for a nebular structure consisting of a dense, 
high-ionization core surrounded by less dense, low-ionization gas. 
Similarly, \citet{harikane25} used the presence of strong rest-optical and 
infrared lines with very different sensitivities to collisional de-excitation 
to argue for a multiphase ISM/density structure. 
This picture is consistent with recent studies of star-forming galaxies with 
very-high ionization emission lines that find that density tends to increase with 
ionization zone, sometimes by orders of magnitude
\citep[e.g.,][]{berg18,berg21,mingozzi22,martinez25}.

In contrast, the close agreement of the $n_e$(\ion{He}{1}) and 
$n_e$([\ion{S}{2}]) densities may suggest that the ionized density 
structure is relatively uniform for the High$-z$ Sample.
This result may be due to sample selection differences, where 
the AURORA sample represents more typical galaxies than the extreme 
very-high-ionization emitters used in the multi-phase density studies.
Alternatively, the similar [\ion{S}{2}] and \ion{He}{1} densities may instead
be biased by their production mechanisms.
The \ion{He}{1} emission lines originate from recombination and so are
relatively insensitive to $T_e$, while the [\ion{S}{2}] and \ion{C}{3}] 
emission lines are produced by collisional excitation such that they 
are exponentially sensitive to $T_e$. 
The much higher excitation energy of \ion{C}{3}] ($E_{\rm exc.}\sim6.5$ eV)
over [\ion{S}{2}] ($E_{\rm exc.}\sim1.8$ eV) means that \ion{C}{3}] emission
is dominated by hotter and, perhaps, denser gas, biasing the measurements. 
Observations of larger samples are needed to better understand how 
densities are related across ionization zones for different samples 
of galaxies; observations of $n_e$(\ion{C}{3}]), $n_e$(\ion{He}{1}), and 
$n_e$([\ion{S}{2}]) in the same galaxies would be especially illuminating.
Our results demonstrate that \ion{He}{1} emission line 
measurements offer a new way to measure densities of moderate- to 
high-ionization gas across a range of redshifts.


\subsection{Total Helium Abundance}\label{sec:tot}
To determine the total He abundance, we must account for all neutral and ionic states 
of helium -- He$^0$, He$^+$, and He$^{+2}$ -- within the ionized H$^+$ gas.
Of the 20 galaxies in the High$-z$ He sample, seven have detections of \ion{He}{2}
\W4687, with strong detections ($>3\sigma$) in only three galaxies. 
We determined the He$^{+2}$/H$^+$ contribution for these three galaxies using
\begin{equation}
    y^{+2} = \frac{\rm He^{+2}}{\rm H^+} = \frac{I_{\rm HeII,\lambda4687}}{I_{\rm H\beta}} 
        \times\frac{\varepsilon_{\rm H\beta}}{\varepsilon_{\rm HeII,\lambda4687}},
\end{equation}
where $I_\lambda$ are the reddening-corrected \ion{He}{2} \W4687 and H$\beta$ fluxes
and $\varepsilon_{\lambda}$ are the respective emissivities at the $T_e$([\ion{O}{3}]) 
and $n_e$([\ion{S}{2}]) measured for each galaxy (see Table~\ref{tbl4}). 
For galaxies without a \ion{He}{2} detection, we assumed the He$^{+2}$/He fraction is zero.

Total He abundances were determined by adding the ionic He abundances:
\begin{equation}
    y = \frac{\rm He}{\rm H} = y^+ + y^{+2} = \frac{\rm He^{+}}{\rm H^+} + \frac{\rm He^{+2}}{\rm H^+}.
\end{equation}
These He abundances were then converted to He mass fractions, $Y$, using:
\begin{equation}
    Y = \frac{4y(1-20\times {\rm O/H})}{1+4y},
\end{equation}	
where O/H is the total direct-method oxygen abundance.
Ionic and total He abundances are reported for the High$-z$ He Sample
in Table~\ref{tbl4}.

Most $z\sim0$ studies argue that He$^0$ contributions are negligible for their high-ionization, 
metal-poor targets and so corrections are not required \citep[e.g.,][]{izotov94,aver10,aver13}.
However, in low-ionization regions, photoionization models suggest that the correction for 
the neutral He contribution to the total He abundance can be to $3-5$\%\ 
\citep[e.g.,][]{pagel92,izotov04}.
As a result, it has become common place to apply a selection cut of O$^+$/O$<0.5$ to ensure 
the He$^0$ correction is negligible \citep[e.g.,][]{aver10,aver13,aver15} or apply an ionization 
correction factor when the ionizing radiation field is sufficiently soft
\citep[e.g.,][]{izotov98}.

\citet{mathis82} and \cite{pagel92} were the first to suggest using photoionization model 
trends to predict the neutral He fraction from the radiation softness 
parameter \citep{vilchez88} characterized as
\begin{equation}
    \eta = \frac{{\rm O}^+/{\rm O}^{+2}}{{\rm S}^+/{\rm S}^{+2}}.
\end{equation}
Using \texttt{cloudy} 23.01 photoionization models \citep{chatzikos23, gunasekera23} 
over a large range of densities with Binary Population and Spectral Synthesis 
\citep[\texttt{BPASS}v2.14]{eldridge17} burst models for the input ionizing radiation field, 
we find that contributions from He$^0$ become non-negligible ($>2$\%) when 
$\log \eta \gtrsim 0.3$ ($\eta \gtrsim 2$) for $0.4\ Z_\odot$ models (the median metallicity 
of the High$-z$ He Sample).
This threshold is consistent with findings of \citet{sauer02}, who report that He ICFs 
are needed to account for neutral helium when $\log \eta \gtrsim 0.2$.

In earlier work, \citet{pagel92} recommended a much larger softness parameter 
of $\log \eta > 1.0$ for which the He ICF deviates from 1. 
This is because the models in \citet{pagel92} used non-local thermodynamic 
equilibrium (NLTE) single star atmosphere models with limited treatment of metal 
line opacities from \citet{mihalas72}, which produce much harder spectra than the 
updated \texttt{BPASS} stellar population models used in this work. 
These hard \citet{mihalas72} spectra efficiently ionize He$^0$ such that  
neutral He contributes only at high softness parameters ($\eta > 10$). 
In contrast, modern models such as \texttt{BPASS}$+$\texttt{cloudy} reveal 
non-negligible He$^0$ fractions at $\eta$ values as low as $\sim1-2$
due to softer instantaneous SEDs (in part due to realistic metal line blanketing) 
and improved treatment of partial ionization zones.
Eleven galaxies in the AURORA High$-z$ He Sample have $\eta \geq 2.0$ (see Table~\ref{tbl4}),
suggesting the correction for the neutral He contribution may be significant.
Therefore, we regard the He abundances for these galaxies with high 
softness parameters as lower limits.

       
\begin{deluxetable*}{rlcccccccR}
\tabletypesize{\small}
\tablecaption{He Abundances for the AURORA He Emission Line Sample}
\tablehead{
\CH{}  & \CH{}      & \CH{$E(B-V)$} & \CH{$T_e$([\ion{O}{3}])} & \CH{$n_e$([\ion{S}{2}])} & \CH{} 
       & \CH{}      & \CH{}         & \CH{} & \CH{}\\ [-2ex]
\CH{\#}& \CH{Name}  & \CH{(mag)}    & \CH{(K)}                 & \CH{(cm$^{-3}$)}         & \CH{$\eta$} 
       & \CH{$y^+$} & \CH{$y^{+2}$} & \CH{$Y$} & \CH{$\Delta Y$}}
\startdata 
\ \ 1. & COSMOS-8442  & 0.125$\pm$0.008 & 9500$\pm$900  &10$^{+30}_{-10}$&0.460 & 0.081$\pm$0.004 & $\ldots$          & 0.241$\pm$0.009 & -0.021 \\
\ \ 2. & COSMOS-4205  & 0.149$\pm$0.004 & 12800$\pm$100 & 170$\pm$60    & 0.621 & 0.098$\pm$0.002 & $\ldots$          & 0.280$\pm$0.004 &  0.023 \\
\ \ 3. & GOODSN-25004 & 0.219$\pm$0.004 &10100$\pm$400  & 350$\pm$60    & 1.154 & 0.088$\pm$0.002 & $\ldots$          & 0.256$\pm$0.004 & -0.010 \\
\ \ 4. & COSMOS-4029  & 0.147$\pm$0.006 & 12200$\pm$200 & 160$\pm$80    & 1.095 & 0.088$\pm$0.002 & $\ldots$          & 0.258$\pm$0.004 &  0.000 \\
\ \ 5. & COSMOS-5283  & 0.103$\pm$0.004 & 11700$\pm$200 & 260$\pm$40    & 1.404 & 0.088$\pm$0.002 & 0.0010$\pm$0.0001 & 0.260$\pm$0.004 & -0.000 \\
\ \ 6. & COSMOS-4156  & 0.112$\pm$0.004 & 14900$\pm$200 & 420$\pm$100   & 2.291 & 0.078$\pm$0.002 & 0.0019$\pm$0.0002 & 0.239$\pm$0.005 & -0.023 \\
\ \ 7. &\bf{GOODSN-30053}
                    & 0.468$\pm$0.006 & 11100$\pm$700 & 300$\pm$60    & 2.265 & 0.112$\pm$0.004 & $\ldots$          & 0.307$\pm$0.008 &  0.045 \\
\ \ 8. & GOODSN-27876 & 0.531$\pm$0.009 &  7300$\pm$700 & 550$\pm$100   & 5.892 & 0.100$\pm$0.007 & $\ldots$          & 0.279$\pm$0.014 & -0.001 \\
\ \ 9. &\bf{COSMOS-5571}
                    & 0.297$\pm$0.008 &  9000$\pm$700 & 160$\pm$50    & 2.731 & 0.113$\pm$0.005 & $\ldots$          & 0.307$\pm$0.009 &  0.039 \\
10.   & GOODSN-19067  & 0.193$\pm$0.011 & 13600$\pm$1200& 100$\pm$100$^{\star}$   
                                                                        & 2.586 & 0.096$\pm$0.006 & $\ldots$          & 0.279$\pm$0.013 &  0.018 \\
11.   & GOODSN-21522  & 0.035$\pm$0.008 & 12800$\pm$600 & 140$\pm$70    & 2.627 & 0.084$\pm$0.004 & $\ldots$          & 0.250$\pm$0.009 & -0.005 \\
12.   & GOODSN-22235  & 0.224$\pm$0.005 & 11300$\pm$200 & 370$\pm$60    & 1.740 & 0.082$\pm$0.002 & $\ldots$          & 0.243$\pm$0.005 & -0.020 \\ 
13.   & GOODSN-30564  & 0.523$\pm$0.005 & 8300$\pm$300  & 550$\pm$50    & 3.011 & 0.108$\pm$0.002 & $\ldots$          & 0.296$\pm$0.004 &  0.022 \\
14.   & GOODSN-19848  & 0.240$\pm$0.008 & 11600$\pm$400 & 1050$\pm$240  & 2.595 & 0.088$\pm$0.003 & $\ldots$          & 0.257$\pm$0.007 & -0.006 \\
15.   &\bf{GOODSN-22384}
                    & 0.246$\pm$0.009 & 11100$\pm$1400& 190$\pm$ 54   & 3.113 & 0.102$\pm$0.004 & 0.0026$\pm$0.0007 & 0.293$\pm$0.008 &  0.036 \\
16.   & GOODSN-21033  & 0.093$\pm$0.006 & 12700$\pm$200 &50$^{+70}_{-50}$&0.673 & 0.082$\pm$0.002 & $\ldots$          & 0.245$\pm$0.005 & -0.009 \\
17.   &\bf{COSMOS-4740}
                    & 0.495$\pm$0.009 & 13300$\pm$1000& 720$\pm$90    & 3.771 & 0.107$\pm$0.003 & $\ldots$          & 0.297$\pm$0.006 &  0.037 \\
18.   & GOODSN-28209  & 0.297$\pm$0.011 & 9400$\pm$500  &1050$\pm$310   & 2.613 & 0.096$\pm$0.004 & $\ldots$          & 0.273$\pm$0.008 &  0.002 \\
19.   & COSMOS-8363   & 0.139$\pm$0.009 & 9600$\pm$400  & 300$\pm$90    & 0.731 & 0.082$\pm$0.003 & $\ldots$          & 0.243$\pm$0.007 & -0.021 \\
20.   & GOODSN-22932  & 0.045$\pm$0.009 & 13200$\pm$500 & 310$\pm$170   & 1.132 & 0.085$\pm$0.004 & $\ldots$          & 0.252$\pm$0.009 & -0.003
\enddata	
\tablecomments{
Columns 3--5 list the nebular reddening, electron temperature ($T_e$), and electron density
($n_e$) derived from the emission-line fits reported in \citet{sanders25}.
Column 6 lists the softness parameter, $\eta=$(O$^+$/O$^{+2}$)/(S$^+$/S$^{+2}$), where high values
indicate a significant contribution of He$^0$ to the total He abundance is likely.
Columns 7 and 8 list the ionic He abundances, $y^+={\rm He}^+/{\rm H}^+$ and 
$y^{+2}={\rm He}^{+2}/{\rm H}^+$ determined from \ion{He}{2} \W4687, 
which were added together to get He/H.
Column 8 lists the He mass fraction, $Y$.
Note that no He$^0$ correction was used, so $Y$ values should be considered lower limits
for galaxies with $\eta\gtrsim 2$.
Column 9 gives the He mass fraction offset, $\Delta Y$, from the $Y-$O/H relationship from \citet{aver22}.
Significant He outliers ($\Delta Y > 0.03$) are bolded in Column 1.\\
$^{\star}$Note that the typical low-density limit of 100 cm$^{-3}$ was adopted for GOODSN-19067; 
good [\ion{S}{2}] line measurements were made, but the ratio is greater than the low-density limit 
theoretical ratio.} \label{tbl4}
\end{deluxetable*}


\begin{figure*}
\begin{center}
	\includegraphics[width=1.0\linewidth, trim=15ex 9ex 15ex 15ex, clip]{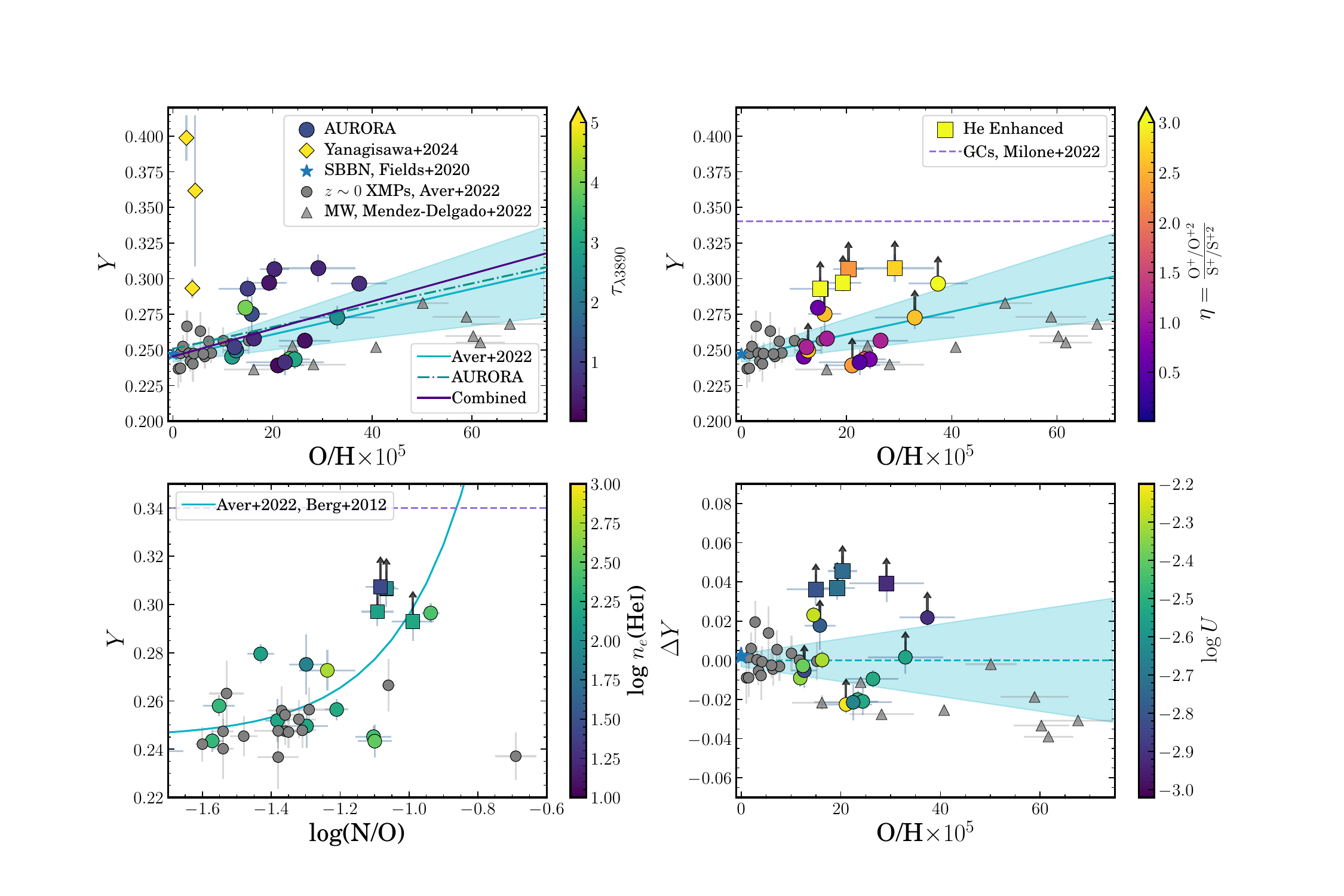} 
\caption{{\it Upper left:} 
Helium abundance (mass fraction) determined from the \ion{He}{1} 
fitting code versus oxygen abundance, color-coded by optical depth at \W3890.
The best fits for the AURORA High$-z$ He Sample are plotted as circular points,
color-coded by the best-fit optical depths.
In comparison, we plot the He abundances and corresponding trend for $z=0$ 
extremely metal-poor galaxies from \citet[][gray points and blue line]{aver22} 
and He abundances for Milky Way \ion{H}{2} regions from 
\citet[][red triangles]{mendez-delgado22}.
Most of the AURORA sample falls within 1$\sigma$ of the \citet{aver22} trend, 
but a few galaxies appear to have enhanced He abundances at low optical depths.
In contrast, the high$-z$ He abundances from \citet{yanagisawa24} are extreme 
outliers, likely due to their high optical depth measurements.
{\it Upper right:} 
Helium mass fraction versus oxygen abundance, now color-coded by the 
softness parameter, $\eta$.
Both of the comparison samples have low softness values ($\eta<1$), but many 
of the AURORA galaxies have high softness values suggesting that they have 
non-negligible He$^0$ contributions that would increase their total He abundances, 
including the highest values.
In comparison, the AURORA He outliers are approaching the enhanced He measurements 
for GCs from \citet{milone22}.
{\it Lower left:} The He mass fraction versus N/O abundance, color-coded by the 
\ion{He}{1} density, shows no clear trends.
{\it Lower right:} 
The offset of the He mass fractions from the \citet{aver22} trend are plotted 
versus O/H abundance.
The largest positive outliers from the AURORA sample all have low ionization 
parameter, consistent with a mixed-age ionizing stellar population and GC 
progenitors. \label{fig:Y}}
\end{center}
\end{figure*}

\section{He/H Evolution}\label{sec:HeEvol}
\subsection{The He/H--O/H Relationship}\label{sec:HeHOH}
With our He abundance measurements for the High$-z$ He Sample now in hand,
we examine the evolution of He/H. 
We plot the results of our He abundance analysis 
as a function of O/H in the upper left-hand plot of Figure~\ref{fig:Y}.
The High$-z$ He Sample spans an oxygen abundance range of 
12+log(O/H)$\approx7.7-8.3$ (or O/H$\times10^5\approx12-45$) and a He mass fraction 
of $Y\approx0.243-0.319$.
To place our He abundance measurements at $z\sim 1.5-3.5$ in context, 
we compare with the standard Big Bang nucleosynthesis (SBBN) primordial He mass 
fraction from the Planck survey \citep[$Y_p=0.2469$;][]{fields20}, the $Y-$O/H 
relation derived for $z\sim0$ extremely-metal poor (XMP) dwarf galaxies 
($Z\leq0.05\ Z_\odot$) by \citet[][gray points and blue line]{aver22}, and 
higher-metallicity Milky Way \ion{H}{2} region measurements from 
\citet[][purple triangles]{mendez-delgado22}.
We find that, on average, the AURORA sample lies systematically above the SBBN 
value, indicating that these galaxies have undergone measurable He enrichment 
by $z\sim2-3$, as expected. 
Surprisingly, the AURORA galaxies also have higher He abundances than the Milky 
Way measurements at a given O/H abundance, which is not consistent with general
expectations of chemical evolution.
However, the Milky Way trend is also inconsistent with the XMP trend and 
Planck SBBN measurement, suggesting the Milky Way values may be underestimated. 

Since O/H abundance is commonly used as a proxy for galaxy evolution, 
He evolution can be characterized by the slope of the $Y$--O/H trend. 
\citet{aver22} report $\Delta Y/\Delta({\rm O/H})=80\pm38$ for their sample 
of $z\sim0$ XMPs.
In comparison, a Monte Carlo least-squares linear fit to only the High$-z$ He 
Sample gives a shallower slope of $\Delta Y/\Delta({\rm O/H})=50\pm34$,
but this trend has a weak, insignificant correlation based on the Spearman's 
rank correlation using the \texttt{scipy.stats.spearmanr} function in
\texttt{python} ($r=0.14$, $p=0.56$).
Visually, the AURORA data appear to represent a scattered extension of the XMP data,
so we also perform a Monte Carlo least-squares linear fit to the XMP$+$AURORA 
combined sample and find a more significant trend ($r=0.46$, $p=0.0004$)
with $\Delta Y/\Delta({\rm O/H})=97\pm22$.
This slope is consistent with the \citet{aver22} trend, but with a slightly
higher offset for the AURORA-only fit, suggesting the AURORA values are on 
average higher in $Y$, yet follow a similar evolutionary trend with metallicity. 

While the AURORA sample paints a picture of He enrichment at $z\sim1.6-3.3$,
we examine the $z\sim6$ He abundance measurements from 
\citet{yanagisawa24} to see if this trend extends to higher redshifts.
We plot the \citet{yanagisawa24} points as diamonds in Figure~\ref{fig:Y}, 
which lie significantly above both the AURORA sample and the local extrapolated 
$Y-$O/H relation. 
These points likely reflect systematic effects related to limited line diagnostics: 
the He abundance analysis in \citet{yanagisawa24} used three \ion{He}{1} lines 
(\W4473, \W5877, and \W7067), where all three were measured at $>3\sigma$ 
significance for RXCJ2248-ID, two lines were significant for GS-NDG-9422, and 
one line was significant for GLASS150008.
Furthermore, the \W4473 and \W5877 \ion{He}{1} lines do not effectively 
constrain the physical parameters of the gas.
For example, \citet{benjamin02} found that the \ion{He}{1} \W4473, \W5877, and \W6680 
line strengths change by less than 1\%\ over a large range in $T_e$, $n_e$, and 
$\tau_{\lambda3890}$.

The limited line set used in the \citet{yanagisawa24} analysis lacks the sensitive 
optical and infrared triplet lines that are crucial for breaking the $T_e-n_e$,
$\tau_{\lambda3890}-n_e$, and $y^+ - n_e$ degeneracies. 
As a result, their models inferred high optical depths ($\tau_{\lambda3890}\sim8$)
that are beyond the calibration range of current He abundance codes, and therefore
lead to inflated He$^+$/H$^+$ values. 
To demonstrate this effect, we color-code the AURORA and \citet{yanagisawa24} points 
by the inferred $\tau_{\lambda3890}$ values. 
The \citet{yanagisawa24} points are clearly outliers in their $\tau_{\lambda3890}$
values, while the lower optical depths of the AURORA sample show no trend with 
derived He abundance. 
\citet{yanagisawa24} address the degeneracy between $n_e$ and $Y$, allowing for the 
possibility that the galaxies modeled have significantly higher $n_e$ and lower $Y$, 
but do not address the additional degeneracy between $\tau_{\lambda3890}$ and $n_e$.
In the absence of stronger observational constraints, particularly from density- and 
optical-depth-sensitive lines like \W10833, such fits are susceptible to compensation 
between $\tau$ and He abundance.
Therefore, robust assessment of the He/H evolution is currently limited to the 
$z\sim0$ and $z\sim1.5-3.5$ AURORA samples, although higher-redshift measurements could 
be achieved by augmenting rest-optical \ion{He}{1} line measurements with JWST/MIRI
observations of the \W10833 line to break the He/H$-n_e$ and 
$\tau_{\lambda3890}-n_e$ degeneracies.\looseness=-2

\subsection{Subsample of He-Enhanced Galaxies}\label{sec:HeEnhance}
While the $z\simeq 1.5-3.5$ AURORA sample is broadly consistent with the $z\sim0$ 
\citet{aver22} trend extrapolated to higher O/H abundances, there are several 
outliers with elevated $Y$ values that may indicate enhanced He production or 
differing chemical evolution pathways in early galaxies.
In particular, there are four galaxies with significantly elevated He/H 
($\Delta Y > 0.03$): GOODSN-30053, GOODSN-5571, GOODSN-22384 and COSMOS-4740.
These four galaxies, plotted as squares in the upper right-hand panel of 
Figure~\ref{fig:Y}, are clustered around $Y\approx0.29-0.31$ and 
O/H$\times10^5\approx15-30$.
Moreover, the four He-enhanced galaxies have high softness parameter values 
($\eta>2$), suggesting that they host an additional He$^0$ contribution that 
would further increase their total He abundance.

We tested for secondary residual trends of $Y-$O/H with a broad range of properties, 
including model parameters ($T_e$, $n_e$, $\tau_{\lambda3890}$),
galaxy properties (SFR, sSFR, $\Sigma_{\rm SFR}$, $M_\star$, ionizing stellar 
population age as parametrized by EW(H$\beta$)),
gas properties (ionization: O$_{32} =$ [\ion{O}{3}] \W5008/[\ion{O}{2}] \W3728 
and He$^{+2}$/He, line width: $v_{\rm FWHM}$(\ion{He}{1})), and
abundances (N/O, N/H, Ne/O, Ne/H, Ar/H, Ar/O, S/H, S/O).
Given the recent discovery of N-enhancement in several high$-z$ galaxies 
\citep[e.g.,][]{pascale23, bunker23, ji24, marques-chaves24, topping24a, schaerer24},
there is great interest in linking N-enhancement with other abundance anomalies.
In the lower left-hand panel of 
Figure~\ref{fig:Y}, we therefore plot $Y$ for the AURORA sample as a function of N/O abundance, 
derived from the optical N$^+$/O$^+$ lines.
We combined the $z\sim0$ direct-method bimodal O/H--N/O relationship from
\citet{berg12} with the $Y$--O/H trend from \citet{aver22} to plot 
$Y$--N/O (blue curve).
The AURORA points are scattered around the $Y$--N/O trend, where significant
scatter is common in samples of N/O abundances. 
Further, while there is a general trend of increasing N/O with He/H, 
the He outliers do not show a similar enhancement in N/O. 
Rather, the high-N/O values of the high He outliers are due to their
relatively-high metallicities ($12+\log({\rm O/H})\gtrsim8.2$).

We plot the AURORA sample outliers as offsets from the $z\sim0$ 
\citet{aver22} $Y-$O/H trend in the lower right-hand panel of Figure~\ref{fig:Y}, 
color-coded by ionization parameter, $\log U$.
While O$_{32}$ is a commonly used proxy for ionization, it also depends on 
other properties such as metallicity (which affects the hardness of the 
ionizing spectrum).
To account for metallicity and ionized gas density, we used the photoionization 
model-based calibration from \citet[][which is a function of the gas-phase O/H 
abundance and O$_{32}$]{berg19a} to determine the dimensionless ionization parameter,
$\log U$, defined as 
\begin{equation*}
    \log U = \log\left(\frac{Q({\rm H)}}{4\pi R^2 n_{\rm H}c}\right).
\end{equation*}
In this equation, $Q({\rm H})$ is the ionizing photon flux (photons s$^{-1}$ 
with $E>13.6$ eV), $R$ is the distance from the ionizing source to the gas (cm),
$n_{\rm H}$ is the total hydrogen number density (cm$^{-3}$), and $c$ is the speed 
of light (cm s$^{-1}$).

We found no strong secondary trends with $\Delta Y$ for any of the 
properties we tested.
This allows us to rule out biases due to optical depth effects, 
modeling systematics, and some nucleosynthetic sources (e.g,
N and He enhancements from Wolf Rayet Nitrogen (WRN) stars).
On the other hand, we find that the AURORA galaxies with elevated He 
abundances do have the following properties in common:
(i) typical abundances of both $\alpha$-elements (e.g., Ne) 
and non-$\alpha$-elements (e.g., N),
(ii) relatively low ionization (O$_{32} < 1.5$ or $\log U < -2.7$), and
(iii) relatively low H$\beta$ equivalent widths (EW(H$\beta$)$\lesssim350$ \AA).
Since Balmer equivalent widths are good diagnostics of the ionizing stellar population
age \citep[e.g.,][]{leitherer99}, the low H$\beta$ equivalent widths for the 
He outliers imply an evolved stellar population is present 
\citep[age $\gtrsim10$ Myr; e.g.,][]{eldridge17}.
On the other hand, the high-excitation energy of the [\ion{O}{3}] \W4363 
line needed for the direct abundance measurements of our sample require 
hot, ionized gas characteristic of young star-forming regions.
Taken together, these properties suggest these galaxies may host 
multiple stellar populations of different ages emitting ionizing photons. 

\section{Chemical Enrichment of He/H}\label{sec:enrichment}
As galaxies evolve, He/H is expected to increase in step with oxygen as 
\begin{align*}
    \Delta Y/\Delta O &\approx \frac{y_{\rm He}^{\rm CC}+y_{\rm He}^{\rm AGB}}{y_{\rm O}^{\rm CC}} \\
    &\text{\citep[see Eqn. 11 of][]{weller25}},
\end{align*}
where the total stellar He yield, $y_{\rm He}$, is an IMF-averaged combination
of massive star ($y_{\rm He}^{\rm CC}$) and asymptotic giant branch ($y_{\rm He}^{\rm AGB}$) contributions: 
\begin{align*}
    y_{\rm He} &= y_{\rm He}^{\rm CC} + y_{\rm He}^{\rm AGB} \approx 0.022 \\
    &\text{\citep[see Eqn. 10 and 12 of][]{weller25}}.
\end{align*}
Massive stars ($M_\star \gtrsim 10\ {\rm M_\odot}$) enrich the ISM with He on short timescales 
($\lesssim 10$ Myr), synthesizing He via the CNO-cycle and ejecting it through stellar 
winds and core-collapse supernovae (CCSNe).
Wind-driven yields are metallicity-dependent, as higher stellar opacities at higher
metallicity enhance mass loss via line-driven winds.
AGB star ($M_\star \approx 1-8\ {\rm M_\odot}$) enrichment is delayed relative to enrichment 
from massive stars, as He is released through thermal pulses and envelope ejection 
roughly 100 Myr -- 1 Gyr after the initial burst.
Thus, if chemical evolution proceeds in a relatively continuous manner across cosmic 
time, we would expect galaxy populations at different redshifts to occupy different 
portions of the same general He/H--O/H trend.
Instead, we have identified a sub-population of galaxies with elevated He mass fractions 
($\Delta Y\gtrsim0.03$) in Section~\ref{sec:HeEnhance}; we discuss the possible
sources of this enhancement below.

\subsection{A High-He Sub-population?}\label{sec:sub}
\begin{figure}
\begin{center}
	\includegraphics[width=1.0\linewidth, trim= 2ex 2ex 0ex 0ex, clip]{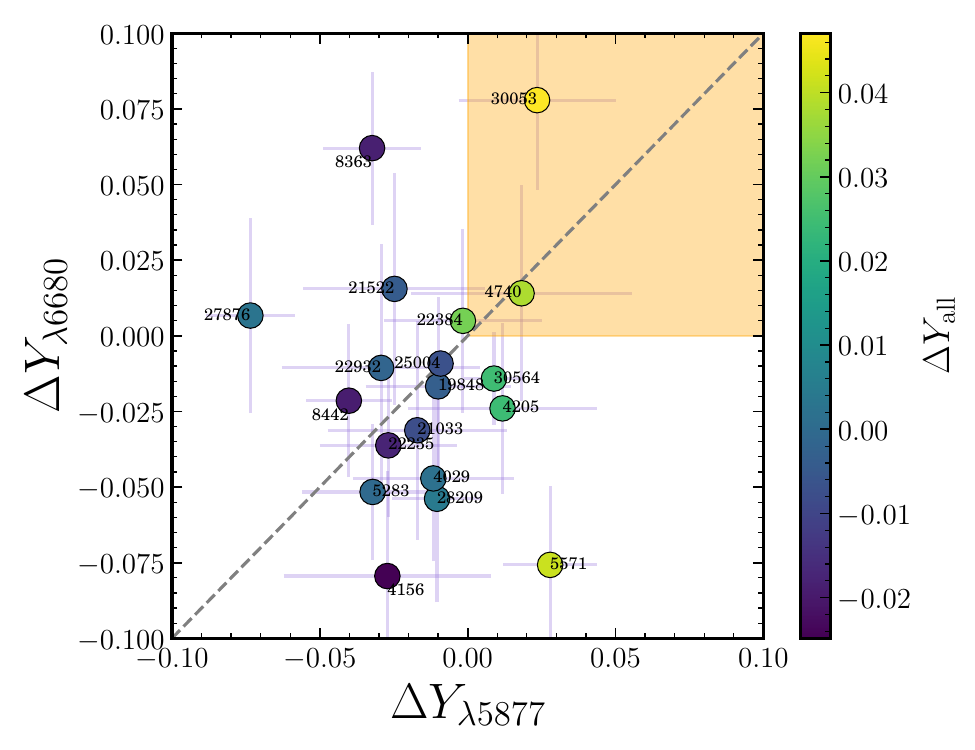} 
\caption{Comparison of the He mass fraction offsets from the $z\sim0$ 
$Y-$O/H trend of \citet{aver22} for single-line He abundance analyses
that are insensitive to $T_e$, $n_e$, and $\tau_{\lambda3890}$. 
The $\Delta Y_{\lambda6680}$, inferred from the \ion{He}{1} \W6680 line only, 
is plotted versus $\Delta Y_{\lambda5877}$, inferred using the \ion{He}{1} 
\W6680 line only. 
The points are color-coded by the $\Delta Y_{\rm all}$ values determined from the 
full multi-line analysis.
In general, the single-line He abundances are lower than the full multi-line
He abundances, but three of the galaxies in the sub-population of high He outliers 
consistently show elevated He across all three methods.
These He-enriched galaxies are highlighted as the three points in the tan colored box.}\label{fig:SingleLines}
\end{center}
\end{figure}

To further test whether the high-He sub-population of AURORA is truly enhanced, 
we determined He abundances for the High$-z$ He Sample based on fits to single 
\ion{He}{1} lines that are relatively insensitive to temperature, density, and 
radiative transfer effects (see Table 2).
The single-line analyses using \W5877 only and \W6680 are presented in 
Appendix~\ref{sec:5877}, with the derived $y^+$ values for each plotted in 
Figure~\ref{fig:SingleLines}.
Figure~\ref{fig:SingleLines} reveals consistently elevated He/H values 
($\Delta Y_{\rm single\ line}>0$) for three of the He outlier AURORA galaxies: both 4740 and 22384
lie close to the 1-to-1 line, while the inferred $Y$ from \W6680 is significantly 
enhanced for 30053 and moderately enhanced from \W5877.
This consistent enhancement, in the absence of strong parameter degeneracies, 
strengthens the case that this sub-population may reflect real He enrichment.
Notably, these galaxies do not stand out in terms of stellar mass, SFR, excitation,
metallicity, or relative abundances.
Rather, they appear normal by conventional diagnostics, 
yet show abnormally high He abundances.

One candidate high-He outlier, COSMOS-5571, does not lie in the "tan" box in 
Figure~\ref{fig:5877}.
Instead, COSMOS-5571 has a high $\Delta Y_{\lambda5877}$ value and low 
$\Delta Y_{\lambda6680}$ value.
The \W6680 singlet line analysis giving a lower He/H abundance suggests that the 
\W5877 triplet line may be affected by optical depth or collisional excitation effects, 
artificially boosting its strength. 
Since \W6680 is less sensitive to such effects, this discrepancy could mean
that the true He/H abundance is not elevated.
However, the \W6680 line is roughly four times weaker than the \W5877 line,
which may indicate that a more significant detection is needed or that a more 
detailed model (e.g., using multiple lines and correcting for optical depth) is 
required.
As an additional check, we also performed a single-line \ion{He}{1} analysis 
on COSMOS-5571 using the \W4473 line and again found significant He-enhancement
($\Delta Y_{\lambda4473}>0.03$).


\subsection{AGB Contributions}
One possibility for the multiple stellar populations of different ages with 
He enhancements is enrichment from AGB stars.
The AGB interpretation is supported by \citet{weller25}, who analyzed 
the time dependence of He production for a range of stellar yield models, 
finding that the net He yield is dominated by contributions on two timescales: 
prompt enrichment from massive stars through winds and supernovae and 
later enrichment from $M_\star \approx 4-8 {\rm M_\odot}$ AGB stars that release 
$\sim50$\%\ of their He yields within $\sim200$~Myr. 
They conclude, therefore, that while the evolution of the He mass fraction, 
$Y$, is expected to be linear with oxygen abundance, deviations from the 
expected $\Delta Y/\Delta$(O/H) -- such as those seen in the AURORA high-He
subpopulation -- could result from variations in the amount of AGB relative to 
CCSNe enrichment.  
Based on the yields from \citet{weller25}, increasing $Y$ by $\Delta Y = 0.05$ requires 
a significant AGB-dominated enrichment episode or episodes that total more than
$10^9 {\rm M_\odot}$ in stellar mass, assuming full retention of ejecta and negligible 
dilution.

The AURORA galaxies that show elevated He/H ratios do not show corresponding 
enhancement in N/O  (Figure~\ref{fig:Y}).
AGB stars, particularly intermediate-mass ones ($\simeq 4-8\,{\rm M_\odot}$), contribute to 
both He and N enrichment, but they are produced through different mechanisms.
He enrichment from AGB stars primarily occurs during the second dredge-up, while N 
enhancement requires hot bottom burning (HBB) which is more likely in more massive 
AGB stars ($\gtrsim~5\ {\rm M_\odot}$). 
Thus, significant He production without N from AGB stars would require a finely
tuned IMF to include numerous second dredge-up stars, but few stars reaching HBB.
While the high-He outliers also have high N/O indicative of substantial past enrichment 
by AGB stars (typical for their O/H metallicities), the scenario laid out above
suggests that AGB stars are unlikely drivers of the {\it excess} He enhancement. 

\subsection{VMS Contributions}\label{sec:VMSs}
Alternatively, the observed decoupling of He and N enrichment in the high-He outliers 
can result from very massive stars (VMSs; $>100\,{\rm M_\odot}$), which can eject large 
amounts of He-rich, N-poor material through stellar winds or binary mass loss before 
significant N production has occurred \citep[e.g.,][]{yusof13,szesci15,crowther16,gotberg20}.
In general, the rotational mixing, stellar winds, and binary stripping processes 
that can occur in VMSs will accelerate He ejection early in the star's life and 
bypass traditional N yields. 
To investigate further, we examined the $\alpha$-element abundance trends in the High$-z$ 
He Sample, finding typical S/H, Ar/H, and Ne/H abundances.
Such an enhancement of He but not heavier $\alpha$-elements is consistent with 
production by VMSs due to the differing timescales of He and heavier $\alpha$-element 
production, where heavier $\alpha$-elements such as Ne are synthesized in the 
later stages of massive star evolution and are returned to the ISM during CCSNe.
Thus, He can be enriched in the ISM rapidly prior to Ne enrichment from CCSNe.
Further, VMSs are likely to directly collapse to black holes rather than explode 
\citep[e.g.,][]{heger03, woosley17}, locking up freshly synthesized material not 
ejected during the wind phase \citep[e.g.,][]{crowther10, spera17}. 

In summary, the most plausible explanation for He enrichment in the high-He outliers 
is He enrichment from VMSs.
Such He enhancement without N or Ne may be a hallmark of prompt enrichment from massive 
stellar populations, consistent with the high star-formation efficiencies, top-heavy IMFs, 
low-metallicity stellar winds, and binary/rotational effects expected in high-redshift 
galaxies. 
Such He enrichment from VMSs can also be consistent with multiple stellar populations 
of different ages emitting ionizing photons, where a first burst of star formation 
(with age $>10$ Myr) is now dominated by B stars that serve to lower the integrated EW 
and ionization parameter, followed by a second burst (with age $<10$ Myr) that is heating 
the gas.
In this scenario, short-lived VMSs from either the first or second burst episode could
have driven the He enrichment.  

By cosmic noon -- the epoch of the AURORA High$-z$ He Sample --
significant enrichment from stars of all ages is expected to have set the general
chemical abundance pattern of a galaxy \citep[e.g.,][]{maiolino19,sanders21}, 
washing out any early signatures from unique enrichment events.
Thus, robust He measurements are needed at even higher redshifts, where larger 
deviations may reveal differences in the IMF, stellar evolution pathways, or 
timescales for metal mixing in the early universe.
For example, a flatter$\Delta Y/\Delta$(O/H) slope might indicate more rapid O 
enrichment relative to He, possibly due to top-heavy IMFs or metal-poor stellar 
populations with low He yields \cite[e.g.,][]{romano10,weller25}. 
A steeper $\Delta Y/\Delta$(O/H) slope, on the other hand, could suggest that He is 
enriched disproportionately early, perhaps due to Population III stars or 
wind phases in VMSs that dominate feedback at early epochs
\citep[e.g,.][]{schaerer03,senchyna24}. 
Alternatively, greater scatter around the enrichment trend may reflect stochastic or 
localized enrichment effects in shallow potential wells, or delayed ISM 
mixing of He and O products from different stellar mass ranges
\citep[e.g.,][]{wise12,krumholz18}.
Such observations for high-redshift galaxies could be obtained by combining deep rest-optical 
\ion{He}{1} observations from NIRSpec with rest-NIR \ion{He}{1} 
observations from MIRI to provide critical leverage on the universality 
of stellar yields and feedback, thus helping to inform the assumptions underlying models 
of reionization-era galaxy populations. \\

\subsection{Globular Cluster Progenitors}\label{sec:GCPs}
Globular clusters (GCs) are dense, gravitationally bound stellar systems typically 
containing $10^4-10^6$ old, metal-poor stars that formed early in a galaxy's history 
and often exhibiting evidence for multiple stellar populations with variations in 
He, C, N, O, and other light elements \citep[e.g.,][]{kraft94,bastian18,cohen05}. 
The unique chemical signatures of GCs and their early formation make them valuable tracers 
of star formation and chemical enrichment in the early Universe.
Direct He measurements using photospheric lines in GC horizontal branch
stars show enhanced He mass fractions of $Y\sim0.34$ \citep[e.g.,][]{milone22}.

Evidence of even more extreme helium enhancement $Y\gtrsim0.40$ has been reported for
red giant branch stars in both NGC 2808 and $\omega$ Centauri using NIR chromospheric
He lines \citep{pasquini11,dupree11}.
The high-He sub-population in Figure~\ref{fig:Y}, with even higher He abundances 
expected due to neutral He contributions for $\eta>2$, bears an intriguing resemblance 
to the second-generation populations in local GCs ($Y\gtrsim0.34$), plotted
as a dashed-purple line in the upper right-hand panel.
We, therefore, speculate that the subpopulation of high-He systems in AURORA may 
represent galaxies caught during an intense, short-duration burst of star formation 
that captures helium-rich ejecta, possibly corresponding to the second generation of 
a GC-forming event.

The maximum He variation of GCs has also been shown to correlate with cluster mass
\citep[][]{milone22}, where higher mass clusters are more likely to form in the 
dense environments at high redshifts.
Additionally, the He abundance enhancement in second-generation GC stars relative 
to first-generation stars correlates with increased N, Na, and Al abundances 
and decreasing C, O, and Mg abundances \citep[e.g.,][]{milone15,marino19}.

The enhanced N/Fe observed in GCs has also been compared to anomalous N/O abundance 
patterns recently observed in high-redshift galaxies.
However, direct measurements of both N and O are rare for GC stars, with large 
variations in the C, N, and O inferred from variations in CH, CN, and NH molecular
bands (e.g., M13, 47 Tuc), making GC N enhancements difficult to link to high-redshift observations.
Furthermore, the lack of enhanced N/O in the AURORA high-He subsample does not
necessarily exclude these galaxies as globular cluster progenitors.
The integrated light of the large-scale ISM and multiple stellar 
populations in the AURORA could dilute a GC-like N/O signature
Alternatively, we could be catching the high-He subsample at an early phase
of GC formation where N enrichment (30-100 Myr from CNO cycling in
intermediate-mass AGB stars or fast-rotating massive stars) has not yet caught 
up to the He enhancement (more instantaneous from, e.g., VMS winds, binary-star
mass loss prior to convective dredge-up of N, or CCSNe).

Enhanced He alone, however, is not a unique diagnostic of GC progenitors. 
Other physical scenarios could lead to elevated He/H observations, such as post-burst 
populations in which massive star ejecta are trapped briefly before mixing, or localized 
ionization structure or geometry effects.
Follow-up observations could help disentangle these scenarios. 
For example, enhanced Na and Al and depleted C abundances could lend further 
support for GC formation, very high-resolution imaging with future large-aperture space
telescopes could reveal compact clusters or density-bound regions, and spatially-resolved 
spectroscopy could probe the spatial distribution of He and other elements.

If confirmed, He-enhancement at intermediate and high redshifts opens the exciting 
possibility of serving as a tracer of globular cluster formation -- a phenomenon which has long 
been predicted to peak at $z\sim2$, but has proved difficult to catch in action.


\section{Lessons Learned from Measuring He Abundances at High\texorpdfstring{$-z$}{}}\label{sec:best}
When measuring He abundances at moderate redshifts, the robustness of the results depend 
critically on whether the \ion{He}{1} \W10833 line is observable. 
At moderate redshifts of $z\lesssim3.6$, where the \ion{He}{1} \W10833 line remains 
within the JWST/NIRSpec wavelength window for $R\geq1000$ spectra ($0.6-5.0\mu$m), 
it is imperative to include this line in He abundance analyses.
It is the strongest \ion{He}{1} emission line within the rest-frame optical and NIR
and it provides an essential constraint on electron density that helps break degeneracies 
with electron temperature and optical depth that would otherwise bias the derived 
He$^+$/H$^+$ abundances. 
Note, however, that despite the relative strength of \W10833, it still fails to 
reach high enough equivalent widths to enable robust detections in lower resolution 
spectra (e.g., $R=100$ prism spectra).
We found that the inclusion of \W10833 in AURORA also enabled \ion{He}{1} 
density measurements that are independent from the commonly-used low-ionization 
[\ion{S}{2}] density diagnostic.
It also ensures more reliable optical depth corrections. 

We showed that the omission of the \ion{He}{1} \W10833 line from He abundance analyses 
can lead to underestimates of electron density, which can have a significant 
impact on the derived He$^+$/H$^+$ abundances. 
Because the optical \ion{He}{1} lines are moderately sensitive to density through 
collisional and optical depth effects, underestimating the density 
typically leads to an overestimate of the helium abundance by several percent 
\citep[e.g.,][]{izotov94}.
Specifically, in the low-density regime, neglecting density corrections can bias 
He$^+$/H$^+$ upward by $\sim1-3$\%, while at higher densities ($n_e \gtrsim 300$ cm$^{-3}$) 
the effect can be larger, reaching $\sim5$\% or more depending on the line set used. 
This is particularly important in precise abundance work, such as primordial helium 
determinations, where percent-level systematics can dominate the error budget. 
Including \ion{He}{1} \W10833 helps ensure that the derived densities — 
and thus the density-dependent optical depth corrections — are properly accounted for, 
yielding more reliable He abundances.

At higher redshifts of $z\gtrsim3.6$, where the \ion{He}{1} \W10833 line is redshifted out 
of NIRSpec’s spectral window, it is worth attempting \W10833 detections with 
deep MIRI spectra, where bright lensed galaxies would serve as ideal targets to
test the observability of this line.
If MIRI observations fail to be sensitive enough to detect \W10833 at $z\gtrsim3.6$,
it will then become crucial to obtain independent intermediate- to high-ionization 
density measurements from non-helium diagnostics, such as the [\ion{Ar}{4}] 
\W\W4711,4741 or \ion{C}{3}] \W\W1907,1909 emission-line doublet, or predict the \ion{He}{1} density from
low-ionization diagnostics like the [\ion{O}{2}] \W\W3727,3729 or [\ion{S}{2}] 
\W\W6718,6733 doublets, and to explicitly incorporate these constraints 
into helium abundance modeling. 
Without such density constraints, MCMC-based He abundance determinations risk drifting 
toward artificially low- or high-density solutions, depending on the optical depth regime, 
leading to systematic biases of several percent in the inferred He$^+$/H$^+$ ratios. 
As a result, combining \ion{He}{1} optical line measurements with robust external density 
and temperature constraints becomes essential for achieving accurate He abundances 
at high redshift and interpreting the evolution of He enrichment.


\section{Conclusions}\label{sec:conclusions}
We have presented the first robust helium abundance measurements in high-redshift 
($1.6\lesssim z\lesssim 3.3$) star-forming galaxies using deep JWST/NIRSpec spectroscopy 
from the AURORA survey. 
We defined a High$-z$ He Sample of 20 galaxies with detections of six or more high-S/N 
($>5\sigma$) \ion{He}{1} emission lines, including the critical NIR \W10833 line. 
This is the first time that the \W10833 line has been leveraged at high-redshifts to 
break the degeneracies between electron density and both optical depth and He$^+$/H$^+$, 
which has historically limited extragalactic He measurements. 
We developed a custom MCMC fitting framework that accounts for optical depth and collisional 
effects using updated $\tau_{\lambda 3890}$ model grid corrections that span densities up to 
$1\times10^6{\rm\ cm}^{-3}$ to match the range of properties observed for high-redshift galaxies.
With this code, we derived direct-method ionic He$^+$/H$^+$ abundances ($y^+$), anchored by 
measured temperature priors ($T_e$([\ion{O}{3}])) and leveraging the full suite of available 
\ion{He}{1} lines.
Ionic He$^{+2}$/H$^+$ abundances ($y^{+2}$) were also derived using the observed \ion{He}{2} 
\W4687 emission for three galaxies and added to $y^+$ to determine the total helium abundances 
($y$) and helium mass fractions, $Y$.

With our derived He abundances for $1.6\lesssim z\lesssim 3.3$ AURORA galaxies, we examined 
the evolution of He/H.
Our key conclusions are:
\begin{itemize}
    \item Typical $1.6\lesssim z\lesssim 3.3$ galaxies follow the expected 
    He/H enrichment trend: 
    Most AURORA galaxies are consistent with the extrapolated $z\sim 0$
    $Y-$O/H relation from \citet{aver22}, confirming that modest 
    He enrichment had already occurred by cosmic noon.
    \item A subsample of high-He galaxies exists at $1.6\lesssim z\lesssim 3.3$: 
    We identify a subpopulation of four galaxies with significantly elevated He mass fractions 
    above the ($\Delta Y > 0.03$), consistent with early He enrichment by very massive stars (VMSs). 
    These high-He outliers do not show any corresponding enhancement in N/O, disfavoring AGB-dominated enrichment which would simultaneously elevate both He and N.
    Their physical properties suggest that multiple ionizing stellar populations with different
    ages are present.    
    \item VMSs are plausible sources of He enrichment: 
    The abundance pattern of high He and normal N/O and $\alpha$-elements favors 
    enrichment by VMSs via stellar winds, rotation, and/or binary stripping. 
    If correct, short-lived VMSs could enrich the ISM with He prior to CCSNe or N production.
    \item The high-He subpopulation are potential globular cluster progenitors: 
    The elevated He levels in the gas of the high-He subpopulation are comparable 
    to those of second-generation stars in local metal-rich GCs. 
    These systems may represent an early phase of GC-formation caught in action, 
    where helium-rich ejecta were captured before significant mixing or secondary N enrichment.
    \item The \ion{He}{1} \W10833 line is essential for robust He abundances: 
    The inclusion of \W10833 provides critical leverage for breaking degeneracies between density, 
    optical depth, and He abundance. 
    Future high-redshift studies at $z\gtrsim3.6$, where \W10833 shifts out of the NIRSpec range, 
    would benefit from deep MIRI observations to extend the current analysis to earlier epochs. 
\end{itemize}
These results demonstrate the power of JWST \ion{He}{1} spectroscopy to trace early 
chemical enrichment and offer a promising new pathway to identify globular cluster progenitors 
and constrain stellar yields in the early universe.

\facilities{JWST (NIRSpec)}
\software{
\texttt{astropy} \citep{astropy:2013, astropy:2018, astropy:2022},
\texttt{calwebb},
\texttt{jupyter} \citep{kluyver16},
\texttt{nsclean},
\texttt{numpy} version 1.26 \citep{harris20},
\texttt{PyNeb} version 1.1.14 \citep{luridiana15},
\texttt{python},
\texttt{scipy.stats.spearmanr}}

\begin{acknowledgements}
This work is based [in part] on observations made with the NASA/ESA/CSA James Webb Space Telescope. The data were obtained from the Mikulski Archive for Space Telescopes at the Space Telescope Science Institute, which is operated by the Association of Universities for Research in Astronomy, Inc., under NASA contract NAS 5-03127 for JWST. These observations are associated with program \#1914.
\end{acknowledgements}
\clearpage
\begin{appendix}
\section{MCMC Code Testing}\label{sec:tests}
Here we discuss the different tests performed on our MCMC He abundance code to 
determine the best method for analyzing the High$-z$ He Sample and 
characterize any biases. 
We examine initial constraints on the temperature and density parameters in
Section~\ref{sec:priors}, the effect of priors on the resulting Y values in 
Section~\ref{sec:Y_priors}, the impact of the \W10833 NIR line on derived 
properties in Section~\ref{sec:10833}, and inferred $y^+$ values using 
single \ion{He}{1} lines that are insensitive to the physical conditions
of the gas in Section~\ref{sec:5877}. 

\subsection{Testing Priors}\label{sec:priors}
Since several \ion{He}{1} lines have a strong-sensitivity to density, and 
density can be degenerate with temperature and secondarily with optical depth, 
constraining the nebular properties in our He$^+$ analysis is very important. 
However, it is not clear whether measured nebular properties provide 
appropriate model guidance.
In other words, it is not obvious whether the commonly-measured high-ionization 
O$^{+2}$ ($35-55$ eV) temperature and low-ionization S$^+$ ($10-23$ eV) density 
are characteristic of the He$^+$ gas ($25-54$ eV) or not.
To assess the robustness of our He model assumptions, we ran four MCMC analyses
with different prior configurations:
\begin{itemize}[leftmargin=5em, itemsep=1pt, parsep=2pt]
    \item[{\it (1)}] No priors on $T_e$(\ion{He}{1}) or $n_e$(\ion{He}{1}).
    \item[{\it (2)}] A Gaussian prior on $T_e$(\ion{He}{1}) centered  
        on the measured or inferred $T_e$([\ion{O}{3}]), with a standard deviation 
        defined by the uncertainty on $T_e$([\ion{O}{3}]).
    \item[{\it (3)}] A Gaussian prior on $\log n_e$(\ion{He}{1}) centered on
    the measured $\log n_e$([\ion{S}{2}]), with a standard deviation 
        defined by the uncertainty on $\log n_e$([\ion{S}{2}]). 
    \item[{\it (4)}] A combination of both the above $T_e$ {\it (2)} and $n_e$ {\it (3)} priors. 
\end{itemize}

\subsubsection{Effect of Priors on \texorpdfstring{$T_e$}{} and \texorpdfstring{$n_e$}{}}
\begin{figure*}[ht]
\begin{center}
	\includegraphics[height=0.4\linewidth, trim=0 4mm 0 0, clip]{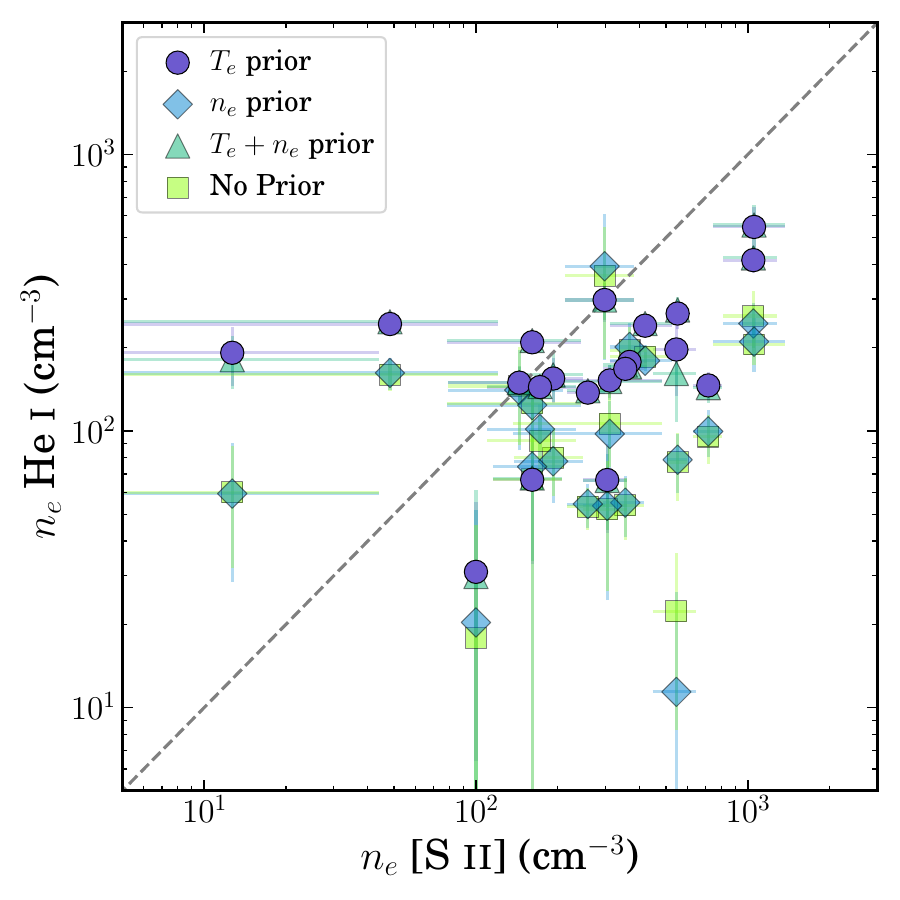} 
    \includegraphics[height=0.4\linewidth, trim=0 4mm 0 0, clip]{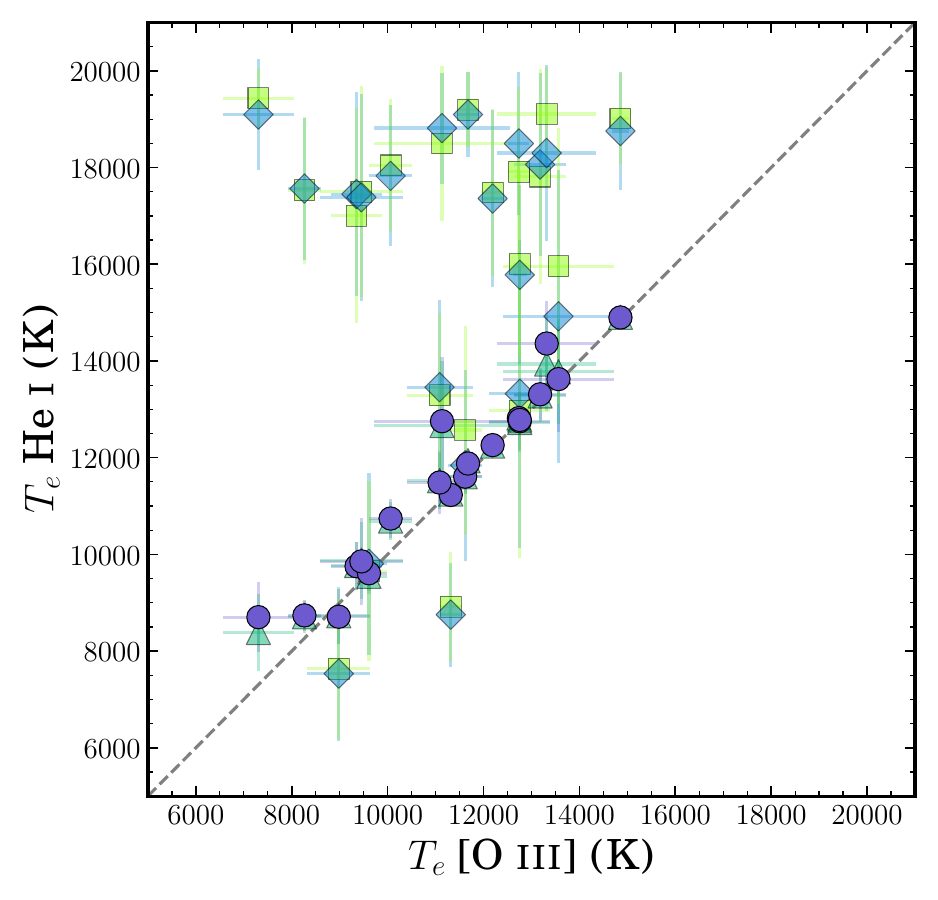}
\caption{Comparison of the electron density ($n_e$) (left) and temperature ($T_e$)
(right) determined from the \ion{He}{1} fitting code versus the properties measured
directly from the [\ion{S}{2}] and [\ion{O}{3}] emission lines, respectively.
The best fits for the AURORA High$-z$ He Sample are plotted as purple points.
In comparison, we plot the densities and temperatures returned using different 
priors of measured $T_e$ (blue diamonds), $n_e$ (green triangles), and 
both $T_e$ and $n_e$ (lime green squares)
Densities are generally well-constrained by modeling the \ion{He}{1} emission lines,
while temperatures are much better constrained by a $T_e$([\ion{O}{3}]) prior.\label{fig:priors}}
\end{center}
\end{figure*}

We plot the density and temperature results from the analyses using the four 
different prior configurations in Figure~\ref{fig:priors} to evaluate how the 
inclusion of prior information influences the inferred physical parameters and 
to identify potential degeneracies between the fit parameters.
In the left-hand panel of Figure~\ref{fig:priors} we compare the MCMC-derived 
$n_e$(\ion{He}{1}) with the measured low-ionization density, $n_e$([\ion{S}{2}]).
While the median inferred \ion{He}{1} density of the sample changes very little 
(median $n_e$(\ion{He}{1}) with and without a $n_e-$prior are $101\pm23$ and $96\pm24$ 
cm$^{-3}$, respectively), there is significant dispersion when no priors are used, 
as characterized by the standard deviation of the inferred \ion{He}{1} densities: 
$\sigma_{n_e,(1)}=250$ cm$^{-3}$.
The dispersion is similar when a density prior is used, with $\sigma_{n_e,(3)}=240$ 
cm$^{-3}$, suggesting that the density is only loosely constrained by the data alone 
and that the inferred \ion{He}{1} density may be incorrectly biased by the [\ion{S}{2}] 
density prior.
Adding a temperature prior reduces the dispersion for both the $T_e$ prior alone and when 
combined with a $n_e$ prior: $\sigma_{n_e,(2)}=\sigma_{n_e,(4)}=200$ cm$^{-3}$.
Note that using both priors together does not reduce the dispersion beyond the 
temperature prior alone. 

In the right-hand panel of Figure~\ref{fig:priors} we compare the MCMC-derived 
$T_e$(\ion{He}{1}) with the measured high-ionization temperature, $T_e$([\ion{O}{3}]).
In the absence of any priors, the \ion{He}{1} temperatures are biased high in most cases, 
with an average dispersion of $\sigma_{T_e,(1)}=4900$ K.
Adding a density prior has little effect on the \ion{He}{1} 
temperatures, retaining in large standard deviation of 
$\sigma_{T_e,(3)}=4800$ K.
The best \ion{He}{1} temperature constraints result from using a temperature prior, 
reducing the dispersion to $\sigma_{T_e,(4)}=370$ K and $\sigma_{T_e,(2)}=340$ K with 
and without a density prior, respectively.
The temperature prior also constrains the characteristic temperature of the High$-z$ 
He Sample; without a $T_e-$prior, the median is $T_e=(1.75\pm0.90)\times10^4$ K, while 
with a $T_e-$prior, the median temperature is lower at $T_e=(1.17\pm600)\times10^4$ K. 

The above analysis suggests that robust $T_e$ measurement priors are useful for breaking 
the degeneracy with density and provide the most reliable He abundance analysis, 
in agreement with past studies.
In particular, $T_e$(O$^{+2}$) is thought to be physically consistent with $T_e$(He$^{+}$) 
\citep[e.g.,][]{peimbert07,peimbert16} and has been argued to reduce degeneracies 
\citep[e.g.,][]{aver11,izotov14,aver15}. 
This also agrees with what we know about the set of \ion{He}{1} emission lines used in the 
fitting (see \S~\ref{sec:sens}) -- $T_e$ is not well constrained by the lines alone, but 
$n_e$ can be.
Since we required the extremely density-sensitive \ion{He}{1} \W10833 line in our He 
analysis, the MCMC is already able to self-consistently constrain the density directly 
from the observations, without the need for an external density prior. 
Adding a $n_e$-prior in combination with \ion{He}{1} \W10833 effectively double-weights 
the density constraints, which can artificially narrow the posterior and under-represent 
the true uncertainty in $n_e$ and other derived parameters. 
Thus, if \ion{He}{1} \W10833 is included, the MCMC is better left to infer the density 
(and, consequently, the optical depth $\tau_{\lambda3890}$) directly from the emission 
line data, keeping the analysis as empirical and unbiased as possible.

In contrast, a $T_e$-prior can be useful because (as discussed in \S~\ref{sec:sens}) the 
\ion{He}{1} lines are only weakly sensitive to temperature.
Further, the tight trend between derived \ion{He}{1} $T_e$ and measured 
$T_e$([\ion{O}{3}]) in Figure~\ref{fig:priors} suggests that the [\ion{O}{3}] temperature 
effectively characterizes the temperature of the He$^{+}$ gas in these galaxies 
\citep[see, also,][]{peimbert07,peimbert16}. 
Thus, $T_e$(\ion{He}{1}) is accurately constrained by the externally-measured 
$T_e$([\ion{O}{3}]), and so we adopt the $T_e$-prior configuration with our High$-z$ He 
Sample, including \W10833, as our best-practice MCMC He$^+$ analysis.
Overall, we find good constraints on all four fit parameters (see Table~\ref{tbl3}).


\subsection{Effect of Priors on \texorpdfstring{$Y$}{}}\label{sec:Y_priors}
\begin{figure*}
\begin{center}
	\includegraphics[width=0.55\linewidth, trim=0 4mm 0 0, clip]{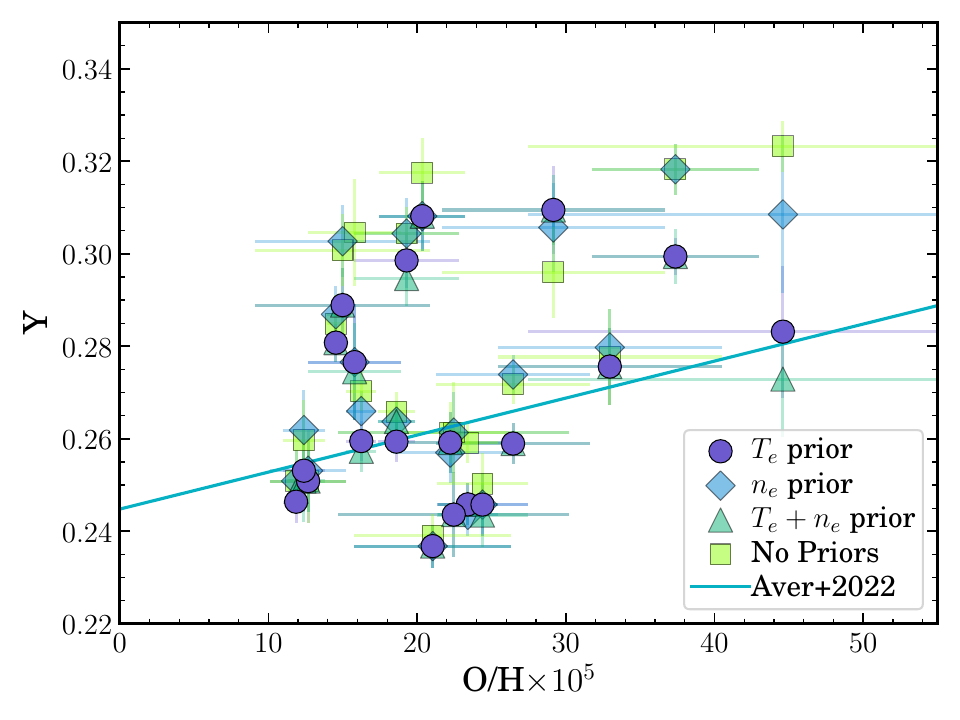}
\caption{Helium abundance (mass fraction) determined from the \ion{He}{1} fitting 
code versus oxygen abundance.
The best fits using a $T_e$ prior for the AURORA High$-z$ He Sample 
are plotted as purple points.
In comparison, we plot the He abundances determined using different priors for the He 
abundance MCMC code of no priors (lime green squares), measured $n_e$ priors (blue 
diamonds), and both $T_e$ and $n_e$ priors (green triangles).
The AURORA sample is generally scattered around the extrapolation of the trend found 
for nearby galaxies by \citet[][blue line]{aver22} to higher O/H abundances for all 
methods, however, the tightest trends are produced using the $T_e$ prior.\label{fig:Ypriors}}
\end{center}
\end{figure*}

We examined the effects of using different priors in our MCMC analysis on the 
derived electron temperatures and densities in Section~\ref{sec:priors}.
Here we expand upon that analysis by exploring how the choice of physical priors influences 
the derived helium mass fraction ($Y$) as a function of oxygen abundance (O/H) in 
the AURORA High$-z$ He sample. 
Figure~\ref{fig:Ypriors} compares the same four MCMC fitting configurations for the 
\ion{He}{1} analysis as Section~\ref{sec:priors}.
For comparison, we also show the $z\sim0$ trend from \citet[][blue line]{aver22}.

Across the range of metallicities seen in the AURORA sample, we find that the use of 
no priors produces the largest scatter in $Y$, with a tendency for up-scatter towards 
higher $Y$ values at all oxygen abundances, resulting in the highest median value of
$Y_{(1)} = 0.271\pm0.016$.
A $n_e$ prior alone has little effect on the distribution with $Y_{(3)} = 0.270\pm0.013$.
Applying a $T_e$ prior alone tightens the relationship, reducing the dispersion, 
especially for the two points at high metallicity, and lowers the median to 
$Y_{(2)} = 0.259\pm0.007$.
The combined $T_e+n_e$ prior yields a similar trend as the $T_e$ prior, with
$Y_{(4)} = 0.261\pm0.006$.
Both trends using the $T_e$ prior are generally consistent with the \citet{aver22} trend 
extrapolated to larger O/H, but with larger dispersion and a few significant outliers 
(the high-He subsample).

The trends in Figure~\ref{fig:Ypriors} illustrate the critical role of temperature priors 
in mitigating degeneracies and producing stable, consistent helium abundance measurements. 
Notably, the average results using a $T_e$ prior align closely with expectations from 
stellar nucleosynthesis and show good agreement with the trend from \cite{aver22}.
Further, despite significant differences in the inferred $T_e$ and $n_e$ values for 
different MCMC prior configurations, the inferred $Y$ values do vary significantly, 
reinforcing the robustness of our adopted methodology where $y^+$ is well constrained by 
the carefully chosen set of eight \ion{He}{1} lines used in this work (see Table 3).
Finally, while our adopted method of using $T_e$ priors only produces similar $Y$ values 
as the joint $T_e+n_e$ priors, it has the important advantage of allowing for real 
differences in the density, optical depth, and He abundance to emerge for individual 
galaxies. 


\subsection{Impact of the \texorpdfstring{\ion{He}{1} \W10833}{} Line}\label{sec:10833}
\begin{figure*}[ht]
\begin{center}
	\includegraphics[width=0.85\linewidth, trim=0 0mm 0 0, clip]{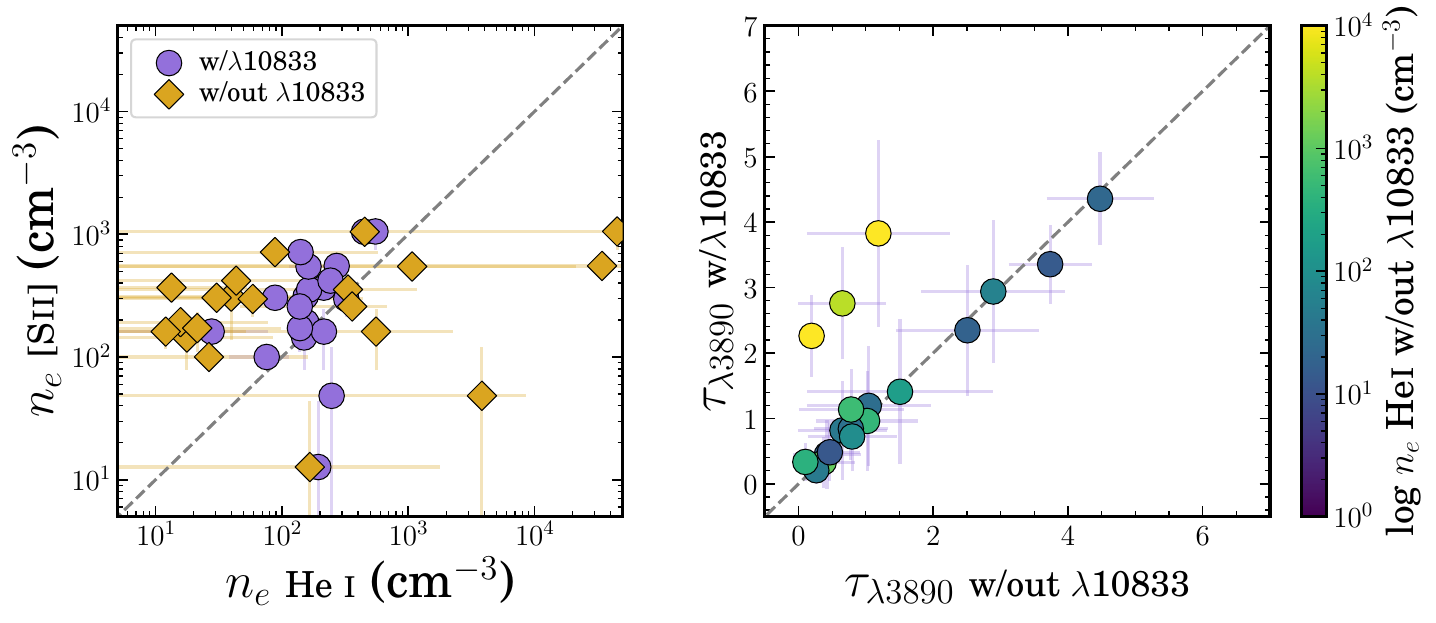}
\caption{Dependence of best-fit parameters on the inclusion of the \ion{He}{1} \W10833 line.
Results from the MCMC He$^+$ abundance code with a $T_e$-prior are shown for the High$-z$ 
He Sample both with the density-sensitive \ion{He}{1} \W10833 line (purple points) 
and without it (gold diamonds).
{\it Left:} Measured [\ion{S}{2}] densities versus best-fit \ion{He}{1} densities show
that the MCMC code generally favors lower-density solutions in the absence of \W10833.
{\it Right:} Derived optical depths, $\tau_{\lambda3890}$, are generally consistent 
from runs with and without \W10833 included. 
The three exceptions shifted to low $\tau_{\lambda3890}$ when \W10833 is excluded
are also offset to high densities, demonstrating the code's biased compensation
mechanism for handling either higher optical depths or densities.
 \label{fig:fitvals}}
\end{center}
\end{figure*}

The inclusion of the \ion{He}{1} \W10833 line in our MCMC analysis is important for 
breaking degeneracies, which in turn can help identify targets with high optical depths, 
and has a significant impact on the inferred electron densities.
Figure~\ref{fig:fitvals} shows the derived $n_e$ and $\tau_{\lambda3890}$ values derived 
from our MCMC code, with the use of a $T_e$ prior, with and without the inclusion 
of the \ion{He}{1} \W10833 line.
When \ion{He}{1} \W10833 is excluded, the \ion{He}{1} density is {\it underestimated} 
for much of the sample compared to cases where the line is included, and the errors are 
significantly larger due to the lack of constraints. 
This is because the optical \ion{He}{1} lines (e.g., \W4471, \W5876, \W6678) are only 
weakly sensitive to density (see Figure~\ref{fig:linesens}), especially at low to moderate 
densities ($n_e \lesssim 300$ cm$^{-3}$), making the interplay between density and optical 
depth in the line formation difficult to disentangle. 

Without a strongly density-sensitive diagnostic like \W10833, the MCMC code 
tends to prefer low-density solutions because it can still reproduce the observed 
optical line ratios by adjusting the optical depth ($\tau_{\lambda3890}$).
This reflects an inherent degeneracy in the \ion{He}{1} model grid: changes in $n_e$ can 
often be compensated by anti-correlated changes in $\tau_{\lambda3890}$, since both affect 
the triplet-to-singlet line ratios in similar ways. 
As a result, the MCMC posterior tends toward the lowest densities allowed by the grid
for much of the sample, effectively biasing the solution toward conditions where the 
optical \ion{He}{1} emissivities are less dependent on density corrections. 
This is especially problematic for precise He abundance determinations, as underestimated 
densities affect the emissivity corrections and, thus, lead to a biased He$^+$/H$^+$ value.

Since we've already broken the primary $T_e-n_e$ degeneracy by using a $T_e$ prior, the 
inclusion of \W10833 helps breaks the $n_e$-$\tau_{\lambda3890}$ and $n_e-$H$^+$ 
degeneracies by providing a direct, collisionally-sensitive handle on the density.
This constraint prevents the MCMC code from artificially sliding toward a more extreme 
combined density/optical-depth solution that fits the optical line ratios but does not 
reflect the true density of the He$^+$ gas.
As shown in the left panel of Figure~\ref{fig:fitvals}, when the \ion{He}{1} 
\W10833 is incorporated, the MCMC code prefers higher \ion{He}{1} densities
for much of the sample.
These \ion{He}{1} densities are also in better agreement with the independent 
[\ion{S}{2}] density measurements. 

The the $n_e$-$\tau_{\lambda3890}$ degeneracy can also swing the other direction to
{\it overestimated} $n_e$ paired with {\it underestimated} $\tau_{\lambda3890}$.
This effect can be seen in the three targets with high \ion{He}{1} densities in 
Figure~\ref{fig:fitvals}. 
This suggests that the triplet lines have significant optical depths in these targets,
but the similar influence of $n_e$ and $\tau$ on the level populations can confuse the 
interpretation.
As a result, the MCMC code may compensate for the lack of optical depth information 
by adjusting density upward to fit the observed line ratios.
This reinforces why \W10833 is so valuable: it anchors the density solution and prevents 
the code from artificially inflating density to absorb optical depth effects, 
and vise versa.

\subsection{Single-Line \texorpdfstring{He$^+$}{} Abundances}\label{sec:5877}
To assess whether our results are biased by density constraints, we 
performed a test using the He$^+$ MCMC code using a $T_e$ prior with 
only a single \ion{He}{1} line, either \W5877 or \W6680, excluding 
other \ion{He}{1} lines,
The utility of this approach lies in isolating the impact of a single, 
strong recombination line that is largely insensitive to density and 
optical depth effects under typical nebular conditions. 
By comparing the He abundances derived from \W5877 alone and from \W6680 
alone to those from the full multi-line analysis, we can evaluate how 
much additional information, and potential bias, the density-sensitive 
lines introduce. 

\begin{figure*}[ht]
\begin{center}
	\includegraphics[width=0.580\linewidth, trim=2mm 100mm -5mm 0, clip]{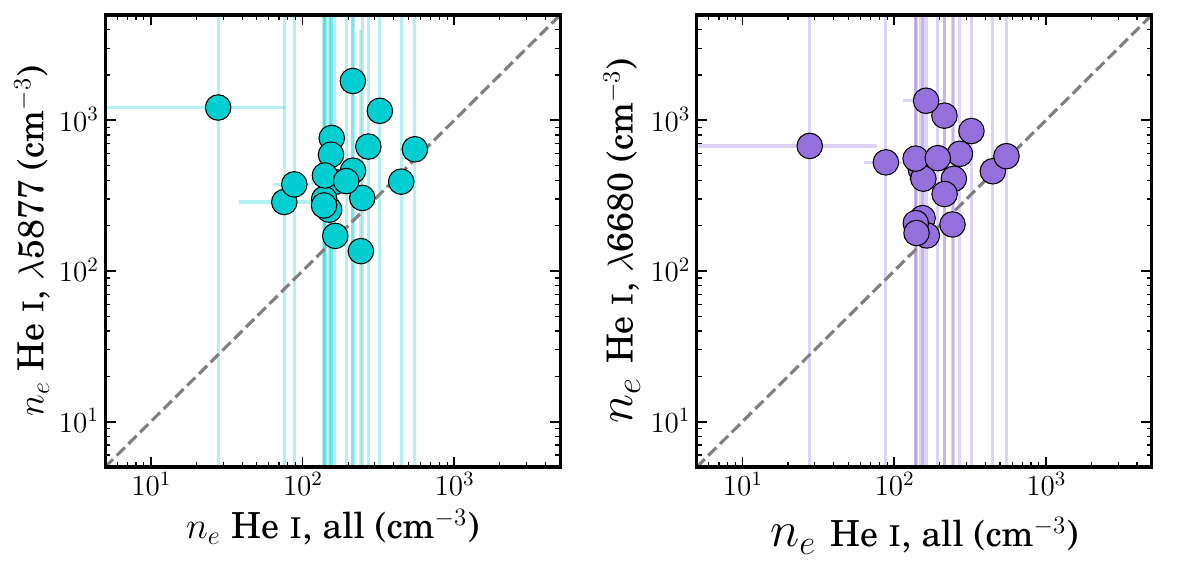} \\
    \hspace{9mm}\includegraphics[height=0.565\linewidth, trim=0 0mm 0 0, clip]{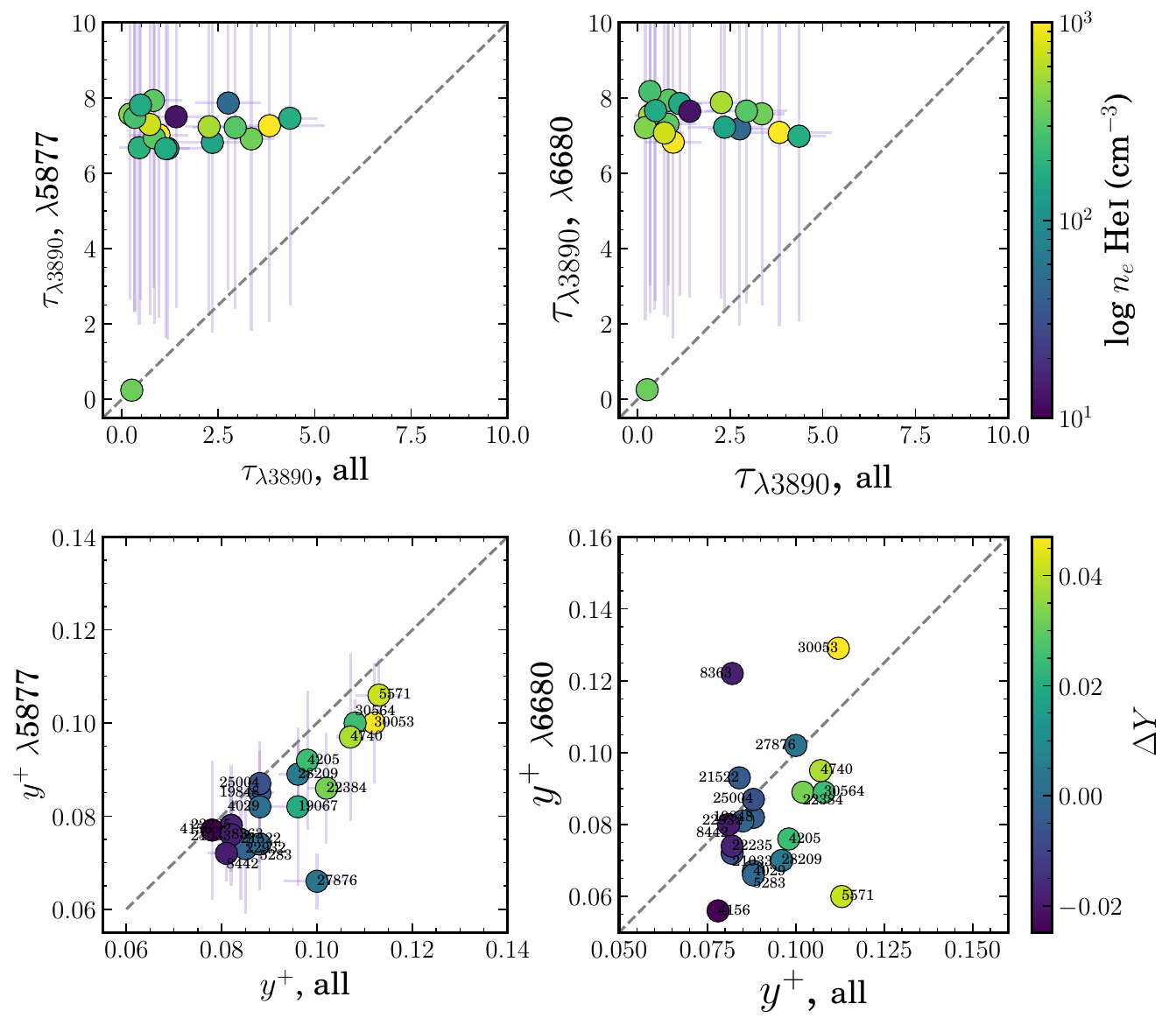} \\
\caption{Comparison of helium abundance and physical parameters derived from the full 
multi-line MCMC analysis (x-axes) versus those derived using only a single 
\ion{He}{1} line (y-axes).
The left column shows results using the \ion{He}{1} \W5877 line, 
while the right column shows results using the \ion{He}{1} \W6680 line.
The \ion{He}{1} \W5877 and \W6680 lines were selected for their emission 
strength and relatively-insensitivity to $T_e$, $n_e$, and $\tau_{\lambda3890}$. 
Each row shows one of the fit parameters from the MCMC analysis: 
electron density $n_e$ (top), optical depth $\tau_{\lambda3890}$ (middle), 
and He$^+$/H$^+$ abundance $y^+$ (bottom). 
The single-line fits produce significantly weaker constraints on $n_e$ and 
$\tau_{\lambda3890}$ due to each line's insensitivity to these parameters, 
while $y^+$ shows better agreement between the two approaches. 
Points are color-coded by the \ion{He}{1} density in the middle row and
by helium mass fraction offset, $\Delta Y$, defined as the deviation of 
the multi-line helium abundance from the \citet{aver22} trend at the 
corresponding O/H value, in the bottom row/
Objects with larger positive $\Delta Y$ exhibit the greatest excesses 
above the expected primordial plus chemical evolution relation. 
This comparison demonstrates that while the single line analyses alone 
can recover approximate $y^+$ values, multi-line fits are essential for 
breaking parameter degeneracies and assessing deviations from established 
$Y$-O/H trends.
\label{fig:5877}}
\end{center}
\end{figure*}

We plot the results of the single line H$^+$ analysis versus the 
multi-line MCMC results, both using a $T_e$ prior, in 
Figure~\ref{fig:5877} for \W5877 in the left column and for 
\W6680 in the right column.
In the top row, we compare the \ion{He}{1} densities returned 
from the MCMC code using a single line versus the full 
multi-line analysis, finding that the electron densities 
inferred from the single line analyses are less constrained 
(large error bars) and offset to higher densities than the
full multi-line results for most objects.
Because \W5877 and \W6680 are largely insensitive to density, 
they are expected to provide weak constraints on $n_e$.

In the middle row of Figure~\ref{fig:5877}, we perform the same
comparison but for the optical depth, $\tau_{\lambda 3890}$.
The both the \W5877-only and the \W6680-only fits produce very 
high and poorly constrained optical depths (clustered near 
the upper limit on the flat prior), while the multi-line fits 
yield better-constrained and more physically plausible 
$\tau_{\lambda 3890}$ values.
The color shows log $n_e$, indicating both density and optical
depth are constrained to more reasonable values when multi-line 
fits are used.
This shows the value of including multiple lines, especially 
optical depth-sensitive triplet lines, which help break 
degeneracies that are otherwise unconstrained in fits using 
only singlet lines.

Finally, in the bottom row of Figure~\ref{fig:5877}, 
we compare the resulting $y^+$ values.
There is generally good agreement, with points clustered
near the 1-to-1 line, but the single-line results tend to 
yield slightly lower $y^+$ than the multi-line fits.
The color bar indicates the difference in inferred helium mass 
fraction, $\Delta Y$, from the \citet{aver22} trend at a given O/H. 
The full multi-line fits provide tighter constraints and 
typically yield higher helium abundances than \W5877- or 
\W6680-only. 
This could reflect the importance of parameter-sensitive lines 
like \W10833 and \W7067 in correcting $y^+$ values 
when $\tau_{\lambda 3890}$ and $n_e$ are unconstrained.

Overall, Figure~\ref{fig:5877} shows that the inclusion of 
multiple \ion{He}{1} lines does not introduce strong biases 
$y^+$, but does improves precision and breaks degeneracies, 
especially for optical depth.
As a result, the multiple line method yields more robust and 
likely more accurate helium abundances ($y^+$).
Meanwhile, single-line fits are valuable benchmarks that 
demonstrate the full MCMC results are not over-constrained 
or skewed by parameter-sensitive lines --
rather, they add important physical constraints.

\end{appendix}

\bibliographystyle{aasjournal}
\bibliography{mybib}

\begin{thebibliography}{}
\expandafter\ifx\csname natexlab\endcsname\relax\def\natexlab#1{#1}\fi
\providecommand{\url}[1]{\href{#1}{#1}}
\providecommand{\dodoi}[1]{doi:~\href{http://doi.org/#1}{\nolinkurl{#1}}}
\providecommand{\doeprint}[1]{\href{http://ascl.net/#1}{\nolinkurl{http://ascl.net/#1}}}
\providecommand{\doarXiv}[1]{\href{https://arxiv.org/abs/#1}{\nolinkurl{https://arxiv.org/abs/#1}}}

\bibitem[{{Aggarwal} \& {Keenan}(1999)}]{aggarwal99}
{Aggarwal}, K.~M., \& {Keenan}, F.~P. 1999, \apjs, 123, 311

\bibitem[{{Asplund} {et~al.}(2021){Asplund}, {Amarsi}, \& {Grevesse}}]{asplund21}
{Asplund}, M., {Amarsi}, A.~M., \& {Grevesse}, N. 2021, \aap, 653, A141, \dodoi{10.1051/0004-6361/202140445}

\bibitem[{{Astropy Collaboration} {et~al.}(2022){Astropy Collaboration}, {Price-Whelan}, {Lim}, {et~al.}}]{astropy:2022}
{Astropy Collaboration}, {Price-Whelan}, A.~M., {Lim}, P.~L., {et~al.} 2022, \apj, 935, 167, \dodoi{10.3847/1538-4357/ac7c74}

\bibitem[{{Astropy Collaboration} {et~al.}(2018){Astropy Collaboration}, {Price-Whelan}, {Sip{\H{o}}cz}, {G{\"u}nther}, {Lim}, {et~al.}}]{astropy:2018}
{Astropy Collaboration}, {Price-Whelan}, A.~M., {Sip{\H{o}}cz}, B.~M., {et~al.} 2018, \aj, 156, 123, \dodoi{10.3847/1538-3881/aabc4f}

\bibitem[{{Astropy Collaboration} {et~al.}(2013){Astropy Collaboration}, {Robitaille}, {Tollerud}, {Greenfield}, {Droettboom}, {et~al.}}]{astropy:2013}
{Astropy Collaboration}, {Robitaille}, T.~P., {Tollerud}, E.~J., {et~al.} 2013, \aap, 558, A33, \dodoi{10.1051/0004-6361/201322068}

\bibitem[{{Aver} {et~al.}(2022){Aver}, {Berg}, {Hirschauer}, \& {others}}]{aver22}
{Aver}, E., {Berg}, D.~A., {Hirschauer}, A.~S., \& {others}. 2022, \mnras, 510, 373, \dodoi{10.1093/mnras/stab3226}

\bibitem[{{Aver} {et~al.}(2021){Aver}, {Berg}, {Olive}, \& {others}}]{aver21}
{Aver}, E., {Berg}, D.~A., {Olive}, K.~A., \& {others}. 2021, \jcap, 2021, 027, \dodoi{10.1088/1475-7516/2021/03/027}

\bibitem[{{Aver} {et~al.}(2013){Aver}, {Olive}, {Porter}, \& {Skillman}}]{aver13}
{Aver}, E., {Olive}, K.~A., {Porter}, R.~L., \& {Skillman}, E.~D. 2013, \jcap, 2013, 017, \dodoi{10.1088/1475-7516/2013/11/017}

\bibitem[{{Aver} {et~al.}(2010){Aver}, {Olive}, \& {Skillman}}]{aver10}
{Aver}, E., {Olive}, K.~A., \& {Skillman}, E.~D. 2010, \jcap, 2010, 003, \dodoi{10.1088/1475-7516/2010/05/003}

\bibitem[{{Aver} {et~al.}(2011){Aver}, {Olive}, \& {Skillman}}]{aver11}
---. 2011, \jcap, 2011, 043, \dodoi{10.1088/1475-7516/2011/03/043}

\bibitem[{{Aver} {et~al.}(2012){Aver}, {Olive}, \& {Skillman}}]{aver12}
---. 2012, \jcap, 2012, 004, \dodoi{10.1088/1475-7516/2012/04/004}

\bibitem[{{Aver} {et~al.}(2015){Aver}, {Olive}, \& {Skillman}}]{aver15}
---. 2015, \jcap, 2015, 011, \dodoi{10.1088/1475-7516/2015/07/011}

\bibitem[{{Bastian} \& {Lardo}(2018)}]{bastian18}
{Bastian}, N., \& {Lardo}, C. 2018, \araa, 56, 83, \dodoi{10.1146/annurev-astro-081817-051839}

\bibitem[{{Benjamin} {et~al.}(2002){Benjamin}, {Skillman}, \& {Smits}}]{benjamin02}
{Benjamin}, R.~A., {Skillman}, E.~D., \& {Smits}, D.~P. 2002, \apj, 569, 288, \dodoi{10.1086/339242}

\bibitem[{{Berg} {et~al.}(2021){Berg}, {Chisholm}, {Erb}, {Skillman}, {Pogge}, \& {Olivier}}]{berg21}
{Berg}, D.~A., {Chisholm}, J., {Erb}, D.~K., {et~al.} 2021, \apj, 922, 170, \dodoi{10.3847/1538-4357/ac141b}

\bibitem[{{Berg} {et~al.}(2018){Berg}, {Erb}, {Auger}, {Pettini}, \& {Brammer}}]{berg18}
{Berg}, D.~A., {Erb}, D.~K., {Auger}, M.~W., {Pettini}, M., \& {Brammer}, G.~B. 2018, \apj, 859, 164

\bibitem[{{Berg} {et~al.}(2019){Berg}, {Erb}, {Henry}, {Skillman}, \& {McQuinn}}]{berg19a}
{Berg}, D.~A., {Erb}, D.~K., {Henry}, R.~B.~C., {Skillman}, E.~D., \& {McQuinn}, K.~B.~W. 2019, \apj, 874, 93, \dodoi{10.3847/1538-4357/ab020a}

\bibitem[{Berg {et~al.}(2012)Berg, Skillman, Marble, {et~al.}}]{berg12}
Berg, D.~A., Skillman, E.~D., Marble, A., {et~al.} 2012, \apj, 754, 98

\bibitem[{{Berg} {et~al.}(2022){Berg}, {James}, {King}, {McDonald}, {Chen}, {Chisholm}, {Heckman}, {Martin}, {Stark}, {Aloisi}, {Amor{\'\i}n}, {Arellano-C{\'o}rdova}, {Bayliss}, {Bordoloi}, {Brinchmann}, {Charlot}, {Chevallard}, {Clark}, {Erb}, {Feltre}, {Gronke}, {Hayes}, {Henry}, {Hernandez}, {Jaskot}, {Jones}, {Kewley}, {Kumari}, {Leitherer}, {Llerena}, {Maseda}, {Mingozzi}, {Nanayakkara}, {Ouchi}, {Plat}, {Pogge}, {Ravindranath}, {Rigby}, {Sanders}, {Scarlata}, {Senchyna}, {Skillman}, {Steidel}, {Strom}, {Sugahara}, {Wilkins}, {Wofford}, {Xu}, \& {Classy Team}}]{berg22}
{Berg}, D.~A., {James}, B.~L., {King}, T., {et~al.} 2022, \apjs, 261, 31, \dodoi{10.3847/1538-4365/ac6c03}

\bibitem[{{Brinchmann}(2023)}]{brinchmann23}
{Brinchmann}, J. 2023, \mnras, 525, 2087, \dodoi{10.1093/mnras/stad1704}

\bibitem[{{Bunker} {et~al.}(2023){Bunker}, {Saxena}, {Cameron}, {et~al.}}]{bunker23}
{Bunker}, A.~J., {Saxena}, A., {Cameron}, A.~J., {et~al.} 2023, \aap, 677, A88, \dodoi{10.1051/0004-6361/202346159}

\bibitem[{{Calzetti} {et~al.}(2000){Calzetti}, {Armus}, {Bohlin}, {et~al.}}]{calzetti00}
{Calzetti}, D., {Armus}, L., {Bohlin}, R.~C., {et~al.} 2000, \apj, 533, 682

\bibitem[{Campbell {et~al.}(1986)Campbell, Terlevich, \& Melnick}]{campbell86}
Campbell, A., Terlevich, R., \& Melnick, J. 1986, \mnras, 223, 811

\bibitem[{Cardelli {et~al.}(1989)Cardelli, Clayton, \& Mathis}]{cardelli89}
Cardelli, J.~A., Clayton, G.~C., \& Mathis, J.~S. 1989, \apj, 345, 245

\bibitem[{{Carnall} {et~al.}(2023){Carnall}, {Begley}, {McLeod}, {Hamadouche}, {Donnan}, {McLure}, {Dunlop}, {Milvang-Jensen}, {Bondestam}, {Cullen}, {Jewell}, \& {Pollock}}]{carnall23}
{Carnall}, A.~C., {Begley}, R., {McLeod}, D.~J., {et~al.} 2023, \mnras, 518, L45, \dodoi{10.1093/mnrasl/slac136}

\bibitem[{{Chabrier}(2003)}]{chabrier03}
{Chabrier}, G. 2003, \pasp, 115, 763, \dodoi{10.1086/376392}

\bibitem[{{Chatzikos} {et~al.}(2023){Chatzikos}, {Bianchi}, {Camilloni}, {Chakraborty}, {Gunasekera}, {Guzm{\'a}n}, {Milby}, {Sarkar}, {Shaw}, {van Hoof}, \& {Ferland}}]{chatzikos23}
{Chatzikos}, M., {Bianchi}, S., {Camilloni}, F., {et~al.} 2023, \rmxaa, 59, 327, \dodoi{10.22201/ia.01851101p.2023.59.02.12}

\bibitem[{{Cohen} \& {Mel{\'e}ndez}(2005)}]{cohen05}
{Cohen}, J.~G., \& {Mel{\'e}ndez}, J. 2005, \aj, 129, 303, \dodoi{10.1086/426369}

\bibitem[{{Conroy} {et~al.}(2009){Conroy}, {Gunn}, \& {White}}]{conroy09}
{Conroy}, C., {Gunn}, J.~E., \& {White}, M. 2009, \apj, 699, 486, \dodoi{10.1088/0004-637X/699/1/486}

\bibitem[{{Cooke} \& {Fumagalli}(2018)}]{cooke18}
{Cooke}, R.~J., \& {Fumagalli}, M. 2018, Nature Astronomy, 2, 957, \dodoi{10.1038/s41550-018-0584-z}

\bibitem[{{Crowther} {et~al.}(2016){Crowther}, {Caballero-Nieves}, {Bostroem}, {Ma{\'\i}z Apell{\'a}niz}, {Schneider}, {et~al.}}]{crowther16}
{Crowther}, P.~A., {Caballero-Nieves}, S.~M., {Bostroem}, K.~A., {et~al.} 2016, \mnras, 458, 624, \dodoi{10.1093/mnras/stw273}

\bibitem[{{Crowther} {et~al.}(2010){Crowther}, {Schnurr}, {Hirschi}, {Yusof}, {Parker}, {et~al.}}]{crowther10}
{Crowther}, P.~A., {Schnurr}, O., {Hirschi}, R., {et~al.} 2010, \mnras, 408, 731, \dodoi{10.1111/j.1365-2966.2010.17167.x}

\bibitem[{{Cyburt} {et~al.}(2016){Cyburt}, {Fields}, {Olive}, \& {Yeh}}]{cyburt16}
{Cyburt}, R.~H., {Fields}, B.~D., {Olive}, K.~A., \& {Yeh}, T.-H. 2016, Reviews of Modern Physics, 88, 015004, \dodoi{10.1103/RevModPhys.88.015004}

\bibitem[{{Dupree} {et~al.}(2011){Dupree}, {Strader}, \& {Smith}}]{dupree11}
{Dupree}, A.~K., {Strader}, J., \& {Smith}, G.~H. 2011, \apj, 728, 155, \dodoi{10.1088/0004-637X/728/2/155}

\bibitem[{{El-Badry} {et~al.}(2019){El-Badry}, {Quataert}, {Weisz}, {Choksi}, \& {Boylan-Kolchin}}]{elbadry19}
{El-Badry}, K., {Quataert}, E., {Weisz}, D.~R., {Choksi}, N., \& {Boylan-Kolchin}, M. 2019, \mnras, 482, 4528, \dodoi{10.1093/mnras/sty3007}

\bibitem[{{Eldridge} {et~al.}(2017){Eldridge}, {Stanway}, {Xiao}, {McClelland}, \& {others}}]{eldridge17}
{Eldridge}, J.~J., {Stanway}, E.~R., {Xiao}, L., {McClelland}, L.~A.~S., \& {others}. 2017, \pasa, 34, e058, \dodoi{10.1017/pasa.2017.51}

\bibitem[{{Ferland} {et~al.}(2017){Ferland}, {Chatzikos}, {Guzm{\'a}n}, {Lykins}, {van Hoof}, {Williams}, {Abel}, {Badnell}, {Keenan}, {Porter}, \& {Stancil}}]{ferland17}
{Ferland}, G.~J., {Chatzikos}, M., {Guzm{\'a}n}, F., {et~al.} 2017, \rmxaa, 53, 385, \dodoi{10.48550/arXiv.1705.10877}

\bibitem[{{Fern{\'a}ndez} {et~al.}(2019){Fern{\'a}ndez}, {Terlevich}, {D{\'\i}az}, \& {Terlevich}}]{fernandez19}
{Fern{\'a}ndez}, V., {Terlevich}, E., {D{\'\i}az}, A.~I., \& {Terlevich}, R. 2019, \mnras, 487, 3221, \dodoi{10.1093/mnras/stz1433}

\bibitem[{{Fields} {et~al.}(2020){Fields}, {Olive}, {Yeh}, \& {Young}}]{fields20}
{Fields}, B.~D., {Olive}, K.~A., {Yeh}, T.-H., \& {Young}, C. 2020, \jcap, 2020, 010, \dodoi{10.1088/1475-7516/2020/03/010}

\bibitem[{{Gonz{\'a}lez Delgado} {et~al.}(2005){Gonz{\'a}lez Delgado}, {Cervi{\~n}o}, {Martins}, {Leitherer}, \& {Hauschildt}}]{gonzalez-delgado05}
{Gonz{\'a}lez Delgado}, R.~M., {Cervi{\~n}o}, M., {Martins}, L.~P., {Leitherer}, C., \& {Hauschildt}, P.~H. 2005, \mnras, 357, 945, \dodoi{10.1111/j.1365-2966.2005.08692.x}

\bibitem[{{Gordon} {et~al.}(2003){Gordon}, {Clayton}, {Misselt}, {Landolt}, \& {Wolff}}]{gordon03}
{Gordon}, K.~D., {Clayton}, G.~C., {Misselt}, K.~A., {Landolt}, A.~U., \& {Wolff}, M.~J. 2003, \apj, 594, 279, \dodoi{10.1086/376774}

\bibitem[{{G{\"o}tberg} {et~al.}(2020){G{\"o}tberg}, {de Mink}, {McQuinn}, {Zapartas}, {Groh}, \& {Norman}}]{gotberg20}
{G{\"o}tberg}, Y., {de Mink}, S.~E., {McQuinn}, M., {et~al.} 2020, \aap, 634, A134, \dodoi{10.1051/0004-6361/201936669}

\bibitem[{{Gunasekera} {et~al.}(2023){Gunasekera}, {van Hoof}, {Chatzikos}, \& {Ferland}}]{gunasekera23}
{Gunasekera}, C.~M., {van Hoof}, P. A.~M., {Chatzikos}, M., \& {Ferland}, G.~J. 2023, Research Notes of the American Astronomical Society, 7, 246, \dodoi{10.3847/2515-5172/ad0e75}

\bibitem[{{Hao} {et~al.}(2011){Hao}, {Kennicutt}, {Johnson}, {Calzetti}, {Dale}, \& {Moustakas}}]{hao11}
{Hao}, C.-N., {Kennicutt}, R.~C., {Johnson}, B.~D., {et~al.} 2011, \apj, 741, 124, \dodoi{10.1088/0004-637X/741/2/124}

\bibitem[{{Harikane} {et~al.}(2025){Harikane}, {Sanders}, {Ellis}, {Jones}, {Ouchi}, {Laporte}, {Roberts-Borsani}, {Katz}, {Nakajima}, {Ono}, \& {Gupta}}]{harikane25}
{Harikane}, Y., {Sanders}, R.~L., {Ellis}, R., {et~al.} 2025, arXiv e-prints, arXiv:2505.09186, \dodoi{10.48550/arXiv.2505.09186}

\bibitem[{Harris {et~al.}(2020)Harris, Millman, van~der Walt, Gommers, Virtanen, Cournapeau, Wieser, Taylor, Berg, Smith, Kern, Picus, Hoyer, van Kerkwijk, Brett, Haldane, del R{\'{i}}o, Wiebe, Peterson, G{\'{e}}rard-Marchant, Sheppard, Reddy, Weckesser, Abbasi, Gohlke, \& Oliphant}]{harris20}
Harris, C.~R., Millman, K.~J., van~der Walt, S.~J., {et~al.} 2020, Nature, 585, 357, \dodoi{10.1038/s41586-020-2649-2}

\bibitem[{{Heger} {et~al.}(2003){Heger}, {Fryer}, {Woosley}, {Langer}, \& {Hartmann}}]{heger03}
{Heger}, A., {Fryer}, C.~L., {Woosley}, S.~E., {Langer}, N., \& {Hartmann}, D.~H. 2003, \apj, 591, 288, \dodoi{10.1086/375341}

\bibitem[{{Heintz} {et~al.}(2025){Heintz}, {Brammer}, {Watson}, {et~al.}}]{heintz2025}
{Heintz}, K.~E., {Brammer}, G.~B., {Watson}, D., {et~al.} 2025, \aap, 693, A60, \dodoi{10.1051/0004-6361/202450243}

\bibitem[{{Hsyu} {et~al.}(2020){Hsyu}, {Cooke}, {Prochaska}, \& {Bolte}}]{hsyu20}
{Hsyu}, T., {Cooke}, R.~J., {Prochaska}, J.~X., \& {Bolte}, M. 2020, \apj, 896, 77, \dodoi{10.3847/1538-4357/ab91af}

\bibitem[{{Hudson} {et~al.}(2012){Hudson}, {Ramsbottom}, \& {Scott}}]{hudson2012}
{Hudson}, C.~E., {Ramsbottom}, C.~A., \& {Scott}, M.~P. 2012, \apj, 750, 65, \dodoi{10.1088/0004-637X/750/1/65}

\bibitem[{{Izotov} \& {Thuan}(1998)}]{izotov98}
{Izotov}, Y.~I., \& {Thuan}, T.~X. 1998, \apj, 500, 188, \dodoi{10.1086/305698}

\bibitem[{{Izotov} \& {Thuan}(2004)}]{izotov04}
---. 2004, \apj, 602, 200, \dodoi{10.1086/380830}

\bibitem[{Izotov \& Thuan(2007)}]{izotov07}
Izotov, Y.~I., \& Thuan, T.~X. 2007, \apj, 665, 1115

\bibitem[{{Izotov} {et~al.}(2014){Izotov}, {Thuan}, \& {Guseva}}]{izotov14}
{Izotov}, Y.~I., {Thuan}, T.~X., \& {Guseva}, N.~G. 2014, \mnras, 445, 778, \dodoi{10.1093/mnras/stu1771}

\bibitem[{{Izotov} {et~al.}(1994){Izotov}, {Thuan}, \& {Lipovetsky}}]{izotov94}
{Izotov}, Y.~I., {Thuan}, T.~X., \& {Lipovetsky}, V.~A. 1994, \apj, 435, 647, \dodoi{10.1086/174843}

\bibitem[{{Ji} {et~al.}(2024){Ji}, {{\"U}bler}, {Maiolino}, {D'Eugenio}, {Arribas}, {Bunker}, {Charlot}, {Perna}, {Rodr{\'\i}guez Del Pino}, {B{\"o}ker}, {Cresci}, {Curti}, {Kumari}, \& {Lamperti}}]{ji24}
{Ji}, X., {{\"U}bler}, H., {Maiolino}, R., {et~al.} 2024, arXiv e-prints, arXiv:2404.04148, \dodoi{10.48550/arXiv.2404.04148}

\bibitem[{{Kisielius} {et~al.}(2009){Kisielius}, {Storey}, {Ferland}, \& {Keenan}}]{kisielius2009}
{Kisielius}, R., {Storey}, P.~J., {Ferland}, G.~J., \& {Keenan}, F.~P. 2009, \mnras, 397, 903, \dodoi{10.1111/j.1365-2966.2009.14989.x}

\bibitem[{Kluyver {et~al.}(2016)Kluyver, Ragan-Kelley, P{\'e}rez, Granger, Bussonnier, Frederic, Kelley, Hamrick, Grout, Corlay, Ivanov, Avila, Abdalla, \& Willing}]{kluyver16}
Kluyver, T., Ragan-Kelley, B., P{\'e}rez, F., {et~al.} 2016, in Positioning and Power in Academic Publishing: Players, Agents and Agendas, ed. F.~Loizides \& B.~Schmidt, IOS Press, 87 -- 90

\bibitem[{{Komatsu} {et~al.}(2011){Komatsu}, {Smith}, {Dunkley}, {Bennett}, {Gold}, {Hinshaw}, {Jarosik}, {Larson}, {Nolta}, {Page}, {Spergel}, {Halpern}, {Hill}, {Kogut}, {Limon}, {Meyer}, {Odegard}, {Tucker}, {Weiland}, {Wollack}, \& {Wright}}]{komatsu11}
{Komatsu}, E., {Smith}, K.~M., {Dunkley}, J., {et~al.} 2011, \apjs, 192, 18, \dodoi{10.1088/0067-0049/192/2/18}

\bibitem[{{Kraft}(1994)}]{kraft94}
{Kraft}, R.~P. 1994, \pasp, 106, 553, \dodoi{10.1086/133416}

\bibitem[{{Kriek} {et~al.}(2009){Kriek}, {van Dokkum}, {Labb{\'e}}, {Franx}, {Illingworth}, {Marchesini}, \& {Quadri}}]{kriek09}
{Kriek}, M., {van Dokkum}, P.~G., {Labb{\'e}}, I., {et~al.} 2009, \apj, 700, 221, \dodoi{10.1088/0004-637X/700/1/221}

\bibitem[{{Kruijssen} {et~al.}(2019){Kruijssen}, {Pfeffer}, {Reina-Campos}, {Crain}, \& {Bastian}}]{kruijssen19}
{Kruijssen}, J.~M.~D., {Pfeffer}, J.~L., {Reina-Campos}, M., {Crain}, R.~A., \& {Bastian}, N. 2019, \mnras, 486, 3180, \dodoi{10.1093/mnras/sty1609}

\bibitem[{{Krumholz} \& {Ting}(2018)}]{krumholz18}
{Krumholz}, M.~R., \& {Ting}, Y.-S. 2018, \mnras, 475, 2236, \dodoi{10.1093/mnras/stx3286}

\bibitem[{{Leitherer} {et~al.}(1999){Leitherer}, {Schaerer}, {Goldader}, {et~al.}}]{leitherer99}
{Leitherer}, C., {Schaerer}, D., {Goldader}, J.~D., {et~al.} 1999, \apjs, 123, 3

\bibitem[{{Luridiana} {et~al.}(2015){Luridiana}, {Morisset}, \& {Shaw}}]{luridiana15}
{Luridiana}, V., {Morisset}, C., \& {Shaw}, R.~A. 2015, \aap, 573, A42

\bibitem[{{Madau} \& {Dickinson}(2014)}]{madau14}
{Madau}, P., \& {Dickinson}, M. 2014, \araa, 52, 415, \dodoi{10.1146/annurev-astro-081811-125615}

\bibitem[{{Magg} {et~al.}(2022){Magg}, {Bergemann}, {Serenelli}, {Bautista}, {Plez}, {Heiter}, {Gerber}, {Ludwig}, {Basu}, {Ferguson}, {Gallego}, {Gamrath}, {Palmeri}, \& {Quinet}}]{magg22}
{Magg}, E., {Bergemann}, M., {Serenelli}, A., {et~al.} 2022, \aap, 661, A140, \dodoi{10.1051/0004-6361/202142971}

\bibitem[{{Maiolino} \& {Mannucci}(2019)}]{maiolino19}
{Maiolino}, R., \& {Mannucci}, F. 2019, \aapr, 27, 3, \dodoi{10.1007/s00159-018-0112-2}

\bibitem[{{Marino} {et~al.}(2019){Marino}, {Milone}, {Renzini}, \& {others}}]{marino19}
{Marino}, A.~F., {Milone}, A.~P., {Renzini}, A., \& {others}. 2019, \mnras, 487, 3815, \dodoi{10.1093/mnras/stz1415}

\bibitem[{{Marques-Chaves} {et~al.}(2024){Marques-Chaves}, {Schaerer}, {Kuruvanthodi}, {Korber}, {Prantzos}, {Charbonnel}, {Weibel}, {Izotov}, {Messa}, {Brammer}, {Dessauges-Zavadsky}, \& {Oesch}}]{marques-chaves24}
{Marques-Chaves}, R., {Schaerer}, D., {Kuruvanthodi}, A., {et~al.} 2024, \aap, 681, A30, \dodoi{10.1051/0004-6361/202347411}

\bibitem[{{Martinez} {et~al.}(2025){Martinez}, {Berg}, {James}, \& {others}}]{martinez25}
{Martinez}, Z., {Berg}, D., {James}, B., \& {others}. 2025, \apj

\bibitem[{{Mathis}(1982)}]{mathis82}
{Mathis}, J.~S. 1982, \apj, 261, 195, \dodoi{10.1086/160330}

\bibitem[{Matteucci \& Tosi(1985)}]{matteucci85}
Matteucci, F., \& Tosi, M. 1985, \mnras, 217, 391

\bibitem[{{M{\'e}ndez-Delgado} {et~al.}(2022){M{\'e}ndez-Delgado}, {Amayo}, {Arellano-C{\'o}rdova}, {Esteban}, {Garc{\'\i}a-Rojas}, {Carigi}, \& {Delgado-Inglada}}]{mendez-delgado22}
{M{\'e}ndez-Delgado}, J.~E., {Amayo}, A., {Arellano-C{\'o}rdova}, K.~Z., {et~al.} 2022, \mnras, 510, 4436, \dodoi{10.1093/mnras/stab3782}

\bibitem[{{M\'endez-Delgado} {et~al.}(2024){M\'endez-Delgado}, {Skillman}, {Aver}, {Morisset}, {Esteban}, {Garc\textbackslash'ia-Rojas}, {Kreckel}, {Rogers}, {Rosales-Ortega}, {Arellano-C\textbackslash'ordova}, {Flury}, {Reyes-Rodr\textbackslash'iguez}, {Orte-Garc\textbackslash'ia}, \& {Tan}}]{mendez-delgado24}
{M\'endez-Delgado}, J.~E., {Skillman}, E.~D., {Aver}, E., {et~al.} 2024, arXiv e-prints, arXiv:2410.17381, \dodoi{10.48550/arXiv.2410.17381}

\bibitem[{{Mihalas}(1972)}]{mihalas72}
{Mihalas}, D. 1972, {Non-LTE model atmospheres for B and O stars.}

\bibitem[{{Milone}(2015)}]{milone15}
{Milone}, A.~P. 2015, \mnras, 446, 1672, \dodoi{10.1093/mnras/stu2198}

\bibitem[{{Milone} \& {Marino}(2022)}]{milone22}
{Milone}, A.~P., \& {Marino}, A.~F. 2022, Universe, 8, 359, \dodoi{10.3390/universe8070359}

\bibitem[{{Milone} {et~al.}(2018){Milone}, {Marino}, {Renzini}, \& {others}}]{milone18}
{Milone}, A.~P., {Marino}, A.~F., {Renzini}, A., \& {others}. 2018, \mnras, 481, 5098, \dodoi{10.1093/mnras/sty2573}

\bibitem[{{Mingozzi} {et~al.}(2022){Mingozzi}, {Berg}, {Chisholm}, {James}, {Heckman}, {et~al.}}]{mingozzi22}
{Mingozzi}, M., {Berg}, D.~A., {Chisholm}, J., {et~al.} 2022, \apj

\bibitem[{{Olive} \& {Skillman}(2004)}]{olive04}
{Olive}, K.~A., \& {Skillman}, E.~D. 2004, \apj, 617, 29, \dodoi{10.1086/425170}

\bibitem[{{Osterbrock} \& {Ferland}(2006)}]{osterbrock06}
{Osterbrock}, D.~E., \& {Ferland}, G.~J. 2006, {Astrophysics of gaseous nebulae and active galactic nuclei} (University Science Books)

\bibitem[{{Pagel} {et~al.}(1992){Pagel}, {Simonson}, {Terlevich}, \& {Edmunds}}]{pagel92}
{Pagel}, B.~E.~J., {Simonson}, E.~A., {Terlevich}, R.~J., \& {Edmunds}, M.~G. 1992, \mnras, 255, 325, \dodoi{10.1093/mnras/255.2.325}

\bibitem[{{Pascale} {et~al.}(2023){Pascale}, {Dai}, {McKee}, \& {Tsang}}]{pascale23}
{Pascale}, M., {Dai}, L., {McKee}, C.~F., \& {Tsang}, B. T.~H. 2023, \apj, 957, 77, \dodoi{10.3847/1538-4357/acf75c}

\bibitem[{{Pasquini} {et~al.}(2011){Pasquini}, {Mauas}, {K{\"a}ufl}, \& {Cacciari}}]{pasquini11}
{Pasquini}, L., {Mauas}, P., {K{\"a}ufl}, H.~U., \& {Cacciari}, C. 2011, \aap, 531, A35, \dodoi{10.1051/0004-6361/201116592}

\bibitem[{{Peimbert} {et~al.}(2016){Peimbert}, {Peimbert}, \& {Luridiana}}]{peimbert16}
{Peimbert}, A., {Peimbert}, M., \& {Luridiana}, V. 2016, \rmxaa, 52, 419, \dodoi{10.48550/arXiv.1608.02062}

\bibitem[{{Peimbert} {et~al.}(2007){Peimbert}, {Luridiana}, \& {Peimbert}}]{peimbert07}
{Peimbert}, M., {Luridiana}, V., \& {Peimbert}, A. 2007, \apj, 666, 636, \dodoi{10.1086/520571}

\bibitem[{{Peimbert} \& {Torres-Peimbert}(1974)}]{peimbert74}
{Peimbert}, M., \& {Torres-Peimbert}, S. 1974, \apj, 193, 327, \dodoi{10.1086/153166}

\bibitem[{{Piotto} {et~al.}(2007){Piotto}, {Bedin}, {Anderson}, {King}, {Cassisi}, {Milone}, {Villanova}, {Pietrinferni}, \& {Renzini}}]{piotto07}
{Piotto}, G., {Bedin}, L.~R., {Anderson}, J., {et~al.} 2007, \apjl, 661, L53, \dodoi{10.1086/518503}

\bibitem[{{Planck Collaboration} {et~al.}(2014){Planck Collaboration}, {Ade}, {Aghanim}, {Armitage-Caplan}, {Arnaud}, {Ashdown}, {Atrio-Barandela}, {Aumont}, {Baccigalupi}, {Banday}, {Barreiro}, {Bartlett}, {Battaner}, {Benabed}, {Beno{\^\i}t}, {Benoit-L{\'e}vy}, {Bernard}, {Bersanelli}, {Bielewicz}, {Bobin}, {Bock}, {Bonaldi}, {Bond}, {Borrill}, {Bouchet}, {Bridges}, {Bucher}, {Burigana}, {Butler}, {Calabrese}, {Cappellini}, {Cardoso}, {Catalano}, {Challinor}, {Chamballu}, {Chary}, {Chen}, {Chiang}, {Chiang}, {Christensen}, {Church}, {Clements}, {Colombi}, {Colombo}, {Couchot}, {Coulais}, {Crill}, {Curto}, {Cuttaia}, {Danese}, {Davies}, {Davis}, {de Bernardis}, {de Rosa}, {de Zotti}, {Delabrouille}, {Delouis}, {D{\'e}sert}, {Dickinson}, {Diego}, {Dolag}, {Dole}, {Donzelli}, {Dor{\'e}}, {Douspis}, {Dunkley}, {Dupac}, {Efstathiou}, {Elsner}, {En{\ss}lin}, {Eriksen}, {Finelli}, {Forni}, {Frailis}, {Fraisse}, {Franceschi}, {Gaier}, {Galeotta}, {Galli}, {Ganga}, {Giard}, {Giardino}, {Giraud-H{\'e}raud},
  {Gjerl{\o}w}, {Gonz{\'a}lez-Nuevo}, {G{\'o}rski}, {Gratton}, {Gregorio}, {Gruppuso}, {Gudmundsson}, {Haissinski}, {Hamann}, {Hansen}, {Hanson}, {Harrison}, {Henrot-Versill{\'e}}, {Hern{\'a}ndez-Monteagudo}, {Herranz}, {Hildebrandt}, {Hivon}, {Hobson}, {Holmes}, {Hornstrup}, {Hou}, {Hovest}, {Huffenberger}, {Jaffe}, {Jaffe}, {Jewell}, {Jones}, {Juvela}, {Keih{\"a}nen}, {Keskitalo}, {Kisner}, {Kneissl}, {Knoche}, {Knox}, {Kunz}, {Kurki-Suonio}, {Lagache}, {L{\"a}hteenm{\"a}ki}, {Lamarre}, {Lasenby}, {Lattanzi}, {Laureijs}, {Lawrence}, {Leach}, {Leahy}, {Leonardi}, {Le{\'o}n-Tavares}, {Lesgourgues}, {Lewis}, {Liguori}, {Lilje}, {Linden-V{\o}rnle}, {L{\'o}pez-Caniego}, {Lubin}, {Mac{\'\i}as-P{\'e}rez}, {Maffei}, {Maino}, {Mandolesi}, {Maris}, {Marshall}, {Martin}, {Mart{\'\i}nez-Gonz{\'a}lez}, {Masi}, {Massardi}, {Matarrese}, {Matthai}, {Mazzotta}, {Meinhold}, {Melchiorri}, {Melin}, {Mendes}, {Menegoni}, {Mennella}, {Migliaccio}, {Millea}, {Mitra}, {Miville-Desch{\^e}nes}, {Moneti}, {Montier}, {Morgante},
  {Mortlock}, {Moss}, {Munshi}, {Murphy}, {Naselsky}, {Nati}, {Natoli}, {Netterfield}, {N{\o}rgaard-Nielsen}, {Noviello}, {Novikov}, {Novikov}, {O'Dwyer}, {Osborne}, {Oxborrow}, {Paci}, {Pagano}, {Pajot}, {Paladini}, {Paoletti}, {Partridge}, {Pasian}, {Patanchon}, {Pearson}, {Pearson}, {Peiris}, {Perdereau}, {Perotto}, {Perrotta}, {Pettorino}, {Piacentini}, {Piat}, {Pierpaoli}, {Pietrobon}, {Plaszczynski}, {Platania}, \& {Pointecouteau}}]{planck14}
{Planck Collaboration}, {Ade}, P.~A.~R., {Aghanim}, N., {et~al.} 2014, \aap, 571, A16, \dodoi{10.1051/0004-6361/201321591}

\bibitem[{{Planck Collaboration} {et~al.}(2020){Planck Collaboration}, {Aghanim}, {Akrami}, {Ashdown}, {Aumont}, {Baccigalupi}, {Ballardini}, {Banday}, {Barreiro}, {Bartolo}, {Basak}, {Battye}, {Benabed}, {Bernard}, {Bersanelli}, {Bielewicz}, {Bock}, {Bond}, {Borrill}, {Bouchet}, {Boulanger}, {Bucher}, {Burigana}, {Butler}, {Calabrese}, {Cardoso}, {Carron}, {Challinor}, {Chiang}, {Chluba}, {Colombo}, {Combet}, {Contreras}, {Crill}, {Cuttaia}, {de Bernardis}, {de Zotti}, {Delabrouille}, {Delouis}, {Di Valentino}, {Diego}, {Dor{\'e}}, {Douspis}, {Ducout}, {Dupac}, {Dusini}, {Efstathiou}, {Elsner}, {En{\ss}lin}, {Eriksen}, {Fantaye}, {Farhang}, {Fergusson}, {Fernandez-Cobos}, {Finelli}, {Forastieri}, {Frailis}, {Fraisse}, {Franceschi}, {Frolov}, {Galeotta}, {Galli}, {Ganga}, {G{\'e}nova-Santos}, {Gerbino}, {Ghosh}, {Gonz{\'a}lez-Nuevo}, {G{\'o}rski}, {Gratton}, {Gruppuso}, {Gudmundsson}, {Hamann}, {Handley}, {Hansen}, {Herranz}, {Hildebrandt}, {Hivon}, {Huang}, {Jaffe}, {Jones}, {Karakci}, {Keih{\"a}nen},
  {Keskitalo}, {Kiiveri}, {Kim}, {Kisner}, {Knox}, {Krachmalnicoff}, {Kunz}, {Kurki-Suonio}, {Lagache}, {Lamarre}, {Lasenby}, {Lattanzi}, {Lawrence}, {Le Jeune}, {Lemos}, {Lesgourgues}, {Levrier}, {Lewis}, {Liguori}, {Lilje}, {Lilley}, {Lindholm}, {L{\'o}pez-Caniego}, {Lubin}, {Ma}, {Mac{\'\i}as-P{\'e}rez}, {Maggio}, {Maino}, {Mandolesi}, {Mangilli}, {Marcos-Caballero}, {Maris}, {Martin}, {Martinelli}, {Mart{\'\i}nez-Gonz{\'a}lez}, {Matarrese}, {Mauri}, {McEwen}, {Meinhold}, {Melchiorri}, {Mennella}, {Migliaccio}, {Millea}, {Mitra}, {Miville-Desch{\^e}nes}, {Molinari}, {Montier}, {Morgante}, {Moss}, {Natoli}, {N{\o}rgaard-Nielsen}, {Pagano}, {Paoletti}, {Partridge}, {Patanchon}, {Peiris}, {Perrotta}, {Pettorino}, {Piacentini}, {Polastri}, {Polenta}, {Puget}, {Rachen}, {Reinecke}, {Remazeilles}, {Renzi}, {Rocha}, {Rosset}, {Roudier}, {Rubi{\~n}o-Mart{\'\i}n}, {Ruiz-Granados}, {Salvati}, {Sandri}, {Savelainen}, {Scott}, {Shellard}, {Sirignano}, {Sirri}, {Spencer}, {Sunyaev}, {Suur-Uski}, {Tauber}, {Tavagnacco},
  {Tenti}, {Toffolatti}, {Tomasi}, {Trombetti}, {Valenziano}, {Valiviita}, {Van Tent}, {Vibert}, {Vielva}, {Villa}, {Vittorio}, {Wandelt}, {Wehus}, {White}, {White}, {Zacchei}, \& {Zonca}}]{planck20}
{Planck Collaboration}, {Aghanim}, N., {Akrami}, Y., {et~al.} 2020, \aap, 641, A6, \dodoi{10.1051/0004-6361/201833910}

\bibitem[{{Porter} {et~al.}(2005){Porter}, {Bauman}, {Ferland}, \& {MacAdam}}]{porter05}
{Porter}, R.~L., {Bauman}, R.~P., {Ferland}, G.~J., \& {MacAdam}, K.~B. 2005, \apjl, 622, L73, \dodoi{10.1086/429370}

\bibitem[{{Porter} {et~al.}(2009){Porter}, {Ferland}, {MacAdam}, \& {Storey}}]{porter09}
{Porter}, R.~L., {Ferland}, G.~J., {MacAdam}, K.~B., \& {Storey}, P.~J. 2009, \mnras, 393, L36, \dodoi{10.1111/j.1745-3933.2008.00593.x}

\bibitem[{{Porter} {et~al.}(2012){Porter}, {Ferland}, {Storey}, \& {Detisch}}]{porter12}
{Porter}, R.~L., {Ferland}, G.~J., {Storey}, P.~J., \& {Detisch}, M.~J. 2012, \mnras, 425, L28, \dodoi{10.1111/j.1745-3933.2012.01300.x}

\bibitem[{{Porter} {et~al.}(2013){Porter}, {Ferland}, {Storey}, \& {Detisch}}]{porter13}
---. 2013, \mnras, 433, L89, \dodoi{10.1093/mnrasl/slt049}

\bibitem[{{Rauscher}(2024)}]{rauscher24}
{Rauscher}, B.~J. 2024, \pasp, 136, 015001, \dodoi{10.1088/1538-3873/ad1b36}

\bibitem[{{Reddy} {et~al.}(2023){Reddy}, {Topping}, {Sanders}, {Shapley}, \& {Brammer}}]{reddy23}
{Reddy}, N.~A., {Topping}, M.~W., {Sanders}, R.~L., {Shapley}, A.~E., \& {Brammer}, G. 2023, \apj, 948, 83, \dodoi{10.3847/1538-4357/acc869}

\bibitem[{{Romano} {et~al.}(2010){Romano}, {Karakas}, {Tosi}, \& {Matteucci}}]{romano10}
{Romano}, D., {Karakas}, A.~I., {Tosi}, M., \& {Matteucci}, F. 2010, \aap, 522, A32, \dodoi{10.1051/0004-6361/201014483}

\bibitem[{{Sanders}(2025)}]{sanders25}
{Sanders}, R.~L. 2025, \apj

\bibitem[{{Sanders} {et~al.}(2021){Sanders}, {Shapley}, {Jones}, {Reddy}, {Kriek}, \& {others}}]{sanders21}
{Sanders}, R.~L., {Shapley}, A.~E., {Jones}, T., {et~al.} 2021, \apj, 914, 19, \dodoi{10.3847/1538-4357/abf4c1}

\bibitem[{{Sanders} {et~al.}(2016){Sanders}, {Shapley}, {Kriek}, \& {others}}]{sanders16}
{Sanders}, R.~L., {Shapley}, A.~E., {Kriek}, M., \& {others}. 2016, \apj, 816, 23

\bibitem[{{Sanders} {et~al.}(2020){Sanders}, {Shapley}, {Reddy}, {Kriek}, {Siana}, \& {others}}]{sanders20}
{Sanders}, R.~L., {Shapley}, A.~E., {Reddy}, N.~A., {et~al.} 2020, \mnras, 491, 1427, \dodoi{10.1093/mnras/stz3032}

\bibitem[{{Sanders} {et~al.}(2024){Sanders}, {Shapley}, {Topping}, {Reddy}, {Berg}, \& {others}}]{sanders24b}
{Sanders}, R.~L., {Shapley}, A.~E., {Topping}, M.~W., {et~al.} 2024, arXiv e-prints, arXiv:2408.05273, \dodoi{10.48550/arXiv.2408.05273}

\bibitem[{{Sauer} \& {Jedamzik}(2002)}]{sauer02}
{Sauer}, D., \& {Jedamzik}, K. 2002, \aap, 381, 361, \dodoi{10.1051/0004-6361:20011567}

\bibitem[{{Schaerer}(2003)}]{schaerer03}
{Schaerer}, D. 2003, \aap, 397, 527

\bibitem[{{Schaerer} {et~al.}(2024){Schaerer}, {Marques-Chaves}, {Xiao}, \& {Korber}}]{schaerer24}
{Schaerer}, D., {Marques-Chaves}, R., {Xiao}, M., \& {Korber}, D. 2024, \aap, 687, L11, \dodoi{10.1051/0004-6361/202450721}

\bibitem[{{Senchyna} {et~al.}(2023){Senchyna}, {Plat}, {Stark}, \& {Rudie}}]{senchyna24}
{Senchyna}, P., {Plat}, A., {Stark}, D.~P., \& {Rudie}, G.~C. 2023, arXiv e-prints, arXiv:2303.04179, \dodoi{10.48550/arXiv.2303.04179}

\bibitem[{{Serenelli} \& {Basu}(2010)}]{serenelli10}
{Serenelli}, A.~M., \& {Basu}, S. 2010, \apj, 719, 865, \dodoi{10.1088/0004-637X/719/1/865}

\bibitem[{{Shapley} {et~al.}(2025{\natexlab{a}}){Shapley}, {Sanders}, {Topping}, {Reddy}, {Berg}, {et~al.}}]{shapley25a}
{Shapley}, A.~E., {Sanders}, R.~L., {Topping}, M.~W., {et~al.} 2025{\natexlab{a}}, \apj, 980, 242, \dodoi{10.3847/1538-4357/adad68}

\bibitem[{{Shapley} {et~al.}(2025{\natexlab{b}}){Shapley}, {Sanders}, {Topping}, {Reddy}, {Pahl}, {et~al.}}]{shapley25b}
---. 2025{\natexlab{b}}, \apj, 981, 167, \dodoi{10.3847/1538-4357/adaf98}

\bibitem[{{Skelton} {et~al.}(2014){Skelton}, {Whitaker}, {Momcheva}, {Brammer}, {van Dokkum}, {Labb{\'e}}, {Franx}, {van der Wel}, {Bezanson}, \& {Da Cunha}}]{skelton2014}
{Skelton}, R.~E., {Whitaker}, K.~E., {Momcheva}, I.~G., {et~al.} 2014, \apjs, 214, 24, \dodoi{10.1088/0067-0049/214/2/24}

\bibitem[{{Spera} \& {Mapelli}(2017)}]{spera17}
{Spera}, M., \& {Mapelli}, M. 2017, \mnras, 470, 4739, \dodoi{10.1093/mnras/stx1576}

\bibitem[{{Steigman}(2007)}]{steigman07}
{Steigman}, G. 2007, Annual Review of Nuclear and Particle Science, 57, 463, \dodoi{10.1146/annurev.nucl.56.080805.140437}

\bibitem[{{Storey} \& {Hummer}(1995)}]{storey95}
{Storey}, P.~J., \& {Hummer}, D.~G. 1995, \mnras, 272, 41, \dodoi{10.1093/mnras/272.1.41}

\bibitem[{{Sz{\'e}csi} {et~al.}(2015){Sz{\'e}csi}, {Langer}, {Yoon}, {Sanyal}, {de Mink}, {Evans}, \& {Dermine}}]{szesci15}
{Sz{\'e}csi}, D., {Langer}, N., {Yoon}, S.-C., {et~al.} 2015, \aap, 581, A15, \dodoi{10.1051/0004-6361/201526617}

\bibitem[{{Tayal}(2011)}]{tayal11}
{Tayal}, S.~S. 2011, \apjs, 195, 12

\bibitem[{{Tayal} \& {Zatsarinny}(2010)}]{tayal2010}
{Tayal}, S.~S., \& {Zatsarinny}, O. 2010, \apjs, 188, 32, \dodoi{10.1088/0067-0049/188/1/32}

\bibitem[{{Topping} {et~al.}(2025){Topping}, {Sanders}, {Shapley}, {Pahl}, {Reddy}, {et~al.}}]{topping25b}
{Topping}, M.~W., {Sanders}, R.~L., {Shapley}, A.~E., {et~al.} 2025, arXiv e-prints, arXiv:2502.08712, \dodoi{10.48550/arXiv.2502.08712}

\bibitem[{{Topping} {et~al.}(2024){Topping}, {Stark}, {Senchyna}, {Plat}, {Zitrin}, \& {others}}]{topping24a}
{Topping}, M.~W., {Stark}, D.~P., {Senchyna}, P., {et~al.} 2024, \mnras, 529, 3301, \dodoi{10.1093/mnras/stae682}

\bibitem[{{Vilchez} {et~al.}(1988){Vilchez}, {Pagel}, {Diaz}, {Terlevich}, \& {Edmunds}}]{vilchez88}
{Vilchez}, J.~M., {Pagel}, B.~E.~J., {Diaz}, A.~I., {Terlevich}, E., \& {Edmunds}, M.~G. 1988, \mnras, 235, 633

\bibitem[{{Walker} {et~al.}(1991){Walker}, {Steigman}, {Schramm}, {Olive}, \& {Kang}}]{walker91}
{Walker}, T.~P., {Steigman}, G., {Schramm}, D.~N., {Olive}, K.~A., \& {Kang}, H.-S. 1991, \apj, 376, 51, \dodoi{10.1086/170255}

\bibitem[{{Weller} {et~al.}(2025){Weller}, {Weinberg}, \& {Johnson}}]{weller25}
{Weller}, M.~K., {Weinberg}, D.~H., \& {Johnson}, J.~W. 2025, \mnras, 538, 1517, \dodoi{10.1093/mnras/staf373}

\bibitem[{{Wise} {et~al.}(2012){Wise}, {Turk}, {Norman}, \& {Abel}}]{wise12}
{Wise}, J.~H., {Turk}, M.~J., {Norman}, M.~L., \& {Abel}, T. 2012, \apj, 745, 50, \dodoi{10.1088/0004-637X/745/1/50}

\bibitem[{{Woosley}(2017)}]{woosley17}
{Woosley}, S.~E. 2017, \apj, 836, 244, \dodoi{10.3847/1538-4357/836/2/244}

\bibitem[{{Yanagisawa} {et~al.}(2024){Yanagisawa}, {Ouchi}, {Watanabe}, {Matsumoto}, {Nakajima}, {Yajima}, {Nagamine}, {Takahashi}, {Nakane}, {Tominaga}, {Umeda}, {Fukushima}, {Harikane}, {Isobe}, {Ono}, {Xu}, \& {Zhang}}]{yanagisawa24}
{Yanagisawa}, H., {Ouchi}, M., {Watanabe}, K., {et~al.} 2024, \apj, 974, 266, \dodoi{10.3847/1538-4357/ad72ec}

\bibitem[{{Yusof} {et~al.}(2013){Yusof}, {Hirschi}, {Meynet}, {Crowther}, {Ekstr{\"o}m}, {Frischknecht}, {Georgy}, {Abu Kassim}, \& {Schnurr}}]{yusof13}
{Yusof}, N., {Hirschi}, R., {Meynet}, G., {et~al.} 2013, \mnras, 433, 1114, \dodoi{10.1093/mnras/stt794}

\end{thebibliography}

\clearpage

\end{document}